\documentclass[12pt]{article}
\usepackage{epsfig}
\usepackage{amsmath}
\usepackage{latexsym}

\def\bra#1{\left<#1\right|}
\def\ket#1{\left|#1\right>}
{\catcode`\|=\active 
  \gdef\Braket#1{\left<\mathcode`\|"8000\let|\bravert {#1}\right>}}
\def\bravert{\egroup\,\vrule\,\bgroup}
\newcommand{\tr}{\mathop{\rm tr}}

\newcommand{\be}{\begin{eqnarray}}
\newcommand{\ee}{\end{eqnarray}}
\newcommand{\bea}{\begin{eqnarray}}
\newcommand{\eea}{\end{eqnarray}}
\newcommand{\ben}{\begin{equation}}
\newcommand{\een}{\end{equation}}

\newcommand{\pderiv}[1]{\frac{\partial}{\partial{#1}}}

\newcommand{\nn}{\nonumber}

\numberwithin{equation}{section}

\begin{document}

\begin{titlepage}

\begin{flushright}
CALT-68-2489\\
PUPT-2115\\
hep-th/0404007
\end{flushright}
\vspace{15 mm}

\begin{center}
{\huge Holography beyond the Penrose limit}
\end{center}

\vspace{12 mm}

\begin{center}
{\large Curtis G.\ Callan, Jr.${}^{a}$,
Tristan McLoughlin${}^{b}$,
Ian Swanson${}^{b}$ }\\

\vspace{3mm}

${}^a$ Joseph Henry Laboratories\\
Princeton University\\
Princeton, New Jersey 08544, USA\\
\vspace{0.5 cm}
${}^b$ California Institute of Technology\\
Pasadena, CA 91125, USA 
\end{center}

\vspace{5 mm}

\begin{center}
{\large Abstract}
\end{center}
\noindent

The flat pp-wave background geometry has been realized as
a particular Penrose limit of $AdS_5 \times S^5$. It describes a 
string that has been infinitely boosted along an equatorial null geodesic 
in the $S^5$ subspace. The string worldsheet Hamiltonian in this
background is free. Finite boosts lead to curvature corrections that
induce interacting perturbations of the string worldsheet Hamiltonian.  
We develop a systematic light-cone gauge quantization of the interacting
worldsheet string theory and use it to obtain the interacting spectrum
of the so-called `two-impurity' states of the string. The
quantization is technically rather intricate and we provide a detailed
account of the methods we use to extract explicit results. We give a 
systematic treatment of the fermionic states and are able to show 
that the spectrum possesses the proper extended supermultiplet structure 
(a non-trivial fact since half the supersymmetry is nonlinearly realized). 
We test holography by comparing the string energy spectrum with 
the scaling dimensions of corresponding gauge theory operators. We confirm
earlier results that agreement obtains in low orders of perturbation
theory, but breaks down at third order. The methods presented
here can be used to explore these issues in a wider context than
is specifically dealt with in this paper.

\vspace{1cm}
\begin{flushleft}
\today
\end{flushleft}
\end{titlepage}
\newpage

\section{Introduction}

The AdS/CFT correspondence encompasses a wide
range of holographic mappings between string theory
and gauge theory.  In its original incarnation, Maldacena's conjecture states
that type IIB superstring theory in $AdS_5\times S^5$ with $N_c$
units of Ramond-Ramond (RR) flux on the sphere is equivalent to 
${\cal N}=4$ supersymmetric $SU(N_c)$ Yang-Mills theory in 
$3+1$ dimensions \cite{Maldacena:1997re}. 
This correspondence is conjectured to be precise for
the identification $g_s = g_{YM}^2$.
The holographically dual gauge theory is defined on the conformal boundary
of $AdS_5 \times S^5$, or ${\bf R} \times S^3$.
Strong evidence in support of this particular correspondence
is the fact that both sides of the duality have the same
symmetry, $PSU(2,2|4)$. 
While a great deal of additional evidence in support of the conjecture
has accumulated in recent years, a direct verification of Maldacena's original
proposal has been elusive.
It is difficult to quantize the superstring theory 
using the RNS formalism in the presence of background RR fields.
The GS formalism naturally accommodates RR backgrounds but,
despite the high degree of symmetry of the $AdS_5\times S^5$ background,
a gauge choice leading to a solvable theory has not been found.
Early studies of the duality have therefore concentrated on
the supergravity approximation to the string theory (and, of course,
yield many impressive results). 

In order to address specifically stringy aspects of the duality,
it has been necessary to consider simplifying limits of the
canonical $AdS_5\times S^5$ background. Metsaev \cite{Metsaev:2001bj} 
showed that, in a certain plane-wave geometry supported by a constant 
RR flux, light-cone gauge worldsheet string theory 
reduces to a free theory with the novel feature  
that the worldsheet bosons and fermions acquire a mass.
This solution was later shown to be a Penrose limit of the familiar 
${AdS}_5 \times S^5$ supergravity solution \cite{Berenstein:2002jq},
and describes the geometry near a null geodesic boosted 
around the equator of the $S^5$ subspace. The energies of Metsaev's
free string theory are thus understood to be those of a string in 
the full ${AdS}_5 \times S^5$ space, in the limit that the states 
are boosted to large angular momentum about an equatorial circle in the $S^5$. 
Callan, Lee, McLoughlin, Schwarz, Swanson and Wu \cite{callan} subsequently
calculated the corrections (in inverse powers of the angular momentum)
to the string spectrum that arise if the string is given a large, but finite,
boost. Comparison of the resulting interacting spectrum with corrections 
(in inverse powers of ${\cal R}$-charge) to the dimensions of the corresponding 
gauge theory operators largely (but not completely) confirms expectations
from AdS/CFT duality (see \cite{callan,Beisert:2003jb} for discussion). 
The purpose of this work is not to present new results, but rather to
describe in fairly complete detail the methods used to obtain the results
presented in \cite{callan} (but only outlined in that paper). Some aspects of
the purely bosonic side of this problem were studied by Parnachev and 
Ryzhov \cite{Parnachev:2002kk}. Although we find no disagreement with them, 
our approach differs from theirs in certain respects, most notably
in taking full account of supersymmetry.

Our approach is to take the GS superstring action on $AdS_5 \times S^5$,
constructed using the formalism of Cartan forms and superconnections on the 
$SU(2,2|4)/(SO(4,1)\times SO(5))$ coset superspace \cite{Metsaev:bj}, 
expand it in powers of the background curvature and
finally eliminate unphysical degrees of freedom
by light-cone gauge quantization. We treat the resulting interaction 
Hamiltonian in first-order degenerate perturbation theory to find the
first corrections to the highly-degenerate pp-wave spectrum. The 
complexity of the problem is such that we are forced to resort to
symbolic manipulation programs to construct and diagonalize the
perturbation matrix. In this paper we give a proof of principle
by applying our methods to the subspace of two-impurity excitations
of the string. We show that the spectrum organizes itself into correct 
extended supersymmetry multiplets whose energies match well 
(if not perfectly) with what is known about gauge theory anomalous 
dimensions.

In section 2 we introduce the problem by considering the 
bosonic sector of the theory alone. We comment on some 
interesting aspects of the theory that arise when
restricting to the point-particle (or zero-mode) subsector.
In section 3 we review the construction of the GS superstring action 
on $AdS_5 \times S^5$ as a nonlinear sigma model on the 
$SU(2,2|4)/(SO(4,1)\times SO(5))$ coset superspace. In sections 4 and 5
we perform a large-radius expansion on the relevant objects in the theory,  
and carry out the light-cone gauge reduction, thereby
extracting explicit curvature corrections to the pp-wave Hamiltonian.
Section 6 presents results on the curvature-corrected energy spectrum,
further expanded to linear order in the modified 't Hooft coupling 
$\lambda^\prime = g_{YM}^2 N_c/J^2$; results from corresponding gauge theory
calculations (at one loop in $\lambda = g_{YM}^2 N_c$) are summarized 
and compared with the string theory. In section 7 we extend the string 
theory analysis to higher orders in $\lambda^\prime $, and compare 
results with what is known about gauge theory operator dimensions 
at higher-loop order. The final section is devoted to discussion and 
conclusions.  

\section{Strings beyond the Penrose limit: General considerations}

To introduce the computation of finite-$J$ corrections
to the pp-wave string spectrum, we begin by discussing the construction of the
light-cone gauge worldsheet Hamiltonian for the bosonic string in the
full $AdS_5 \times S^5$ background. The problem is much more complicated
when fermions are introduced, and we will take up that aspect of the
calculation in a later section. A study of the purely bosonic problem
gives us the opportunity to explain various strategic points in a simpler
context.

In convenient global coordinates, the ${AdS}_5 \times S^5$
metric can be written in the form
\begin{equation}
\label{adsmetricPre}
ds^2 = R^2 ( - {\rm cosh}^2 \rho\ dt^2 + d \rho^2 + {\rm sinh}^2
\rho\ d \Omega_3^2 + {\rm cos}^2  \theta\ d \phi^2 +  d \theta^2 +
{\rm sin}^2 \theta\ d \tilde\Omega_3^2)~,
\end{equation}
where $R$ denotes the radius of both the sphere and the AdS space,
and $d\Omega_3^2$, $d\tilde\Omega_3^2$ denote separate three-spheres.
The coordinate $\phi$ is periodic with period $2\pi$ and, strictly speaking, 
so is the time coordinate $t$. In order to
accommodate string dynamics, it is necessary to pass to the
covering space in which time is {\sl not} taken to be periodic.
This geometry, supplemented by an RR field with $N_c$ units of flux
on the sphere, is a consistent, maximally supersymmetric type IIB
superstring background, provided that $R^4 =  g_s N_c
(\alpha^{\prime})^2$ (where $g_s$ is the string coupling). 

In its initial stages, development of the AdS/CFT correspondence
focused on the supergravity approximation to string theory in
$AdS^5\times S^5$. Recently, attention has turned to the problem of
evaluating truly stringy physics in this background and studying its
match to gauge theory physics. The obstacles to such a program, of
course, are the general difficulty of quantizing strings in curved
geometries, and the particular problem of defining the superstring in
the presence of RR background fields. As noted above, the string
quantization problem is partly solved by looking at the dynamics
of a string that has been boosted to lightlike momentum along
some direction, or, equivalently, by quantizing the string in the
background obtained by taking the Penrose limit of the original
geometry using the lightlike geodesic corresponding to the boosted
trajectory. The simplest choice is to boost along the equator of 
the $S^5$ or, equivalently, to take the Penrose limit with respect 
to the lightlike geodesic $\phi=t,~\rho=\theta=0$ and to quantize
the system in the appropriate light-cone gauge.

To quantize about the lightlike geodesic at $\rho=\theta=0$, 
it is helpful to make the reparameterizations
\begin{eqnarray}
    \cosh\rho  =  \frac{1+z^2/4}{1-z^2/4} \qquad
    \cos\theta  =  \frac{1-y^2/4}{1+ y^2/4}\ ,
\end{eqnarray}
and work with the metric
\begin{equation}
\label{metric}
ds^2  =  R^2
\biggl[ -\left({1+ \frac{1}{4}z^2\over 1-\frac{1}{4}z^2}\right)^2dt^2
        +\left({1-\frac{1}{4}y^2\over 1+\frac{1}{4}y^2}\right)^2d\phi^2
    + \frac{d z_k dz_k}{(1-\frac{1}{4}z^2)^{2}}
    + \frac{dy_{k'} dy_{k'}}{(1+\frac{1}{4}y^2)^{2}} \biggr]~.
\end{equation}
The $SO(8)$ vectors spanning the eight directions
transverse to the geodesic are broken into two $SO(4)$ subgroups 
parameterized by $z^2 = z_k z^k$ with $k=1,\dots,4$, and $y^2 = y_{k'} y^{k'}$ 
with $k'=5,\dots,8$.
This form of the metric is well-suited for the present calculation: the spin
connection, which will be important for the superstring action, turns out to
have a simple functional form and the $AdS_5$ and $S^5$ subspaces appear
nearly symmetrically.  This metric has the full $SO(4,2) \times SO(6)$
symmetry associated with $AdS_5 \times S^5$, but only the translation symmetries
in $t$ and $\phi$ and the $SO(4)\times SO(4)$ symmetry of the transverse
coordinates remain manifest.
The translation symmetries mean that string states have a conserved energy
$\omega$, conjugate to $t$, and a conserved (integer) angular momentum $J$,
conjugate to $\phi$. Boosting along the equatorial geodesic is equivalent
to studying states with large $J$, and the lightcone Hamiltonian gives
eigenvalues for $\omega-J$ in that limit. On the gauge theory side, the $S^5$
geometry is replaced by an $SO(6)$ ${\cal R}$-symmetry, and $J$ corresponds to the
eigenvalue of an $SO(2)$ ${\cal R}$-symmetry generator. The AdS/CFT correspondence
implies that string energies in the boosted limit should match operator
dimensions in the limit of large ${\cal R}$-charge (a limit in which perturbative 
evaluation of operator dimensions becomes legitimate).

On dimensional grounds, taking the $J\to\infty$ limit on the string states is
equivalent to taking the $R\to\infty$ limit on the metric (in the
right coordinates). The coordinate redefinitions
\begin{eqnarray}
\label{rescalePre}
    t \rightarrow x^+ - \frac{x^-}{2 R^2}
\qquad
    \phi \rightarrow x^+ + \frac{x^-}{2 R^2}
\qquad
    z_k \rightarrow \frac{z_k}{R}
\qquad
    y_{k'} \rightarrow \frac{y_{k'}}{R}\ 
\end{eqnarray}
make it possible to take a smooth $R\to\infty$ limit. Expressing
the metric (\ref{metric}) in these new coordinates, we obtain the
following expansion in powers of $1/R^2$:
\begin{eqnarray}
\label{expndmet}
ds^2 & \approx &
2\,{dx^+}{dx^-} + {dz }^2 + {dy }^2  -
        \left( {z }^2 + {y }^2 \right) ({dx}^+)^2 + \nonumber \\
& &     \left[ 2\left( z^2  - y^2  \right) dx^- dx^+
    + z^2 dz^2 - y^2 dy^2 -
    \left( {z }^4 - {y }^4 \right) (dx^+)^2 \right]
    \frac{1}{2 R^2} \nonumber \\
& &     + {\cal O}(1/R^4)\ .
\end{eqnarray}
The leading $R$-independent part is the well-known pp-wave metric. 
The coordinate $x^+$ is dimensionless, $x^-$ has dimensions of
length squared, and the transverse coordinates now have dimensions of length.
Since it is quadratic in the eight transverse bosonic coordinates, the pp-wave limit
leads to a quadratic (and hence  soluble) Hamiltonian for the bosonic string.
The $1/R^{2}$ corrections to the metric are what will eventually concern
us: they will add quartic interactions to the light-cone Hamiltonian and
lead to first-order shifts in the energy spectrum of the string.

After introducing lightcone coordinates $x^\pm$ according to (\ref{rescalePre}),
the general $AdS_5\times S^5$ metric can be cast in the form
\begin{equation} 
\label{genmetric}
ds^2 = 2 G_{+-} dx^+ dx^- + G_{++} dx^+ dx^+ + G_{--} dx^- dx^- + G_{AB}
dx^A dx^B\ ,
\end{equation}
where $x^A$ ($A = 1,\ldots,8$) labels the eight transverse directions, the metric components
are functions of the $x^A$ only, and the components $G_{+A}$ and $G_{-A}$
are not present. This simplifies even further for the pp-wave metric, 
where $G_{--} =0$ and $G_{+-} = 1$. 
We will use (\ref{genmetric}) as the starting point for constructing 
the light-cone gauge worldsheet Hamiltonian 
(as a function of the transverse $x^A$ and their
conjugate momenta $p_A$) and for discussing its expansion about the free
pp-wave Hamiltonian.

The general bosonic Lagrangian density has a simple expression in
terms of the target space metric:
\begin{equation}
{\cal L} = \frac{1}{2}h^{ab} G_{\mu\nu}
\partial_{a}x^{\mu} \partial_{b}x^{\nu}\ ,
\end{equation}
where $h$ is built out of the worldsheet metric $\gamma$ according to
$h^{ab}=\sqrt{-{\rm det}\, \gamma}\gamma^{ab}$ and the indices $a,b$ 	
label the worldsheet coordinates $\sigma,\tau$. Since
${\rm det} \, h = -1$, there are only two independent components of
$h$. The canonical momenta (and their inversion in terms of velocities) are
\begin{eqnarray}
\label{canmomenta}
p_{\mu} = h^{\tau a} G_{\mu\nu} \partial_{a} x^{\nu}\ ,
\qquad
\dot x^{\mu} = \frac{1 }{h^{\tau\tau} } G^{\mu\nu} p_{\nu} -
\frac{h^{\tau\sigma} }{h^{\tau\tau} } x^{\prime\mu}\ .
\end{eqnarray}
The Hamiltonian density ${\cal H} = p_{\mu} \dot x^{\mu} - {\cal L}$ is
\begin{equation}\label{hamilton}
{\cal H} = \frac{1 }{2 h^{\tau\tau} } ( p_{\mu} G^{\mu\nu} p_{\nu}
+ x^{\prime \mu} G_{\mu\nu} x^{\prime \nu} ) -
\frac{h^{\tau\sigma}}{h^{\tau\tau} } (x^{\prime\mu}p_{\mu})\ .
\end{equation}
As is usual in theories with general coordinate invariance (on the
worldsheet in this case), the Hamiltonian is a sum of constraints times
Lagrange multipliers built out of metric coefficients
(${1 }/{h^{\tau\tau} }$ and ${h^{\tau\sigma}}/{h^{\tau\tau}}$).

One can think of the dynamical system we wish to solve as
being defined by ${\cal L} = p_{\mu} \dot x^{\mu} - {\cal H}$
(a phase space Lagrangian) regarded as a function of the coordinates
$x^{\mu}$, the momenta $p_{\mu}$ and the components $h^{ab}$ of
the worldsheet metric. To compute the quantum path integral, the
exponential of the action constructed from this Lagrangian is functionally
integrated over each of these variables. For a spacetime geometry like
(\ref{genmetric}), one finds that with a suitable gauge choice
for the worldsheet coordinates $(\tau,\sigma)$, the
functional integrations over all but the transverse (physical)
coordinates and momenta can be performed, leaving an
effective path integral for these physical variables.
This is the essence of the light-cone approach to quantization.

The first step is to eliminate integrations over $x^+$ and
$p_-$ by imposing the light-cone gauge conditions $x^+ = \tau$ and
$p_-= {\rm const}$.  (At this level of analysis, which is
essentially classical, we will not be concerned
with ghost determinants arising from this gauge choice.)
As noted above, integrations over the worldsheet metric
cause the coefficients ${1 }/{h^{\tau\tau} }$ and
${h^{\tau\sigma}}/{ h^{\tau\tau} }$ to act as Lagrange multipliers,
generating delta functions that impose two constraints:
\begin{eqnarray}
\label{hamconstrnts}
    x^{\prime -} p_- + &&\kern-18pt x^{\prime A} p_A = 0
\nonumber \\
    G^{++} p_+^2 + 2 G^{+-} p_+ p_- +  G^{--} p_-^2+ p_A &&\kern-20pt G^{AB} p_B +
    x^{\prime A} G_{AB} x^{\prime B} + G_{--}
    \frac{(x^{\prime A} p_A)^2 }{p_-^2} = 0\ .
\end{eqnarray}
When integrations over $x^-$ and $p_+$ are performed, the
constraint delta functions serve to evaluate $x^-$ and $p_+$ in
terms of the dynamical transverse variables (and the constant
$p_-$). The first constraint is linear in $x^-$ and yields
$x^{\prime -}=-x^{\prime \, A} p_A/p_-$. Integrating this over
$\sigma$ and using the periodicity of $x^-$ yields the standard
level-matching constraint, without any modifications. The second
constraint is quadratic in $p_+$ and can be solved explicitly
for $p_+= - {\cal H}_{lc}(x^A,p_A)$. The remaining transverse
coordinates and momenta have dynamics which follow from the
phase space Lagrangian
\begin{equation}
{\cal L}_{\rm ps} = p_+ + p_- \dot x^- + p_A \dot x^A \sim p_A \dot
x^A - {\cal H}_{lc}(x^A,p_A)\ ,
\end{equation}
where we have eliminated the $p_-$ term by integrating by parts in time
and imposing that $p_-$ is constant. The essential result is that
$- p_+= {\cal H}_{lc}$ is the Hamiltonian that generates evolution of
the physical variables $x^A, ~ p_A$ in worldsheet time $\tau$. This is,
of course, dynamically consistent with the light-cone gauge identification
$x^+=\tau$ (which requires worldsheet and target space time translation
to be the same).

We can solve the quadratic constraint equation (\ref{hamconstrnts})
for $p_+= -{\cal H}_{\rm lc}$ explicitly, obtaining the uninspiring result
\begin{equation}
\label{hamiltonianLC}
{\cal H}_{\rm lc} = -\frac{p_- G_{+-}}{ G_{--} } -
\frac{p_- \sqrt{G}}{ {G_{--}} } \sqrt{ 1 + \frac{G_{--}}{p_-^2}
( p_A G^{AB} p_B + x^{\prime A} G_{AB} x^{\prime B} ) +
\frac{G_{--}^2}{p_-^4}(x^{\prime A} p_A)^2 }\ ,
\end{equation}
where
\begin{equation}
G \equiv G_{+-}^2 - G_{++} G_{--}\ .
\end{equation}
This is not very useful as it stands, but we can put it in more manageable
form by expanding it in powers of $1/R^{2}$. 
We can actually do slightly better by observing
that the constraint equation (\ref{hamconstrnts}) becomes a linear
equation for $p_+$ if $G_{--}=0$ (which is equivalent to $G^{++}=0$).
Solving the linear equation for $p_+$ gives
\begin{equation} \label{Hlimit}
{\cal H}_{\rm lc} = \frac{p_- G_{++} }{2 G_{+-} } + \frac{ G_{+-}}{ 2p_- }
( p_A G^{AB} p_B + x^{\prime A} G_{AB} x^{\prime B} )~,
\end{equation}
a respectable non-linear sigma model Hamiltonian. 
In the general $AdS_5\times S^5$
metric (\ref{adsmetricPre}) we cannot a convenient set of coordinates such that $G_{--}$ 
identically vanishes.
Using (\ref{rescalePre}), however, we can find coordinates where
$G_{--}$ has an expansion which begins at ${\cal O}(1/R^{4})$, while the
other metric coefficients have terms of all orders in $1/R^{2}$. Therefore,
if we expand in $1/R^{2}$ and keep terms of at most ${\cal O}(1/R^{2})$, we may
set $G_{--}=0$ and use (\ref{Hlimit}) to construct the expansion of the
lightcone Hamiltonian to that order. The leading ${\cal O}(R^{0})$ terms in
the metric reproduce (as they should) the bosonic pp-wave Hamiltonian
\begin{eqnarray}\label{ppHam}
{\cal H}_{\rm lc}^{pp} & = & \frac{1}{2}\left[ (\dot p^A)^2
    + ({x'}^A)^2 + (x^A)^2 \right]\ ~ ,
\end{eqnarray}
(choosing $p_-=1$ for the conserved worldsheet momentum density). 
The ${\cal O}(1/R^{2})$ terms
generate a perturbing Hamiltonian density which is quartic in fields
and quadratic in worldsheet time and space derivatives:
\begin{equation}\label{pertham}
{\cal H}_{\rm lc}^{R^{-2}} =
\frac{1}{4R^2}(y^{2}  p_z^{2}- z^{2} p_y^{2})
+\frac{1}{4R^2}((2 z^2- y^2)({z}^{\prime})^2  -
            (2 y^2-z^2)({y}^{\prime})^2)~.
\end{equation}
This is the bosonic part of the perturbing Hamiltonian we wish to derive.
If we express it in terms of the creation and annihilation operators of
the leading quadratic Hamiltonian (\ref{ppHam}) we can see that its
matrix elements will be of order $1/J$, as will be the first-order
perturbation theory shifts of the string energy eigenvalues. We
defer the detailed discussion of this perturbation theory until we
have the fermionic part of the problem in hand.
Note that this discussion implies that if we wanted to determine
the perturbed energies to higher orders in $1/R^{2}$, we would
have the very unpleasant problem of dealing with the square root form
of the Hamiltonian (\ref{hamiltonianLC}).


We have to this point been discussing a perturbative approach to
finding the effect of the true geometry of the $AdS_5\times S^5$
background on the string spectrum. Before proceeding with this program,
however, it is instructive to study a different limit in which the
kinematics are unrestricted (no large-$J$ limit is taken) but only
modes of the string that are independent of the worldsheet coordinate
(the zero-modes of the string) are kept in the Hamiltonian. This is
the problem of quantizing the superparticle of the underlying
supergravity in the $AdS_5\times S^5$ background, a problem which has 
been solved many times (for references, see \cite{bigreview}).
A remarkable fact, which seems not to have been explicitly observed
before, is that the spectrum of the zero-mode Hamiltonian is {\em exactly}
a sum of harmonic oscillators: the curvature corrections we
propose to compute actually vanish on this special subspace. This fact
is important to an understanding of the full problem, so we
will make a brief digression to explain the solution to this
toy problem.

The quantization of the superparticle in a supergravity background
is equivalent to finding the eigensolutions of certain Laplacians,
one for each spin that occurs in the superparticle massless
multiplet. The point of interest to us can be made by analyzing
the dynamics of the scalar particle and its associated scalar
Laplacian, which only depends on the background metric. 
With apologies, we will adopt another
version of the $AdS_5\times S^5$ metric, chosen because the scalar
Laplacian is very simple in these coordinates:
\bea\label{zm_metric} ds^2 = -dt^2(R^2+  z^2)+d\phi^2(R^2-
y^2)+\qquad\qquad \nonumber\\
    dz^j\left(\delta_{jk}-\frac{z^j z^k}{R^2+ z^2}\right)dz^k +
    dy^{j'}\left(\delta_{j'k'}+\frac{y^{j'}y^{k'}}{R^2- y^2}\right)dy^{k'}~.
\eea 
As before, the coordinates $z^k$ and $y^{k'}$ parameterize the
two $SO(4)$ subspaces, and the indices $j,k$ and $j',k'$ run over
$j,k=1,\ldots,4,$ and $j',k' = 5,\ldots,8$.
This is a natural metric for
analyzing fluctuations of a particle (or string) around the
lightlike trajectory $\phi=t$ and $\vec z=\vec y=0$. Because the
metric components depend neither on $t$ nor on $\phi$, and because
the problem is clearly separable in $\vec z$ and $\vec y$, it
makes sense to look for solutions of the form
$\Phi=e^{-i\omega t}e^{i J\phi}F(\vec z)G(\vec y)$. The
scalar Laplacian for $\phi$ in the above metric then reduces to
\bea
\label{scalrlap} 
\Bigl[-\frac{\omega^2}{R^2+\vec z^2}+\frac{J^2}{R^2-\vec y^2}-
\pderiv{x^j}\Bigl(\delta^{jk}+\frac{z^j z^k}{R^2}\Bigr)\pderiv{z^k}-
\qquad\qquad\nonumber\\
\pderiv{y^{j'}}\Bigl(\delta^{j'k'}-\frac{y^{j'} y^{k'}}{R^2}\Bigr)
\pderiv{y^{k'}}~ \Bigr]F(z)G(y)=0~. 
\eea 
The radius $R$ disappears
from the equation upon rescaling the transverse coordinates by $z\to
z/R$ and $y\to y/R$, so we can set $R=1$ in what follows and use
dimensional analysis to restore $R$ if it is needed. The scalar
Laplacian is essentially the light-cone Hamiltonian constraint
(\ref{hamconstrnts}) for string coordinates $z^k, y^{k'}$ and string
momenta $p_z^k=-i \pderiv{z^k}$ and $p_y^{k'}=-i \pderiv{y^{k'}}$
(projected onto their zero modes). This implies that we
can use the structure of the Laplacian to correctly order
operators in the string Hamiltonian.

The periodicity $\phi\equiv\phi+2\pi$ means that the angular momentum $J$
is integrally quantized. The allowed values of $\omega$ then follow
from the solution of the eigenvalue problem posed by (\ref{scalrlap}).
As the trial function $\Phi$ indicates, (\ref{scalrlap})
breaks into separate problems for $\vec z$ and $\vec y$:
\bea\label{separated}
{\cal H}_{AdS_5} F(\vec z) = 
	\left[ p^z_j(\delta^{jk}+z^j z^k)p^z_k +
    \omega^2\frac{z_k z^k}{1+(z_k z^k)^2} \right] F(\vec z)
        = A(\omega)  F(\vec z) \nonumber\\
{\cal H}_{S^5} G(\vec y) = \left[ p^y_{j'}(\delta^{j'k'}-y^{j'} y^{k'})p^y_{k'}+
    J^2\frac{y_{k'} y^{k'} }{1-(y_{k'}y^{k'})^2} \right] G(\vec y) = B(J) G(\vec y)\ ,
\eea
where $\omega^2 - J^2 = A + B$.
The separation eigenvalues $A,B$ depend on their respective parameters
$\omega,J$, and we determine the energy eigenvalues $\omega$ by finding
the roots of the potentially complicated equation
$\omega^2 - J^2 - A - B=0$.  The scalar Laplacian (\ref{scalrlap})
is equivalent to the constraint equation (\ref{hamconstrnts})
projected onto string zero modes, and we are once again seeing
that the constraint doesn't directly give the Hamiltonian but rather an
equation (quadratic or worse) to be solved for the Hamiltonian.

The ${\cal H}_{S^5}$ equation is just a
repackaging of the problem of finding the eigenvalues of the $SO(6)$
Casimir invariant (another name for the scalar Laplacian on $S^5$)
and ${\cal H}_{AdS_5}$ poses the corresponding problem for $SO(4,2)$. The
$SO(6)$ eigenvalues are obviously discrete, and the $SO(4,2)$ problem
also turns out to be discrete when one imposes the condition of
finiteness at $z^2\to\infty$ on the eigenfunctions (this is a natural restriction
in the context of the AdS/CFT correspondence; for a detailed discussion
see \cite{bigreview}). Thus we expect $\omega$ to have a purely discrete
spectrum, with eigenvalues labeled by a set of integers. The simplest way
to solve for the spectrum is to expand $F(\vec z)$ and $G(\vec y)$ in
$SO(4)$ harmonics (since this symmetry is explicit), recognize that
the radial equation is, in both cases, an example of Riemann's
differential equation and then use known properties of the hypergeometric
function to find the eigenvalues and eigenfunctions of (\ref{separated}).
Since it takes three integers to specify an $SO(4)$ harmonic and one
to specify a radial quantum number, we expect each of the two separated
equations to have a spectrum labeled by four integers. The exact
results for the separation eigenvalues turn out to be remarkably simple:
\bea
\label{eignvls}
A & = & 2\omega\sum_1^4\left(n_i+\frac{1}{2}\right) -
        \left[\sum_1^4\left(n_i+\frac{1}{2}\right)\right]^2 +4
    \qquad n_i=0,1,2,\ldots
\nonumber\\
B & = & 2J\sum_1^4\left(m_i+\frac{1}{2}\right) +
        \left[\sum_1^4\left(m_i+\frac{1}{2}\right)\right]^2 +4
            \qquad m_i=0,1,2,\ldots
\eea
Different eigenfunctions correspond to different choices of the collection
of eight integers $\{n_i,m_i\}$, and the fact that the energies depend only
on $\Sigma n_i$ and $\Sigma m_i$ correctly accounts for the degeneracy of
eigenvalues.  The special form of $A$ and $B$ means that the equation
for the energy eigenvalue, $\omega^2-J^2-A-B=0$, can be factored as
\be
&&
\left[ \omega -J-\sum_1^4\left(n_i+\frac{1}{2}\right)-
    \sum_1^4\left(m_i+\frac{1}{2}\right)\right] \nn\\
&&\kern+150pt
	\times
	\left[ \omega+J-\sum_1^4\left(n_i+\frac{1}{2}\right)+
    \sum_1^4\left(m_i+\frac{1}{2}\right)\right] = 0 ~.
\nn\\
\ee
For obvious reasons, we retain the root that assigns only positive
values to $\omega$, the energy conjugate to the global time $t$:
\be
\omega-J~=~\sum_1^4\left(n_i+\frac{1}{2}\right)~+~ 
	\sum_1^4\left(m_i+\frac{1}{2}\right) ~.
\ee
From the string point of view, $\omega$ catalogs the eigenvalues 
of the string worldsheet Hamiltonian restricted to the zero-mode subspace. 
Quite remarkably, it is an exact `sum of harmonic oscillators', 
independent of whether $J$ (and $\omega$) are large or not. This is
simply to say that the eigenvalues of the string Hamiltonian restricted
to the zero-mode sector receive no curvature corrections and could
have been calculated from the pp-wave string Hamiltonian (\ref{ppHam}).
We have only shown this for the massless bosons of the theory, but
we expect the same thing to be true for all the massless fields
of type IIB supergravity. The implication for a perturbative account 
of the string spectrum is that states created using only zero-mode 
oscillators (of any type) will receive no curvature corrections. This 
feature will turn out to be a useful consistency check on our 
quantization procedure. It is of course not true for a general
classical background and is yet another manifestation of the special
nature of the $AdS_5\times S^5$ geometry.

\section{GS superstring action on $AdS_5 \times S^5$}

The $AdS_5\times S^5$ target space can be realized as the coset superspace 
\be
G/H = \frac{SU(2,2|4)}{ SO(4,1)\times SO(5)  }\ .
\ee
The bosonic reduction of this coset is precisely
$SO(4,2)\times SO(6)/SO(4,1)\times SO(5)\equiv AdS_5\times S^5$. 
To quantize the theory, we will expand the action about a classical trajectory 
which happens to be invariant under the stabilizer group
$H$. There is a general strategy for constructing a non-linear 
sigma model on a super-coset space in terms of the Cartan one-forms 
and superconnections of the super-coset manifold.
In such a construction, the symmetries of the stabilizer subgroup
remain manifest in the action while the remaining symmetries are 
nonlinearly realized (see, e.g.,~\cite{Kallosh:1998zx,Metsaev:2000yf, 
Metsaev:bj, Kallosh:1998ji, Metsaev:1999gz, Metsaev:1998it}).
Metsaev and Tseytlin \cite{Metsaev:2000yf} carried out this construction
for the $AdS_5\times S^5$ geometry, producing a $\kappa$-symmetric,
type IIB superstring action possessing the full $PSU(2,2|4)$ 
supersymmetry of $AdS_5\times S^5$. Their action is conceptually simple, 
comprising a kinetic term and a Wess--Zumino term built out of Cartan 
(super)one-forms on the super-coset manifold in the following way 
(this form was first presented in \cite{Grisaru:fv}):
\begin{eqnarray}
\label{genlagrangian}
{\cal S} & = & 
	-\frac{1}{2}\int_{\partial M_3} d^2\sigma\ h^{ab} L_a^\mu L_b^\mu
	+ i\int_{M_3} s^{IJ} L^\mu \wedge \bar L^I \Gamma^\mu \wedge L^J\ .
\end{eqnarray}
Repeated upper indices are summed over a Minkowskian inner product.
The indices $a,b$ are used to indicate the worldsheet coordinates
$(\tau,\sigma)$, and we use the values $a,b=0$ to indicate the
worldsheet time direction $\tau$, and $a,b=1$ to specify the $\sigma$ direction.
The matrix $s^{IJ}$ is defined by $s^{IJ} \equiv {\rm diag}(1,-1)$, where
$I,J = 1,2$.
The Wess-Zumino term appears as an integral over a 3-manifold $M_3$, while
the kinetic term is integrated over the 2-dimensional boundary $\partial M_3$.
The left-invariant Cartan forms are
defined in terms of the coset space representative $G$ by
\begin{eqnarray}
G^{-1}dG = 
	L^\mu P^\mu + L^\alpha  &&\kern-18pt  \bar{Q}_\alpha +  \bar{L}^\alpha Q_\alpha
	+ {1\over 2} L^{\mu\nu}J^{\mu\nu} 
\nonumber \\
L^N = dX^M L^N_M
\qquad 
L_a^N =  L^N_M  &&\kern-18pt  \partial_a X^M
\qquad
X^M = (x^\mu,\theta^\alpha,\bar{\theta}^\alpha)\ ,
\end{eqnarray}

The explicit expansion of this action in terms of independent fermionic degrees
of freedom is rather intricate. One starts with two 32-component Majorana-Weyl
spinors in 10 dimensions: $\theta^I$, where $I=1,2$ labels the two spinors.
In a suitably-chosen representation for the $32\times 32$ ten-dimensional
gamma matrices $\Gamma^\mu$, the Weyl projection reduces to picking
out the upper 16 components of $\theta$ and the surviving spinors can
combined into one complex 16-component spinor $\psi$:
\begin{eqnarray}
\theta^I  =  \left( { \theta^\alpha \atop 0} \right)^I
\qquad\qquad
\label{psi}
\psi^\alpha  =  \sqrt{2} \left[ (\theta^\alpha)^1 + i (\theta^\alpha)^2 \right]\ .
\end{eqnarray}
The following representation for $\Gamma^\mu$ (which has the desired property that 
$\Gamma_{11}=({\bf 1}_8,-{\bf 1}_8)$) allows us to express their action on 
$\psi$ in terms of real $16\times 16$ $\gamma$-matrices:
\begin{eqnarray}
\label{16gamma}
\Gamma^\mu = \left( \begin{array}{cc}
    0   &   \gamma^\mu  \\
    \bar\gamma^\mu &    0
    \end{array}  \right)\
& \qquad &
\gamma^\mu \bar\gamma^\nu
+ \gamma^\nu \bar\gamma^\mu = 2\eta^{\mu\nu}
\nonumber \\
	\gamma^\mu = (1,\gamma^A, \gamma^9)
& \qquad &
	\bar\gamma^\mu = (-1,\gamma^A, \gamma^9)\ .
\end{eqnarray}
The indices  $\mu,\nu,\rho = 0,\dots,9$ denote $SO(9,1)$ vectors,
and  we will denote the corresponding spinor indices by 
$\alpha,\beta,\gamma,\delta = 1,\dots,16$ (we also use the convention 
that upper-case indices $A,B,C,D = 1,\dots,8$ indicate vectors of $SO(8)$,
while $i,j,k = 1,\dots,4$ ($i',j',k' = 5,\dots,8$) indicate vectors
from the $SO(3,1) \cong SO(4)$ ($SO(4)$) subspaces associated with
$AdS_5$ and $S^5$ respectively). The matrix $\gamma^9$ is formed by 
taking the product of the eight $\gamma^A$. A representation of
$\gamma^A$ matrices which will be convenient for explicit calculation
is given in Appendix A. We also note that in the course of quantization
we will impose the fermionic lightcone gauge fixing condition 
$\bar\gamma^9 \psi = \psi$. This restricts the worldsheet fermions
to lie in the $8_s$ representation of $SO(8)$ (and projects out the
$8_c$ spinor), thus reducing the number of independent components
of the worldsheet spinor from 16 to 8. The symmetric matrix
\begin{equation}
\label{Pidef}
\Pi \equiv \gamma^1 \bar\gamma^2 \gamma^3 \bar\gamma^4\ 
\end{equation}
appears in a number of places in the expansion of the action, so
we give it an explicit definition. Since $\Pi^2=1$, it has eigenvalues 
$\pm 1$ which turn out to provide a useful sub-classification of the 8
components of the $8_s$ worldsheet spinor into two groups of 4. 
The quantity $\tilde\Pi = \Pi \gamma_9$ also appears, but does not
require a separate definition because $\Pi \psi = \tilde\Pi \psi$ 
for spinors satisfying the lightcone gauge restriction to the
$8_s$ representation.

Kallosh, Rahmfeld and Rajaraman presented in \cite{Kallosh:1998zx}  
a general solution to the supergravity constraints (Maurer-Cartan equations) 
for coset spaces exhibiting a superconformal isometry 
algebra of the form
\begin{eqnarray}
\label{genalgebra}
\left[ B_\mu, B_\nu \right] & = & f_{\mu\nu}^\rho B_\rho \nonumber \\
\left[ F_\alpha, B_\nu \right] & = & f_{\alpha \nu}^\beta F_\beta \nonumber \\
\{ F_\alpha, F_\beta \} & = & f_{\alpha\beta}^\mu B_\mu\ ,
\end{eqnarray}
with $B_\mu$ and $F_\alpha$ representing bosonic and fermionic generators,
respectively.  In terms of these generators, the Cartan forms $L^\mu$ and superconnections $L^\alpha$
are determined completely by the structure constants
$f_{\alpha \mu}^J$ and $f_{\alpha\beta}^\mu$:
\begin{eqnarray}
\label{kallosh1}
L_{at}^\alpha & = &
	\left( \frac{\sinh t{\cal M}}{\cal M} \right)^\alpha_\beta 
	({\cal D}_a \theta)^\beta
 \\
\label{kallosh2}
L_{at}^\mu & = & e^\mu_{\phantom{\mu}\nu}\partial_a x^\nu + 2\theta^\alpha f_{\alpha\beta}^\mu
	\left( \frac{\sinh^2 (t {\cal M}/2)}{{\cal M}^2}\right)^\beta_\gamma 
	({\cal D}_a \theta)^\gamma
 \\
({\cal M}^2)^\alpha_\beta & = & -\theta^\gamma f^\alpha_{\gamma\mu} 
	\theta^\delta f^\mu_{\delta\beta}\ .
\end{eqnarray}
The dimensionless parameter $t$ is used here to define ``shifted'' Cartan
forms and superconnections where, for example, $L_a^\mu = {L_{at}^\mu}|_{t=1}$.
In the case of $AdS_5 \times S^5$, the Lagrangian takes the form
\begin{eqnarray}
\label{lagrangiank}
{\cal L}_{\rm Kin} & = & -\frac{1}{2} h^{ab} L_a^\mu L_b^\mu  
\\
\label{lagrangianwz}
{\cal L}_{\rm WZ} & = & -2i\epsilon^{ab} \int_0^1 dt\, L_{at}^\mu s^{IJ}
    \bar\theta^I \Gamma^\mu L_{bt}^J\ .
\end{eqnarray}

In the context of eqns.~(\ref{kallosh1},\ref{kallosh2}), 
it will be useful to choose a 
manifestation of the spacetime metric that yields a compact form of 
the spin connection.  The form appearing in eqn.~(\ref{metric}) 
is well suited to this requirement; the $AdS_5$ and $S^5$ subspaces
are represented in (\ref{metric}) nearly symmetrically, and the
spin connection is relatively simple:
\begin{eqnarray}
\omega^{t\,z_k}_{\phantom{t\,z_k}\,t}  =
         {z_k \over 1-{1\over 4}z^2} & \qquad &
\omega^{z_j\,z_k}_{\phantom{z_j\,z_k}\,z_j}  =
        {{1\over 2} z_k \over 1-{1\over 4}z^2}
\nonumber \\
\omega^{\phi\,y_{k'}}_{\phantom{\phi\,y_{k'}}\,\phi}  =
        -{y_{k'} \over 1+{1\over4}y^2} & \qquad &
\omega^{y_{j'}\,y_{k'}}_{\phantom{y_{j'}\,y_{k'}}\,y_{j'}}  =
        -{{1\over2}y_{k'}\over 1+{1\over4}y^2}\ .
\end{eqnarray}
%
Upon moving to the light-cone coordinate system in (\ref{rescalePre}), the $x^+$
direction remains null ($G_{--} = 0$) to ${\cal O}(1/R^4)$ in this
expansion.

By introducing dimensionless contraction parameters $\Lambda$ and
$\Omega$ \cite{Blau:2002mw}, one may express the $AdS_5 \times S^5$ isometry
algebra keeping light-cone directions explicit:
\begin{eqnarray}
\left[P^+,P^k\right] = {\Lambda^2 \Omega^2}J^{+k}
    & \quad &
\left[P^+,P^{k'}\right] = -{\Lambda^2 \Omega^2}J^{+k'} \nonumber \\
\left[P^+,J^{+k}\right] = -{\Lambda^2}P^{k}
    & \quad &
\left[P^+,J^{+k'}\right] = {\Lambda^2}P^{k'} \nonumber \\
\left[P^-,P^{A}\right] = { \Omega^2}J^{+A}
    & \quad &
\left[P^-,J^{+A}\right] = P^{A} \nonumber \\
\left[P^j,P^k\right] = {\Lambda^2 \Omega^2}J^{jk}
    & \quad &
\left[P^{j'},P^{k'}\right] = -{\Lambda^2 \Omega^2}J^{j'k'} \nonumber \\
\left[J^{+j},J^{+k}\right] = {\Lambda^2}J^{jk}
    & \quad &
\left[J^{+j'},J^{+k'}\right] = -{\Lambda^2}J^{j'k'} \nonumber \\
\left[P^j,J^{+k}\right] = -\delta^{jk}(P^+ - \Lambda^2\,P^-)
    & \quad &
\left[P^{r},J^{+s}\right] = -\delta^{rs}(P^+ + \Lambda^2\,P^-) \nonumber \\
\left[P^{i},J^{{j}{k}}\right] = \delta^{ij}P^{k} - \delta^{ik}P^j
    & \quad &
\left[P^{i'},J^{{j'}{k'}}\right] = \delta^{i'j'}P^{k'}- \delta^{i'j'}P^{k'}  \nonumber \\
\left[J^{+{i}},J^{{j}{k}}\right] = \delta^{ij}J^{+k} - \delta^{ik}J^{+j}
    & \quad &
\left[J^{+{i'}},J^{j'k'}\right] = \delta^{i'j'}J^{+k'}- \delta^{i'j'}J^{+k'}  \nonumber \\
\left[J^{ij},J^{kl}\right]
    = \delta^{jk}J^{il} + 3\ {\rm terms}
    & \quad &
\left[J^{i'j'},J^{k'l'}\right]
    = \delta^{j'k'}J^{i'l'} + 3\ {\rm terms}\ .
\end{eqnarray}
The bosonic sector of the algebra relevant to (\ref{genalgebra}) takes the form
\begin{eqnarray}
\left[J^{ij},Q_{\alpha}\right] & = &
    {1\over2}Q_{\beta}(\gamma^{ij})^\beta_{\ \alpha}
    \nonumber \\
\left[J^{i'j'},Q_{\alpha}\right] & = &
    {1\over2}Q_{\beta}(\gamma^{i'j'})^\beta_{\ \alpha}
    \nonumber \\
\left[J^{+i},Q_{\alpha}\right] & = &
    {1\over2}Q_{\beta}(\gamma^{+i}-\Lambda^2\gamma^{-i})^\beta_{\ \alpha}
    \nonumber \\
\left[J^{+i'},Q_{\alpha}\right] & = &
    {1\over2}Q_{\beta}(\gamma^{+i'}+\Lambda^2\gamma^{-i'})^\beta_{\ \alpha} \nonumber
    \nonumber \\
\left[P^\mu,Q_{\alpha}\right] & = &
    {i \Omega\over 2}Q_{\beta}(\Pi \gamma^{+}\bar{\gamma}^\mu)^\beta_{\ \alpha}
    -{i\Lambda^2 \Omega\over 2}Q_{\beta}(\Pi \gamma^{-}\bar{\gamma}^\mu)^\beta_{\ \alpha}\ .
\end{eqnarray}
The fermi-fermi anticommutation relations are
\begin{eqnarray}
\{Q_\alpha,\bar Q_\beta\} & = &
    -2i\gamma^\mu_{\alpha\beta}P^\mu
    -{2 \Omega}(\bar\gamma^k \Pi)_{\alpha\beta}J^{+k}
    -{2 \Omega}(\bar\gamma^{k'}\Pi)_{\alpha\beta}J^{+k'} \nonumber \\
& &     + { \Omega}(\bar\gamma^+\gamma^{jk}\Pi)_{\alpha\beta}J^{jk}
    + {\Omega}(\bar\gamma^+\gamma^{j'k'}\Pi)_{\alpha\beta}J^{j'k'} \nonumber \\
& &     -{\Lambda^2 \Omega}(\bar\gamma^-\gamma^{jk}\Pi)_{\alpha\beta}J^{jk}
    +{\Lambda^2  \Omega}(\bar\gamma^-\gamma^{j'k'}\Pi)_{\alpha\beta}J^{j'k'}\ .
\end{eqnarray}
This form of the superalgebra has the virtue that one can
easily identify the flat space $(\Omega\rightarrow 0)$
and plane-wave $(\Lambda\rightarrow 0)$ limits. 
The Maurer-Cartan equations in this coordinate system take the form
\begin{eqnarray}
\label{diffeq1}
dL^\mu & = & -L^{\mu\nu}L^\nu - 2i\bar{L}\bar{\gamma}^\mu L \nonumber \\
dL^\alpha & = & -{1\over 4}L^{\mu\nu}(\gamma^{\mu\nu})^\alpha_{\ \beta}L^\beta
    + {i\Omega \over 2}L^\mu(\Pi\gamma^+\bar{\gamma}^\mu)^\alpha_{\ \beta}L^\beta
    - {i\Lambda^2\Omega \over 2}L^\mu(\Pi\gamma^-\bar{\gamma}^\mu)^\alpha_{\ \beta}L^\beta 
\nonumber \\
d\bar{L}^\alpha & = & -{1\over 4}L^{\mu\nu}(\gamma^{\mu\nu})^\alpha_{\ \beta}\bar{L}^\beta
    - {i\Omega\over 2}L^\mu(\Pi\gamma^+\bar\gamma^\mu)^\alpha_{\ \beta}\bar{L}^\beta
    + {i\Lambda^2\Omega \over 2}L^\mu(\Pi\gamma^-\bar\gamma^\mu)^\alpha_{\ \beta}\bar{L}^\beta\ ,
\end{eqnarray}
where wedge products (\ref{genlagrangian}) 
are understood to be replaced by the following rules:
\begin{equation}
    L^\mu L^\nu = -L^\nu L^\mu
\qquad  L^\mu L^\alpha = -L^\alpha L^\mu
\qquad  L^\alpha L^\beta = L^\beta L^\alpha\ .
\end{equation}
Upon choosing a parameterization of the
coset representative $G$
\begin{eqnarray}
G(x,\theta) = f(x)g(\theta) \qquad
g(\theta) = \exp(\theta^\alpha \bar{Q}_\alpha + \bar{\theta}^\alpha Q_\alpha)\ ,
\end{eqnarray}
one derives a set of coupled differential equations for the shifted 
Cartan forms and superconnections:
\begin{eqnarray}
\partial_t L_t & = & d\theta + {1\over 4}L_t^{\mu\nu}\gamma^{\mu\nu}\theta
    - {i\Omega\over 2}L_t^\mu \Pi\gamma^+ \bar\gamma^\mu\theta
    + {i\Lambda^2\Omega \over 2}L_t^\mu \Pi\gamma^-\bar\gamma^\mu \theta 
\nonumber \\
\partial_t L_t^\mu & = & -2i\theta \bar\gamma^\mu \bar L_t
    - 2i\bar\theta\bar\gamma^\mu L_t 
\nonumber \\
\partial_t L_t^{-i} & = & 
	{2\Omega}(\theta \bar\gamma^i\Pi \bar L_t)
	- {2\Omega}(\bar\theta \bar\gamma^i\Pi L_t) 
\nonumber \\
\partial_t L_t^{-r} & = & 
	{2\Omega}(\theta \bar\gamma^{r}\Pi \bar L_t)
	- {2\Omega}(\bar\theta \bar\gamma^{r}\Pi L_t) 
\nonumber \\
\partial_t L_t^{ij} & = &
	-{2\Omega}(\theta \bar\gamma^+ \gamma^{ij} \Pi \bar L_t)
    + {2\Omega} (\bar\theta \bar \gamma^+ \gamma^{ij} \Pi L_t)
    + {2\Lambda^2\Omega} (\theta \bar\gamma^- \gamma^{ij} \Pi \bar L_t)
    - {2\Lambda^2\Omega}(\bar\theta \bar\gamma^- \gamma^{ij} \Pi L_t )
\nonumber \\
\partial_t L_t^{i'j'} & = &
    -{2\Omega}(\theta \bar\gamma^+ \gamma^{i'j'} \Pi \bar L_t)
    + {2\Omega} (\bar\theta \bar \gamma^+ \gamma^{i'j'} \Pi L_t)
    - {2\Lambda^2\Omega} (\theta \bar\gamma^- \gamma^{i'j'} \Pi \bar L_t)
\nonumber \\
& &     + {2\Lambda^2\Omega}(\bar\theta \bar\gamma^- \gamma^{i'j'} \Pi L_t)\ .
\label{diffeq}
\end{eqnarray}
These coupled equations are subject to the following boundary conditions:
\begin{eqnarray}
L_\pm(t=0) = 0 \quad L_{t=0}^\mu = e^\mu \quad L_{t=0}^\pm = e^\pm  \nonumber \\
L_{t=0}^{\mu\nu} = \omega^{\mu\nu} \qquad L_{t=0}^{-\mu} = \omega^{-\mu}\ .
\end{eqnarray}  
The generators $J^{-\mu}$ and $J^{kk'}$ are not present in the superalgebra, so
the conditions
\begin{equation}
L^{+\mu} = 0 \qquad L^{kk'} = 0
\end{equation}
are imposed as constraints.

To employ the general solution to the Maurer-Cartan equations 
(\ref{kallosh1},\ref{kallosh2}), the relevant sectors of the superalgebra
may be rewritten in the more convenient 32-dimensional notation
(setting $\Lambda = 1$ and $\Omega = 1$):
\begin{eqnarray}
\label{superalg}
\left[ Q_I , P^\mu \right] & = &
    \frac{i}{2} \epsilon^{I J}Q_J \Gamma_* \Gamma^\mu
\nonumber \\
\left[ Q_I , J^{\mu\nu} \right] & = &
    -\frac{1}{2} Q_I  \Gamma^{\mu\nu}
\nonumber \\
\{ (Q_I)^\mu, (Q_J)_\mu \} & = &
    -2i\delta_{IJ}\Gamma^0 \Gamma^\rho P_\rho
    +  \epsilon^{IJ}\left( -\Gamma^0 \Gamma^{jk}\Gamma_* J_{jk}
    + \Gamma^0 \Gamma^{j'k'} {\Gamma'}_* J_{j'k'} \right)\ ,
\end{eqnarray}
where
\begin{eqnarray}
\Gamma_* \equiv i\Gamma_{01234}
\qquad
{\Gamma'}_* \equiv i\Gamma_{56789}\ .
\end{eqnarray}
The Cartan forms and superconnections
then take the following form:
\begin{eqnarray}
\label{sol}
L_{bt}^J  =  \frac{\sinh t{\cal M}}{{\cal M}} {\cal D}_b \theta^J
    & \qquad &
L_{at}^\mu  =  e^\mu_{\phantom{\mu}\rho}\partial_a x^\rho
    - 4i\bar\theta^I \Gamma^\mu \left( \frac{\sinh^2 (t{\cal M}/2)}{{\cal M}^2}
    \right){\cal D}_a \theta^I\ ,
\end{eqnarray}
where the covariant derivative is given by
\begin{eqnarray}
({\cal D}_a \theta)^I
    & = & \left( \partial_a \theta + {1\over 4}
    \left(\omega^{\mu\,\nu}_{\phantom{\mu\,\nu}\,\rho}\,\partial_a x^\rho \right)
    \Gamma^{\mu\nu} \theta\right)^I
    -{i\over 2}\epsilon^{IJ}
    e^\mu_{\phantom{\mu}\,\rho}\,\partial_a x^\rho \Gamma_* \Gamma^\mu \theta^J\ .
\end{eqnarray}
The object ${\cal M}$ is a $2\times 2$ matrix which, for convenience, is 
defined in terms of its square:
\begin{eqnarray}
({\cal M}^2)^{IL} & = &
    -\epsilon^{IJ}(\Gamma_* \Gamma^\mu \theta^J \bar\theta^L \Gamma^\mu)
    +\frac{1}{2}\epsilon^{KL}(-\Gamma^{jk}\theta^I \bar\theta^K \Gamma^{jk}\Gamma_*
        + \Gamma^{j'k'}\theta^I \bar\theta^K \Gamma^{j'k'} {\Gamma'}_*)\ .
\end{eqnarray}

At this point, the GS action on $AdS_5\times S^5$ (\ref{lagrangiank},\ref{lagrangianwz}) 
may be expanded to arbitrary 
order in fermionic and bosonic fields.  In the present calculation, 
the parameters $\Omega$ and $\Lambda$ remain set to unity, and 
the action is expanded in inverse powers of the target-space radius $R$, 
introduced in the rescaled light-cone coordinates in eqn.~(\ref{rescalePre}). 
The fact that supersymmetry must be protected
at each order in the expansion determines a rescaling prescription
for the fermions.  Accordingly, the eight transverse bosonic directions $x^A$
and the corresponding fermionic fields $\psi^\alpha$ receive a rescaling
coefficient proportional to $R^{-1}$.
The first curvature correction
away from the plane-wave limit therefore occurs at quartic order in both
bosonic and fermionic fluctuations.
The particular light-cone coordinate system chosen in (\ref{rescalePre}), however,
gives rise to several complications.  The $x^\pm$ coordinates given by 
\be
\label{lccoords}
t  =  x^+ - \frac{x^-}{2R^2}  \qquad\qquad
\phi  =  x^+ + \frac{x^-}{2R^2}
\ee
have conjugate momenta (in the language of BMN)
\be
-p_+ & = & i\partial_{x^+} = i(\partial_t + \partial_\phi) = \Delta - J \\
-p_- & = & i\partial_{x^-} = \frac{i}{2R^2}(\partial_\phi - \partial_t) 
	= -\frac{1}{2R^2}(\Delta + J)\ ,
\ee
with $\Delta = E = i\partial_t$ and $J = -i\partial_\phi$. 
The light-cone Hamiltonian is ${\cal H} = -p_+$, so with $\Delta = J-p_+$ one
may schematically write
\be
p_- & = & \frac{1}{2R^2}(2J - p_+) \nn\\
& = & \frac{J}{R^2}+ \frac{{\cal H}}{2R^2} \nn\\
& = & \frac{J}{R^2}\left( 1 + \frac{1}{2J}\sum N \omega \right)\ .
\ee
This result appears to be incorrect in the context of 
the light-cone gauge condition $\partial_\tau t = p_-$.  
To compensate for this, one must set the constant worldsheet density 
$p_-$ equal to something different from 1 (and non-constant) if the parameter 
length of the worldsheet is to be proportional to $J$. 
This operation introduces an additional ${\cal O}(1/R^2)$ shift in the energy 
of the string oscillators.
This is acceptable because, in practice, we wish to consider only degenerate subsets of 
energy states for comparison between the gauge theory and string theory results.  
Because of the compensation between corrections to
$J$ and the Hamiltonian contribution from $p_-$, the eigenvalues of $J$ will remain
constant within these degenerate subsets.  Therefore, while it may seem incorrect to introduce
operator-valued corrections to $p_-$, one could proceed pragmatically 
with the intent of restricting oneself to these degenerate subsets.\footnote{
When such a program is carried out, however, the resulting theory is subject to 
normal-ordering ambiguities; we instead use a coordinate
system that is free of these complications. }

A different choice of light-cone coordinates allows us
to avoid this problem completely. 
By choosing
\be
\label{newcoords}
t & = & x^+ \nn\\
\phi & = & x^+ + \frac{x^-}{R^2}\ ,
\ee
we have
\be
\label{RJdefs}
-p_+ & = & \Delta - J \\
-p_- & = & i\partial_{x^-} = \frac{i}{R^2}\partial_\phi = -\frac{J}{R^2}\ ,
\ee
such that $p_-$ appears as a legitimate expansion parameter in the theory.
In this coordinate system, the curvature expansion of the metric becomes
\be
ds^2 & \approx & 2dx^+ dx^- - (x^A)^2 (dx^+)^2 + (dx^A)^2 
\nonumber \\
& & 	+ \frac{1}{R^2}\left[ 
	-2y^2 dx^+ dx^- + \frac{1}{2}(y^4-z^4) (dx^+)^2 + (dx^-)^2
	+\frac{1}{2}z^2 dz^2 - \frac{1}{2}y^2 dy^2 \right] 
\nonumber \\
\label{expmetric2}
& & 	+ {\cal O}\left(R^{-4}\right).
\ee
The operator-valued terms in $p_-$ that appear under the first
coordinate choice (\ref{lccoords}) are no longer present.
However, it will be shown that this new coordinate system induces 
correction terms to the spacetime curvature of the worldsheet metric.  
Furthermore, the appearance of a nonvanishing $G_{--}$ component,
and the loss of many convenient symmetries between terms associated with the
$x^+$ and $x^-$ directions bring some additional complications into the analysis.
The advantage is that the results will be unambiguous in the end (and free
from normal-ordering ambiguities).

\section{Curvature corrections to the Penrose limit}

In this section we expand the GS superstring action on $AdS_5 \times S^5$
in powers of $1/R^2$.  We begin by constructing various quantities
including combinations of Cartan 1-forms relevant to the worldsheet 
Lagrangian.  Spacetime curvature corrections to the 
worldsheet metric will be calculated by 
analyzing the $x^-$ equation of motion and the covariant gauge constraints
order-by-order.  

We introduce the notation
\begin{eqnarray}
\Delta_n^\mu & \equiv & \bar\theta^I \Gamma^\mu {\cal D}_0^n \theta^I \\
{\Delta'}_n^\mu & \equiv & \bar\theta^I \Gamma^\mu {\cal D}_1^n \theta^I~,
\end{eqnarray}
where the covariant derivative is expanded in powers of $(1/R)$:
\begin{eqnarray}
{\cal D}_a & = & {\cal D}_a^0 + \frac{1}{R}{\cal D}_a^1
    + \frac{1}{R^2}{\cal D}_a^2 + {\cal O}(R^{-3})\ .
\end{eqnarray}
Terms in the Wess-Zumino Lagrangian are encoded using a similar notation:
\be
\Box_n^\mu & \equiv & s^{IJ}\bar\theta^I\Gamma^\mu{\cal D}_0^n \theta^J \\
{\Box'}_n^\mu & \equiv & s^{IJ}\bar\theta^I\Gamma^\mu{\cal D}_1^n \theta^J~.
\ee
The subscript notation $(\Delta_n^\mu)_{\theta^4}$ will be used
to indicate the quartic fermionic term involving ${\cal M}^2$:
\be
(\Delta_n^\mu)_{\theta^4} \equiv \frac{1}{12}\bar\theta^I({\cal M}^2){\cal D}_0^n\theta^I~.
\ee
For the present, it will be convenient to remove an overall factor of $R^2$ from the
definition of the vielbeins $e^\mu_{\phantom{\mu}\nu}$.  
In practice, this choice makes it easier to recognize terms that contribute to the 
Hamiltonian at the order of interest, and, in the end, 
allows us to avoid imposing an additional rescaling operation on the fermions. 
We proceed by keeping terms to ${\cal O}(1/R^4)$, with the understanding that 
an extra factor of $R^2$ must be removed in the final analysis.  
The covariant derivative
\be
{\cal D}_a\theta^I = \partial_a \theta^I + \frac{1}{4}\partial_a x^\mu \omega^{\nu\rho}_\mu
	\Gamma_{\nu\rho}\theta^I - \frac{i}{2}\epsilon^{IJ}\Gamma_*\Gamma_\mu
	e^\mu_{\phantom{\mu}\nu}\partial_a x^\nu \theta^J
\ee
may then be expanded to ${\cal O}(1/R^2)$ (we will not need ${\cal O}(1/R^3)$ terms, 
because the covariant derivative always appears left-multiplied by a spacetime spinor
$\bar\theta$):
\be
{\cal D}_0\theta^I & = &  
	\biggl[\partial_0\theta^I - p_-\epsilon^{IJ}\Pi\theta^J\biggr] 
	+\frac{1}{R}\biggl[ \frac{p_-}{4}  
	\left( z_j\Gamma^{-j} 
	- y_{j'}\Gamma^{-j'}\right)\theta^I
	+\frac{1}{4}\epsilon^{IJ}\Gamma^-\Pi(\dot x^A\Gamma^A)\theta^J\biggr] 
\nonumber \\
&&\kern-0pt	
	+\frac{1}{R^2}
	\biggl[\frac{1}{4}(\dot z_j z_k\Gamma^{jk} 
	- \dot y_{j'}y_{k'}\Gamma^{j'k'})\theta^I
	+ \frac{p_-}{4}\epsilon^{IJ}\Pi(y^2-z^2)\theta^J 
	- \frac{1}{2}\epsilon^{IJ}(\dot x^-) \Pi\theta^J \biggr] 
\nn\\
&&\kern+300pt	+{\cal O}(R^{-3}) \\
{\cal D}_1\theta^I &= & 
	\partial_1\theta^I 
	+\frac{1}{4R}\epsilon^{IJ}\Gamma^-\Pi ({x'}^A \Gamma^A)\theta^J \nonumber \\
& & 	+\frac{1}{R^2}\left[\frac{1}{4}(z'_jz_k\Gamma^{jk}-y'_{j'}y_{k'}\Gamma^{j'k'})\theta^I
	-\frac{1}{2}\epsilon^{IJ}({x'}^-)\Pi\theta^J\right] +{\cal O}(R^{-3}) ~.
\ee
Note that we have not rescaled the spinor field $\theta$ in the above expansion.  
This allows us to isolate the bosonic scaling contribution from the covariant derivative 
when combining various terms in the Lagrangian. Subsequently, the fermionic rescaling is
performed based on the number of spinors appearing in each term (two spinors for each
$\Delta^\mu$ or $\Box^\mu$, and four for each $(\Delta^\mu)_{\theta^4}$). 
The worldsheet derivative notation is given by $\partial_\tau x = \partial_0 x = \dot x$ 
and $\partial_\sigma x = \partial_1 x = x'$.

The various sectors of the worldsheet Lagrangian 
are assembled keeping $x^-$ and its derivatives explicit;
these will be removed by imposing the covariant gauge 
constraints.  From the supervielbein and superconnection
\be
L_{at}^\mu & = & 
	 e^\mu_{\phantom{\mu}\nu}\partial_a x^\nu
  	  - 4i\bar\theta^I \Gamma^\mu \left( \frac{\sinh^2 (t{\cal M}/2)}{{\cal M}^2}
  	  \right){\cal D}_a \theta^I \nonumber \\
& \approx & 
	e^\mu_{\phantom{\mu}\nu}\partial_ax^\nu 
	- i\bar\theta^I \Gamma^\mu\left(t^2 + \frac{t^4{\cal M}^2}{12}\right){\cal D}_a\theta^I \\
L_{at}^I & = & \frac{\sinh t{\cal M}}{{\cal M}}{\cal D}_a\theta^I 
	\approx \left( t+\frac{t^3}{6}{\cal M}^2\right){\cal D}_a\theta^I~, 
\ee
we form the following objects:
\be
\label{L0L0}
L_0^\mu L_0^\mu & = & 
	\frac{1}{R^2}\left\{
	2p_-\dot x^- - p_-^2 (x^A)^2 + (\dot x^A)^2 - 2ip_-\Delta_0^-\right\} 
\nonumber \\
& & 	+ \frac{1}{R^4}\biggl\{
	(\dot x^-)^2 - 2p_-y^2\dot x^- + \frac{1}{2}(\dot z^2 z^2 - \dot y^2 y^2)
	+ \frac{p_-^2}{2}(y^4-z^4) 
\nonumber \\
& & \kern-50pt	-2i\left[\frac{1}{2}\dot x^-\Delta_0^- + p_-\Delta_2^-
	+p_-(\Delta_0^-)_{\theta^4} - \frac{p_-}{4}(y^2-z^2)\Delta_0^- +\dot x^A \Delta_1^A
	\right] \biggr\} +{\cal O}(R^{-6})
\ee
\be
\label{L1L1}
L_1^\mu L_1^\mu & = & 
	\frac{1}{R^2}({x'}^A)^2 
\nn\\
&&	+ \frac{1}{R^4}\biggl\{
	\frac{1}{2}({z'}^2 z^2 - {y'}^2 y^2) + ({x'}^-)^2 - 2i{x'}^A {\Delta'}_1^A
	-i{x'}^-{\Delta'}_0^- \biggr\}+{\cal O}(R^{-6})
\ee
\be 
\label{L0L1}
L_0^\mu L_1^\mu & = & 
	\frac{1}{R^2}\left\{ p_-{x'}^- + \dot x^A{x'}^A - ip_-{\Delta'}_0^-\right\}
\nonumber \\
& &  	+ \frac{1}{R^4}\biggl\{ {x'}^- \dot x^- - p_-y^2{x'}^- + \frac{1}{2}(z^2 \dot z_k z'_k
	-y^2 \dot y_{k'}{y'}_{k'}) 
 	-i p_-{\Delta'}_2^- -i p_-({\Delta'}_0^-)_{\theta^4} 
\nonumber \\
& & 	-i \frac{p_-}{4}(z^2 - y^2){\Delta'}_0^- 
 	- \frac{i}{2}\dot x^-{\Delta'}_0^-
	-i\dot x^A {\Delta'}_1^A -i {x'}^A \Delta_1^A - \frac{i}{2}{x'}^-\Delta_0^- \biggr\}
	+{\cal O}(R^{-6})~.
\nn\\
&&
\ee

It will be advantageous
to enforce the light-cone gauge condition $x^+ = \tau$ at all orders in 
the theory.\footnote{ This differs from the approach presented in \cite{Parnachev:2002kk}. }   
When fermions are included, this choice allows us to 
keep the $\kappa$-symmetry condition $\Gamma^+ \theta = 0$  exact.
In the pp-wave limit, keeping the worldsheet metric flat 
in this light-cone gauge is consistent with the equations of motion.  
Beyond leading order, however, we are forced to consider curvature 
corrections to the worldsheet metric that appear in both the 
conformal gauge constraints and the worldsheet Hamiltonian.
In the purely bosonic case described in section 2 above, these corrections
are kept implicit by defining gauge constraints
in terms of canonical momenta.    
In the supersymmetric theory, we must explicitly calculate these corrections.
The strategy is to expand the $x^-$ equations of motion in rescaled coordinates
(\ref{newcoords}) and solve for the components of the worldsheet 
metric order-by-order.  By varying $x^-$ in the full Lagrangian we obtain
\be
\label{xmeom}
\frac{\delta {\cal L}}{\delta \dot x^-} & = & 
\frac{1}{2}h^{00} \left\{ \frac{2p_-}{R^2} + \frac{1}{R^4}\left[2\dot x^-
	-2p_-y^2 - i\bar\theta^I\Gamma^- \partial_0 \theta^I
	+2ip_-\bar\theta^I\Gamma^- \epsilon^{IJ}\Pi\theta^J\right]\right\} 
\nonumber \\
& & 	+\frac{i}{2R^4}s^{IJ}\bar\theta^I \Gamma^-\partial_1\theta^J+ {\cal O}(R^{-6}).
\ee
The worldsheet metric is taken to be flat at leading order, so there is no contribution
from $L_0^\mu L_1^\mu$ here.  
To obtain corrections to $h^{ab}$ entirely in terms of physical variables,
however, we must eliminate all instances of $x^-$ 
(or its derivatives) from the above variation.
We can solve the conformal gauge constraints at leading order 
to remove $\dot x^-$ from (\ref{xmeom}).  These constraints are obtained by varying the
Lagrangian with respect to the worldsheet metric itself:
\be
\label{Tab}
T_{ab} = L_a^\mu L_b^\mu - \frac{1}{2}h_{ab}h^{cd}L_c^\mu L_d^\mu~,
\ee
yielding a symmetric traceless tensor with two independent components.
To leading order in $1/R$, we find 
\be
\label{T00}
T_{00} & = & \frac{1}{2}(L_0^\mu L_0^\mu + L_1^\mu L_1^\mu) + \dots = 0
\nonumber \\
	& = & \frac{1}{2R^2}\left( 2p_-\dot x^- - p_-^2(x^A)^2 + (\dot x^A)^2
		- 2ip_- \Delta_0^- + ({x'}^A)^2 \right) + {\cal O}(R^{-4}) \\
\label{T01}
T_{01} & = & L_0^\mu L_1^\mu + \dots = 0
\nonumber \\
	& = & p_-{x'}^- + \dot x^A {x'}^A - ip_-{\Delta'}_0^- + {\cal O}(R^{-4})~.
\ee
Expanding $\dot x^-$ and ${x'}^-$ in the same fashion,
\be
\dot x^- = \sum_n \frac{a_n}{R^n} \qquad {x'}^- = \sum_n \frac{a'_n}{R^n}~,
\ee
we use (\ref{T00}) and (\ref{T01}) to obtain
\be
\label{a0}
a_0 & = & \frac{p_-}{2}(x^A)^2 - \frac{1}{2p_-}\left[(\dot x^A)^2 + ({x'}^A)^2\right]
	+ i\bar\theta^I\Gamma^- \partial_0 \theta^I
	-ip_-\epsilon^{IJ} \bar\theta^I\Gamma^- \Pi\theta^J \\
\label{a1}
a'_0 & = & -\frac{1}{p_-}\dot x^A {x'}^A + i\bar\theta^I\Gamma^-\partial_1\theta^I~.
\ee
By substituting back into (\ref{xmeom}), and performing the analogous operation
for the ${x'}^-$ variation, these leading-order solutions provide the following 
expansions for the objects that enter into the $x^-$ equation of motion:
\be
\label{var1}
\frac{\delta {\cal L}}{\delta \dot x^-} & = & 
	\frac{1}{2}h^{00}\left\{ \frac{2p_-}{R^2} 
	+ \frac{1}{R^4}\left[p_-(z^2-y^2) - \frac{1}{p_-}\left[(\dot x^A)^2 + ({x'}^A)^2\right]
	+ i\bar\theta^I\Gamma^-\partial_0\theta^I \right] \right\} \nonumber \\ 
& & 	+ \frac{i}{2R^4}s^{IJ} \bar\theta^I\Gamma^-\partial_1\theta^J + {\cal O}(R^{-6})
\nonumber \\
\frac{\delta {\cal L}}{\delta {x'}^-} & = & 
	\frac{h^{01}p_-}{R^2} + \frac{h^{11}}{R^4}\left( -\frac{1}{p_-}\dot x^A {x'}^A
	+ \frac{i}{2}\bar\theta^I\Gamma^- \partial_1\theta^I \right)
	-\frac{i}{2R^4}s^{IJ}\bar\theta^I\Gamma^-\partial_0\theta^J+ {\cal O}(R^{-6})~.
\nn\\
&&
\ee
It is obvious from these expressions that the $x^-$ equation of motion will not
be consistent with the standard choice of flat worldsheet metric 
($h^{00}=-h^{11}=1,h^{01}=0$). We therefore expand $h^{ab}$ in powers
of $R^{-1}$, taking it to be flat at leading order and allowing the higher-order 
terms (the $\tilde h^{ab}$) to depend on the physical variables in some way:
\be
h^{00}  =  -1 + \frac{\tilde h^{00}}{R^2} + {\cal O}(R^{-4})\qquad
h^{11}  =  1 + \frac{\tilde h^{11}}{R^2} + {\cal O}(R^{-4}) \qquad
h^{01}  =  \frac{\tilde h^{01}}{R^2} + {\cal O}(R^{-4})~.
\ee
Using (\ref{a0}) and (\ref{a1}), we find that the specific metric choice
\be
\label{h00}
\tilde h^{00} & = & \frac{1}{2}(z^2-y^2) 
	- \frac{1}{2p_-^2}\left[(\dot x^A)^2 + ({x'}^A)^2\right] 
	+ \frac{i}{2p_-}\bar\theta^I\Gamma^-\partial_0\theta^I
	-\frac{i}{2p_-}s^{IJ}\bar\theta^I\Gamma^-\partial_1\theta^J \\
\label{h01}
\tilde h^{01} & = & \frac{1}{p_-^2}\dot x^A {x'}^A 
	- \frac{i}{2p_-}\bar\theta^I \Gamma^- \partial_1\theta^I
	+\frac{i}{2p_-}s^{IJ}\bar\theta^I\Gamma^-\partial_0\theta^J~
\ee
simplifies the expressions of (\ref{var1}) to
\be
\frac{\delta {\cal L}}{\delta \dot x^-} = 1 + {\cal O}(R^{-4}) \qquad
	\frac{\delta {\cal L}}{\delta {x'}^-} = {\cal O}(R^{-4})~.
\ee
The $x^-$ equation of motion is then consistent with the standard 
light-cone gauge choice $\dot x^+ = p_-$ to ${\cal O}(1/R^2)$
(with no corrections to $p_-$, which must remain constant).
Note that $\tilde h^{00} = -\tilde h_{00}$ and $\tilde h_{00} = \tilde h_{11}$.
The fact that these curvature corrections have bi-fermionic contributions is
ultimately due to the presence of a non-vanishing $G_{--}$ term in the expanded 
metric (\ref{expmetric2}).  

Since the worldsheet metric is known to ${\cal O}(1/R^2)$, $x^-$ 
can now be determined to this order from the covariant gauge constraints
(\ref{Tab}).  By invoking the leading-order solutions (\ref{T00},\ref{T01}), 
we can simplify the equations to some extent:
\be
\label{tcorr00}
T_{00} & = & \frac{1}{2}\left( L_0^\mu L_0^\mu + L_1^\mu L_1^\mu \right)
	+ \frac{\tilde h^{00}}{R^2}L_1^\mu L_1^\mu + {\cal O}(R^{-3}) = 0 \\
T_{01} & = & L_0^\mu L_1^\mu - \frac{\tilde h_{01}}{R^2}L_1^\mu L_1^\mu +{\cal O}(R^{-3}) = 0~.
\ee
Equation (\ref{tcorr00}) may be expanded to solve for $a_2$, the first subleading
correction to $\dot x^-$:
\be
\label{a2}
T_{00} & = & 2p_- a_2 + a_0^2 - 2p_- y^2 a_0 + {a'}_0^2 + \frac{1}{2}(\dot z^2 z^2 - \dot y^2 y^2)
	+ \frac{p_-^2}{2}(y^4 - z^4) + \frac{1}{2}({z'}^2 z^2 - {y'}^2 y^2) 
\nonumber \\
& & 	+ (z^2 - y^2)({x'}^A)^2
	-\frac{1}{p_-^2}\left[(\dot x^A)^2 + ({x'}^A)^2\right]({x'}^A)^2
	+\frac{i}{p_-}({x'}^A)^2\bar\theta^I\Gamma^-\partial_0\theta^I
\nonumber \\
& & 	-\frac{i}{p_-}({x'}^A)^2 s^{IJ}\bar\theta^I\Gamma^-\partial_1\theta^J
	-ia_0 \Delta_0^- - 2ip_-\Delta_2^- - 2ip_-(\Delta_0^-)_{\theta^4}
	+\frac{ip_-}{2}(y^2-z^2)\Delta_0^- 
\nonumber \\
& & 	- 2i(\dot x^A\Delta_1^A + {x'}^A\Delta_1^A)
	-ia'_0{\Delta'}_0^- = 0~.
\ee

The remaining independent component $T_{01}$ is the current associated with
translation symmetry on the closed-string worldsheet.  Enforcing the
constraint $T_{01} = 0$ is equivalent to imposing the level-matching condition
on physical string states.  This condition can be used to fix 
higher-order corrections to ${x'}^-$, as is required by conformal 
invariance on the worldsheet.  However, since our goal is to examine curvature
corrections to the pp-wave limit using first-order perturbation theory,
we will only need to enforce the level-matching condition on 
string states that are eigenstates of the pp-wave theory.
We therefore need only consider the equation $T_{01} = 0$ 
to leading order in the expansion, which yields (\ref{a1}) above.
If we were interested in physical eigenstates of the
geometry corrected to ${\cal O}(1/R^2)$ (ie.~solving the theory exactly to
this order), we would be forced to solve $T_{01}=0$ to ${\cal O}(1/R^2)$.

With solutions to the $x^-$ equations of motion and an expansion
of the worldsheet metric to the order of interest, we may proceed
with expressing the Hamiltonian as the generator of light-cone
time translation: $p_+ = \delta{\cal L}/\delta\dot x^+$.
It is helpful to first vary 
$\Delta^\mu$ with respect to $\partial_0 t$ and $\partial_0\phi$:
\be
\frac{\delta \Delta^\mu}{\delta(\partial_0 t)}
	& = & \bar\theta^I\Gamma^\mu\left[
	-\frac{1}{2R^3}z_j\Gamma^{0j}\theta^I 
	- \frac{1}{2}\epsilon^{IJ}\Pi\left(\frac{1}{R^2}+\frac{z^2}{2R^4}\right)\theta^J\right]
	+ {\cal O}(R^{-6})
\\
\frac{\delta \Delta^\mu}{\delta(\partial_0 \phi)}
	& = & \bar\theta^I\Gamma^\mu\left[
	-\frac{1}{2R^3}y_{j'}\Gamma^{9j'}\theta^I - \frac{1}{2}\epsilon^{IJ}\Pi
	\left(\frac{1}{R^2} - \frac{y^2}{2R^4}\right)\theta^J\right]
	+ {\cal O}(R^{-6})~.
\ee
The kinetic term in the Lagrangian (\ref{lagrangiank}) yields
\be
\frac{\delta {\cal L}_{\rm Kin}}{\delta \dot x^+} & = & 
	\frac{1}{R^2}\left\{
	p_-(x^A)^2 - \dot x^- + i\Delta_0^- 
	- ip_-\bar\theta^I\Gamma^-\epsilon^{IJ}\Pi\theta^J \right\} 
\nonumber \\
& & 	+\frac{1}{R^4}\biggl\{-\frac{p_-}{2}(y^4-z^4) + y^2\dot x^- + i\Delta_2^-
	+ i(\Delta_0^-)_{\theta^4} + \frac{i}{4}(z^2-y^2)\Delta_0^-
\nonumber \\
& & 	-\frac{ip_-}{2}(z^2-y^2)\bar\theta^I\Gamma^-\epsilon^{IJ}\Pi\theta^J
	-\frac{ip_-}{12}\bar\theta^I\Gamma^-({\cal M}^2)^{IJ}\epsilon^{JL}\Pi\theta^L
\nonumber \\
& & 	+\frac{i}{4}\dot x^A\bar\theta^I\Gamma^A\left(z_k\Gamma^{-k}-y_{k'}\Gamma^{-k'}\right)\theta^I
	-\frac{i}{2}(\dot x^-)\bar\theta^I\Gamma^-\epsilon^{IJ}\Pi\theta^J
	+\biggl[ -\frac{1}{2}(z^2-y^2)
\nonumber \\
& & 	+\frac{1}{2p_-^2}\left[ (\dot x^A)^2 + ({x'}^A)^2\right]
	-\frac{i}{2p_-}\bar\theta^I\Gamma^-\partial_0\theta^I
	+\frac{i}{2p_-}s^{IJ}\bar\theta^I\Gamma^-\partial_1\theta^J\biggr]
	\biggl[ p_-(x^A)^2 - \dot x^- 
\nonumber \\
& & 	+ i\Delta_0^- 
	- ip_-\bar\theta^I\Gamma^-\epsilon^{IJ}\Pi\theta^J\biggr]
 	+\biggl[\frac{1}{p_-^2}\dot x^A{x'}^A
	-\frac{i}{2p_-}\bar\theta^I\Gamma^-\partial_1\theta^I
\nonumber \\
& & 	+\frac{i}{2p_-}s^{IJ}\bar\theta^I\Gamma^-\partial_0\theta^J\biggr]
	\left({x'}^- - i{\Delta'}_0^-\right) \biggr\}+ {\cal O}(R^{-6})~,
\ee
while the Wess-Zumino term (\ref{lagrangianwz}) gives
\be
\frac{\delta {\cal L}_{\rm WZ}}{\delta \dot x^+} & = & 
	\frac{i}{R^2}s^{IJ}\bar\theta^I\Gamma^-\partial_1\theta^J
	+\frac{1}{R^4}\biggl\{
	\frac{i}{4}s^{IJ}\bar\theta^I\Gamma^-(z'_jz_k\Gamma^{jk}-y'_{j'}y_{k'})\theta^J
	+\frac{i}{12}s^{IJ}\bar\theta^I\Gamma^-({\cal M}^2)^{JL}\partial_1\theta^L
\nonumber \\
& & 	-\frac{i}{4}(y^2-z^2)s^{IJ}\bar\theta^I\Gamma^-\partial_1\theta^J
	+\frac{i}{4}{x'}^A s^{IJ}\bar\theta^I\Gamma^A(y_{j'}\Gamma^{-j'}-z_j\Gamma^{-j})\theta^J
	\biggr\}+ {\cal O}(R^{-6})~.
\ee
The variation is completed prior to any gauge fixing (with the worldsheet metric
held fixed).  After computing the variation, 
the light-cone coordinates $x^\pm$ and the worldsheet metric corrections
$\tilde h^{00}, \tilde h^{01}$ are to be replaced with dynamical 
variables according to the 
$x^-$ equations of motion and the gauge conditions $x^+ = \tau$ and $T_{ab} = 0$.
Hence, using $a_0$ and $a_2$ determined from the
covariant gauge constraints (\ref{a0},\ref{a2}), we remove
$x^-$ ($x^+$ has already been replaced with $p_-\tau$ in the above variations)
and restore proper powers of $R$ in the vielbeins 
(so that the desired corrections enter at ${\cal O}(1/R^2)$). 
As expected, the pp-wave Hamiltonian emerges at leading order:
\be
{\cal H}_{pp} = 
	\frac{p_-}{2}(x^A)^2 + \frac{1}{2p_-}\left[(\dot x^A)^2 + ({x'}^A)^2\right]
	-ip_-\bar\theta^I\Gamma^-\epsilon^{IJ}\Pi\theta^J
	+is^{IJ}\bar\theta^I\Gamma^-\partial_1\theta^J~.
\ee
The first curvature correction to the pp-wave limit is found to be
\be
{\cal H}_{\rm int} & = & 
	\frac{1}{R^2}
	\biggl\{
	\frac{1}{4p_-}\left[y^2(\dot z^2 - {z'}^2 - 2{y'}^2)+
	z^2(-\dot y^2 + {y'}^2+2{z'}^2) \right]
\nonumber \\
& & 	+\frac{1}{8p_-^3}\left[3(\dot x^A)^2-({x'}^A)^2\right]
	\left[(\dot x^A)^2 + ({x'}^A)^2\right] 
	+\frac{p_-}{8}\left[(x^A)^2\right]^2
	-\frac{1}{2p_-^3}(\dot x^A{x'}^A)^2
\nonumber \\
& & 	-\frac{i}{4p_-}\sum_{a=0}^1\bar\theta^I (\partial_a x^A \Gamma^A )\epsilon^{IJ}\Gamma^-
	\Pi (\partial_a x^B \Gamma^B)\theta^J
	-\frac{i}{2}p_-(x^A)^2\bar\theta^I\Gamma^-\epsilon^{IJ}\Pi\theta^J
\nonumber \\
& & 	-\frac{i}{2p_-^2}(\dot x^A)^2\bar\theta^I\Gamma^-\partial_0\theta^I
	-\frac{ip_-}{12}\bar\theta^I\Gamma^-({\cal M}^2)^{IJ}\epsilon^{JL}\Pi\theta^L
	-\frac{p_-}{2}(\bar\theta^I\Gamma^-\epsilon^{IJ}\Pi\theta^J)^2
\nonumber \\
& & 	-\frac{i}{2p_-^2}(\dot x^A {x'}^A)s^{IJ}\bar\theta^I\Gamma^-\partial_0\theta^J
	-\frac{i}{4}(y^2-z^2)s^{IJ}\bar\theta^I\Gamma^-\partial_1\theta^J
\nonumber \\
& & 	+\frac{i}{4}{x'}^A s^{IJ}\bar\theta^I\Gamma^A(y_{j'}\Gamma^{-j'}-z_j\Gamma^{-j})\theta^J
	+\frac{i}{4}s^{IJ}\bar\theta^I\Gamma^-(z'_j z_k \Gamma^{jk} - y'_{j'}y_{k'}\Gamma^{j'k'})\theta^J
\nonumber \\
& & 	+\frac{i}{4p_-^2}\left[(\dot x^A)^2-({x'}^A)^2\right]s^{IJ}\bar\theta^I\Gamma^-\partial_1\theta^J
	+\frac{i}{12}s^{IJ}\bar\theta^I\Gamma^-({\cal M}^2)^{JL}\partial_1\theta^L
\nonumber \\
& & 	+\frac{1}{2}(s^{IJ}\bar\theta^I\Gamma^-\partial_1\theta^J)(\bar\theta^K\Gamma^-\epsilon^{KL}
	\Pi\theta^L) + \frac{i}{4}(x^A)^2 s^{IJ}\bar\theta^I\Gamma^-\partial_1\theta^J
	\biggr\}~.
\ee

The full Lagrangian (\ref{lagrangiank},\ref{lagrangianwz}) 
can also be expressed to this order.  
In terms of the quantities found in equations
(\ref{L0L0},\ref{L1L1},\ref{L0L1},\ref{h00},\ref{h01}), 
the kinetic term ${\cal L}_{\rm Kin} = -\frac{1}{2}h^{ab}L_a^\mu L_b^\mu$ 
can be written schematically as
\be
\label{Lkin}
{\cal L}_{\rm Kin} & = & 
	\frac{1}{2} \left(L_0^\mu L_0^\mu - L_1^\mu L_1^\mu \right)_2
	+\frac{1}{2R^2} \left(L_0^\mu L_0^\mu - L_1^\mu L_1^\mu \right)_4
	- \frac{1}{2R^2}\tilde h^{00}\left(L_0^\mu L_0^\mu \right)_2 
\nn\\
& & 	+ \frac{1}{2R^2}\tilde h^{00}\left(L_1^\mu L_1^\mu \right)_2
	- \frac{1}{R^2}\tilde h^{01}\left(L_0^\mu L_1^\mu \right)_2+ {\cal O}(R^{-4})~,
\ee
where external subscripts indicate quadratic or quartic order in fields.  
The Wess-Zumino term is given explicitly by:
\be
\label{WZfull}
{\cal L}_{\rm WZ} & = & 
	-2i\epsilon^{ab}\int_0^1 dt L_{at}^\mu s^{IJ}\bar\theta^I \Gamma^\mu L_{bt}^J 
\nonumber \\
& \approx &  -{ip_-}\left( s^{IJ}\bar\theta^I\Gamma^- \partial_1\theta^J\right) 
	- \frac{i}{R^2}\biggl\{ 
	p_-{\Box'}_2^- + p_-({\Box'}_0^-)_{\theta^4}
	+\frac{p_-}{4}(z^2-y^2){\Box'}_0^- + \frac{1}{2}\dot x^- {\Box'}_0^-
\nonumber \\
& &  	-\frac{1}{2}{x'}^-\Box_0^- + \dot x^A {\Box'}_1^A - {x'}^A\Box_1^A \biggr\}
	+ {\cal O}(R^{-4})~.
\ee
It will be useful to recast both the Hamiltonian and Lagrangian 
in 16-component notation (details may be found in Appendix A):
\be
\label{hamiltonian16}
{\cal H} & = &
	\frac{1}{2p_-} \left( (\dot x^A)^2
    + ({x'}^A)^2 + p_-^2(x^A)^2 \right) 
    -  p_- \psi^\dagger \Pi \psi
    + \frac{i}{2}( \psi {\psi'}
    + \psi^\dagger {\psi'}^\dagger )
\nonumber \\
& &     +\frac{1}{R^2}\biggl\{
     \frac{z^2}{4p_-} \left[{y'}^2
    + 2{z'}^2- \dot y^2 \right]
    - \frac{y^2}{4p_-} \left[{z'}^2 + 2{y'}^2
    - \dot z^2 \right]
	-\frac{1}{2p_-^3}(\dot x^A{x'}^A)^2
\nonumber \\
& & 	+\frac{1}{8p_-^3}\left[3(\dot x^A)^2-({x'}^A)^2\right]
	\left[(\dot x^A)^2+({x'}^A)^2\right]
	+\frac{p_-}{8}\left[(x^A)^2\right]^2
\nonumber \\
& &     + \frac{i}{8} \psi \left(
    	z_k z'_j \gamma^{jk}
    	- y_{k'}  y'_{j'} \gamma^{j'k'}
    	+  {x'}^A (z_k \bar\gamma^{A}\gamma^k - y_{k'} \bar\gamma^{A}\gamma^{k'})
    	\right) \psi
	- \frac{i}{4p_-^2}(\dot x^A)^2\left[\psi\dot\psi^\dagger + \psi^\dagger\dot\psi\right]
\nonumber \\
& &     + \frac{i}{8} \psi^\dagger \left(z_k z'_j \gamma^{jk}
    	- y_{k'} y'_{j'} \gamma^{j'k'}
    	+  {x'}^A (z_k \bar\gamma^{A}\gamma^k - y_{k'} \bar\gamma^{A}\gamma^{k'})
    	\right) \psi^\dagger
\nn\\
&&	+\frac{1}{2p_-}\left( \dot z^i \dot y^{j'} + {z'}^i {y'}^{j'}\right)
	\psi^\dagger \gamma^{ij'}\Pi \psi
    + \frac{i}{8}(z^2 - y^2) (\psi \psi'
        + \psi^\dagger  {\psi'}^\dagger )
\nn\\
&&\kern-20pt 	- \frac{1}{4p_-} \left[
    	(\dot z^2 - \dot y^2)
    	+ ({z'}^2 - {y'}^2 ) \right]
    	\psi^\dagger \Pi \psi
 	+ \frac{i}{8}\left[
	\frac{1}{p_-^2}\left( (\dot x^A)^2 - ({x'}^A)^2\right) 
	+(x^A)^2\right](\psi\psi'+\psi^\dagger{\psi'}^\dagger)
\nn\\
&&	-\frac{p_-}{2}(x^A)^2(\psi^\dagger\Pi\psi)
 	- \frac{i}{4p_-^2}(\dot x^A{x'}^A)(\psi\dot\psi+\psi^\dagger\dot\psi^\dagger)
	+\frac{p_-}{48}
   	(\psi^\dagger \gamma^{jk} \psi )( \psi^\dagger \gamma^{jk} \psi)
\nn\\
&&	-\frac{p_-}{48}
    	(\psi^\dagger \gamma^{j'k'} \psi )( \psi^\dagger \gamma^{j'k'} \psi)
     - \frac{i}{192}(\psi\gamma^{jk} \psi
    	+ \psi^\dagger\gamma^{jk}\psi^\dagger)
    	(\psi^\dagger \gamma^{jk} \Pi\psi' - \psi \gamma^{jk} \Pi{\psi'}^\dagger)
\nn\\
&&	+\frac{p_-}{2}(\psi^\dagger\Pi\psi)(\psi^\dagger\Pi\psi)
     + \frac{i}{192}(\psi\gamma^{j'k'} \psi
    	+ \psi^\dagger\gamma^{j'k'}\psi^\dagger)
	(\psi^\dagger \gamma^{j'k'} \Pi\psi' - \psi \gamma^{j'k'} \Pi{\psi'}^\dagger)
\nn\\
&&\kern+150pt
	- \frac{i}{4}(\psi\psi'+\psi^\dagger{\psi'}^\dagger)(\psi^\dagger\Pi\psi)
	\biggr\}+ {\cal O}(R^{-4})~.
\ee
One could scale the length of the worldsheet such that all $p_-$ are absorbed into
the upper limit on worldsheet integration over $d\sigma$. To organize correction terms by their 
corresponding coupling strength in the gauge theory, however, we find it convenient to
keep factors of $p_-$ explicit in the above expression.  
The Lagrangian can be computed from (\ref{Lkin},\ref{WZfull}), giving
\be
\label{Lkinfinal}
{\cal L}_{\rm Kin} & = & 
	p_-\dot x^- 
	- \frac{1}{2}\left[ {p_-^2} (x^A)^2 - (\dot x^A)^2 + ({x'}^A)^2 \right]
	- \frac{ip_-}{2}(\psi\dot\psi^\dag + \psi^\dag\dot\psi)
	- p_-^2 \psi\Pi\psi^\dag
\nn\\
& & 	+ \frac{1}{2R^2}\biggl\{
	(\dot x^-)^2 - 2p_-y^2 \dot x^- + \frac{1}{2}(\dot z^2 z^2 - \dot y^2 y^2)
	+ \frac{p_-^2}{2}(y^4 - z^4)
\nn\\
& & 	-\frac{ip_-}{4}(\dot z_j z_k)(\psi\gamma^{jk} \psi^\dag + \psi^\dag \gamma^{jk} \psi)
	+\frac{ip_-}{4}(\dot y_{j'} y_{k'})(\psi\gamma^{j'k'} \psi^\dag + \psi^\dag \gamma^{j'k'} \psi)
\nn\\
& & 	-\frac{ip_-}{48}(\psi\gamma^{jk}\psi^\dag)(\psi\gamma^{jk}\Pi\dot\psi^\dag 
		- \psi^\dag\gamma^{jk}\Pi\dot\psi)
	+\frac{ip_-}{48}(\psi\gamma^{j'k'}\psi^\dag)(\psi\gamma^{j'k'}\Pi\dot\psi^\dag 
		- \psi^\dag\gamma^{j'k'}\Pi\dot\psi)
\nn\\
& & 	+ \frac{i}{2}\left[ \frac{p_-}{2}(y^2 - z^2) - \dot x^-\right](\psi\dot\psi^\dag + \psi^\dag\dot\psi)
 	-p_-\left[2\dot x^- - p_-(y^2 - z^2)\right]\psi\Pi\psi^\dag
\nn\\
& & 	-\frac{p_-^2}{24}(\psi^\dag\gamma^{jk}\psi)^2
	+\frac{p_-^2}{24}(\psi^\dag\gamma^{j'k'}\psi)^2
	+\frac{ip_-}{4}(\dot x^A z_j)(\psi\gamma^A\bar\gamma^j\psi^\dag + \psi^\dag\gamma^A\bar\gamma^j\psi)
\nn\\
& & 	-\frac{ip_-}{4}(\dot x^A y_{j'})
		(\psi\gamma^A\bar\gamma^{j'}\psi^\dag + \psi^\dag\gamma^A\bar\gamma^{j'}\psi)
	+\frac{1}{4}(\dot x^A\dot x^B)(\psi^\dag\gamma^A\Pi\bar\gamma^B\psi 
		- \psi\gamma^A\Pi\bar\gamma^B\psi^\dag )
\nn\\
& & 	-\frac{1}{2}({z'}^2z^2 - {y'}^2y^2)-({x'}^-)^2
	+\frac{i}{2}{x'}^-(\psi{\psi'}^\dag + \psi^\dag\psi')
\nn\\
& & 	-\frac{1}{4}({x'}^A{x'}^B)(\psi^\dag\gamma^A\Pi\bar\gamma^B\psi - \psi\gamma^A\Pi\bar\gamma^B\psi^\dag)
	-\tilde h^{00}\bigl[ 2p_-\dot x^- - p_-^2 (x^A)^2 + (\dot x^A)^2 - ({x'}^A)^2
\nn\\
& & 	- ip_-(\psi\dot\psi^\dag + \psi^\dag\dot\psi)
	-2 p_-^2 \psi\Pi\psi^\dag \bigr]
	-2\tilde h^{01}\left[ p_-{x'}^- + \dot x^A {x'}^A -\frac{ip_-}{2}(\psi{\psi'}^\dag + \psi^\dag\psi')
	\right] \biggr\}
\nn\\
&&\kern+320pt + {\cal O}(R^{-4})~,
\ee
and
\be
\label{Lwzfinal}
{\cal L}_{\rm WZ} & = & 
	-\frac{ip_-}{2}(\psi\psi' + \psi^\dag{\psi'}^\dag )
	-\frac{i}{R^2}\biggl\{
	\frac{p_-}{8}(z'_j z_k)(\psi\gamma^{jk}\psi + \psi^\dag\gamma^{jk}\psi^\dag)
\nn\\
& & 	-\frac{p_-}{8}(y'_{j'}y_{k'})(\psi\gamma^{j'k'}\psi + \psi^\dag\gamma^{j'k'}\psi^\dag)
	+\frac{1}{4}\left[ \dot x^- + \frac{p_-}{2}(z^2 - y^2)\right](\psi\psi' + \psi^\dag{\psi'}^\dag)
\nn\\
& & 	-\frac{1}{4}({x'}^-)(\psi\dot\psi + \psi^\dag\dot\psi^\dag )
	+ \frac{i}{8}({x'}^A \dot x^B + \dot x^A {x'}^B)(\psi^\dag\gamma^A\Pi\bar\gamma^B\psi^\dag
		- \psi\gamma^A\Pi\bar\gamma^B\psi )
\nn\\
& & 	+ \frac{p_-}{8}({x'}^A z_j )(\psi^\dag\gamma^A\bar\gamma^j\psi^\dag
		+ \psi\gamma^A\bar\gamma^j\psi )
	- \frac{p_-}{8}({x'}^A y_{j'} )(\psi^\dag\gamma^A\bar\gamma^{j'}\psi^\dag
		+ \psi\gamma^A\bar\gamma^{j'}\psi )
\nn\\
& & 	+ \frac{p_-}{8}(\psi\gamma^{jk}\psi + \psi^\dag\gamma^{jk}\psi^\dag)
		(\psi\gamma^{jk}\Pi{\psi'}^\dag - \psi^\dag\gamma^{jk}\Pi\psi') 
\nn\\
& & 	- \frac{p_-}{8}(\psi\gamma^{j'k'}\psi + \psi^\dag\gamma^{j'k'}\psi^\dag)
		(\psi\gamma^{j'k'}\Pi{\psi'}^\dag - \psi^\dag\gamma^{j'k'}\Pi\psi') 
	\biggr\}+ {\cal O}(R^{-4})~.
\ee
For later convenience, the Lagrangian is not fully gauge fixed, though we set $\dot x^+$
to $p_-$ for simplicity and ignore any $\ddot x^+$ that arise through
partial integration (since we will ultimately choose the light-cone gauge 
$x^+ = p_-\tau$). 
As noted above, sending $h^{00} \rightarrow -1 + {\tilde h^{00}}/{R^2}$
simply rewrites the function $h^{00}$, 
and does not amount to a particular gauge choice 
for the worldsheet metric.

\section{Quantization of the lightcone gauge Hamiltonian}

Our goal is to calculate explicit energy corrections due to the
rather complicated perturbed Hamiltonian derived in the last section.
To explain our strategy, we begin with a review of the pp-wave 
energy spectrum in the Penrose limit. This limit is obtained by keeping
only the leading term in $R^{-1}$ in the Hamiltonian expansion of
(\ref{hamiltonian16}) and leads to linear equations of motion for the 
fields. The eight bosonic transverse string coordinates obey the equation
\begin{eqnarray}
{\ddot{x}}^A - {x''}^A + p_-^2 x^A = 0\ .
\end{eqnarray}
This is solved by the usual expansion in terms of Fourier modes
\begin{eqnarray}
x^A(\sigma,\tau) & = &
    \sum_{n=-\infty}^\infty x_n^A(\tau) e^{-i k_n \sigma}
\nonumber \\
\label{adef}
x_n^A(\tau) & = & \frac{i}{\sqrt{2 \omega_n}} (a_n^A e^{-i \omega_n \tau}
    - {a_{-n}^{A\dagger}} e^{i \omega_n \tau} )\ ,
\end{eqnarray}
where $k_n = n$ (integer), $\omega_n = \sqrt{ p_-^2 + k_n^2}$,
and the raising and lowering operators obey the commutation relation
$ [ a_m^A, {a_n^B}^\dagger ] = \delta_{mn}\delta^{AB}$.
The bosonic piece of the pp-wave Hamiltonian takes the form
\begin{eqnarray}
{\cal H}_{\rm pp}^B & = & \frac{1}{p_-} \sum_{n=-\infty}^\infty \omega_n
    \left( {a_n^A}^\dagger a_n^A + 4 \right)\ .
\end{eqnarray}
The fermionic equations of motion are
\begin{eqnarray}
(\dot\psi^\dagger + \psi') + i{p_-}\Pi \psi^\dagger = 0 \\
(\dot\psi + {\psi'}^\dagger) - i{p_-}\Pi \psi  = 0\ ,
\end{eqnarray}
where $\psi$ is a 16-component complex SO(9,1) Weyl spinor. As mentioned
earlier, $\psi$ is further restricted by a light-cone gauge fixing condition
$\bar\gamma^9\psi=\psi$ which reduces the number of spinor components to 8
(details are given in the Appendix). In what follows, $\psi$, and the various 
matrices acting on it, should therefore be regarded as 8-dimensional. The
fermionic equations of motion are solved by
\begin{eqnarray}
\label{Fourierfermi}
\psi & = &
    \sum_{n=-\infty}^\infty \psi_n(\tau)  e^{-i k_n \sigma}
\\
\label{bdef1}
\psi_n(\tau) & = & 
	\frac{1}{2\sqrt{p_-}}\left( 
	A_n b_n e^{-i\omega_n\tau} + B_n b_{-n}^\dagger e^{i\omega_n\tau} \right)e^{-ik_n\sigma}
\\
\label{bdef2}
\psi_n^\dagger(\tau) & = & 
	\frac{1}{2\sqrt{p_-}}\left( 
	\Pi B_n b_n e^{-i\omega_n\tau} 
	- \Pi A_n b_{-n}^\dagger e^{i\omega_n\tau} \right) e^{-ik_n\sigma}~,
\end{eqnarray}
where we have defined
\begin{eqnarray}
\label{Amatdef}
A_n & \equiv & \frac{1}{\sqrt{\omega_n}}\left(\sqrt{\omega_n - k_n} - \sqrt{\omega_n + k_n}\Pi\right) \\
\label{Bmatdef}
B_n & \equiv & \frac{1}{\sqrt{\omega_n}}\left(\sqrt{\omega_n + k_n} + \sqrt{\omega_n - k_n}\Pi\right)~.
\end{eqnarray}
The anticommuting mode operators $b_n,b_n^\dagger$ carry a spinor index which takes
8 values. In the gamma matrix representation described in the Appendix, the matrix 
$\Pi$ is diagonal and assigns eigenvalues $\pm 1$ to the mode operators.
The fermionic canonical momentum is $\rho = {ip_-}\psi^\dagger$, which implies that
the fermionic creation and annihilation operators obey the anticommutation rule
$\{ b_m^\alpha, {b_n^\beta}^\dagger \} =  \delta^{\alpha\beta}\delta_{mn}$.
The fermionic piece of the pp-wave Hamiltonian can be written in terms of
these operators as
\begin{eqnarray}
{\cal H}_{\rm pp}^F & = & \frac{1}{p_-}\sum_{n=-\infty}^\infty
   \omega_n \left( b_n^{\alpha\dagger} b_n^\alpha - 4 \right)\ .
\end{eqnarray}
Given our earlier conventions, it is necessary to invoke 
the coordinate reflection $x^\mu \rightarrow -x^\mu$ 
(Metsaev studied a similar operation on the pp-wave Hamiltonian in
\cite{Metsaev:2001bj}).
Such a transformation is, at this stage, 
equivalent to sending $x^A \rightarrow -x^A$, $p_- \rightarrow -p_-$,
and ${\cal H} \rightarrow -{\cal H}$. 
In essence, this operation allows us to choose the positive-energy 
solutions to the fermionic equations of motion while maintaining
our convention that $b^{\alpha^\dag}$ represent a creation operator and $b^{\alpha}$
denote an annihilation operator. The total pp-wave Hamiltonian
\be
{\cal H}_{\rm pp} = \frac{1}{p_-}\sum_{n=-\infty}^\infty
   \omega_n \left( {a_n^A}^\dagger a_n^A + b_n^{\alpha\dagger} b_n^\alpha  \right)\ 
\ee
is just a collection of free, equal mass fermionic and bosonic oscillators.

Canonical quantization requires that we express the Hamiltonian in terms
of physical variables and conjugate momenta. At leading order in $1/R^2$, $\dot x^A$ 
is canonically conjugate to $x^A$ and can be expanded in terms of creation and
annihilation operators.  Beyond leading order, however, the conjugate variable 
$p_A = \delta {\cal L} /\delta \dot x^A $ differs from $\dot x^A$ by terms
of ${\cal O}(1/R^2)$. Substituting these ${\cal O}(1/R^2)$ corrected 
expressions for canonical momenta into the pp-wave Hamiltonian
\be
{\cal H}_{\rm pp} \sim (\dot x^A)^2 + \psi^\dagger \Pi \psi + \psi^\dagger {\psi'}^\dagger
\ee 
to express it as a function of canonical variables
will yield indirect ${\cal O}(1/R^2)$ corrections to the Hamiltonian 
(to which we must add the contribution of explicit ${\cal O}(1/R^2)$ 
corrections to the action). For example, bosonic momenta in 
the $SO(4)$ descending from the $AdS_5$ subspace
take the following corrections:
\be
p_k & = & \dot z_k + \frac{1}{R^2}\biggl\{
	\frac{1}{2} y^2 p_k + \frac{1}{2p_-^2}\left[ (p_A)^2 + ({x'}^A)^2 \right]p_k
 	- \frac{1}{p_-^2}(p_A {x'}^A ){z'}_k
	- \frac{i}{2p_-}p_k \bar\theta^I \Gamma^-\partial_0\theta^I
\nn\\
& & 	+ \frac{i}{2p_-}p_k s^{IJ}\bar\theta^I\Gamma^-\partial_1\theta^J
	- \frac{ip_-}{4}\bar\theta^I\Gamma^- z_j \Gamma_k^{\phantom{k}j}\theta^I
	- \frac{ip_-}{4}\bar\theta^I \Gamma^k \left(
		z_j\Gamma^{-j} - y_{j'}\Gamma^{-j'}\right)\theta^I
\nn\\
& & 	+ \frac{i}{4}p_A \epsilon^{IJ}\bar\theta^I\Gamma^- \left(
		\Gamma_k \Pi \Gamma^A  + \Gamma^A \Pi \Gamma_k \right ) \theta^J
  	+ \frac{i}{2p_-}{z'}_k\bar\theta^I\Gamma^-\partial_1 \theta^I
	- \frac{i}{2p_-}{z'}_k s^{IJ}\bar\theta^I\Gamma^-\partial_0\theta^J
\nn\\
& & 	+ \frac{i}{4}{x'}^A s^{IJ}\epsilon^{JK}\bar\theta^I\Gamma^-
		\left( \Gamma_k\Pi\Gamma^A - \Gamma^A \Pi \Gamma_k \right) \theta^K
	\biggr\}+{\cal O}(R^{-4})~.
\ee
The leading-order relationship $p_k = \dot z_k$ has
been substituted into the correction term at ${\cal O}(1/R^2)$, and the light-cone gauge choice 
$x^+ = p_-\tau$ has been fixed after the variation.

To compute fermionic momenta $\rho = {\delta {\cal L}}/{\delta \dot \psi}$,
it is convenient to work with complex 16-component spinors. 
Terms in ${\cal L}$ relevant to the fermionic momenta $\rho$ are as follows:
\be
\label{Lferm}
{\cal L} & \sim & -ip_- \left( \psi^\dagger \dot\psi \right)
	- \frac{i}{R^2}\biggl\{
	\frac{1}{4}\left[\dot x^- + \frac{p_-}{2}(z^2 - y^2)\right]
		\left(\psi\dot \psi^\dagger + \psi^\dagger\dot\psi\right)
	-\frac{p_-\tilde h^{00}}{2}\left(\psi\dot\psi^\dagger + \psi^\dagger\dot\psi\right)
\nn\\
& & \kern-25pt	+ \frac{p_-}{96}\left(\psi\gamma^{jk}\psi^\dagger\right)
		\left(\psi\gamma^{jk}\Pi\dot\psi^\dagger - \psi^\dagger\gamma^{jk}\Pi\dot\psi\right)
 	-\frac{{x'}^-}{4}\left(\psi\dot\psi + \psi^\dagger\dot\psi^\dagger\right) 
	- (j,k \rightleftharpoons j',k')
	\biggr\}+{\cal O}(R^{-4})~.
\nn\\
& & 
\ee
This structure can be manipulated to simplify the
subsequent calculation.  Using partial integration,
we can make the following replacement at leading order:
\be
\label{IBP}
\frac{ip_-}{2}\left( \psi^\dagger \dot\psi + \psi\dot\psi^\dagger\right) = 
	ip_- \left(\psi^\dagger \dot \psi\right) + {\rm surface\ terms}.
\ee
Operations of this sort have no effect on the $x^-$ equation of motion or 
the preceding calculation of $\delta {\cal L} / \delta \dot x^+$, for example.
Similarly, terms in ${\cal L}$ containing the matrix $({\cal M})^2$ 
may be transformed according to
\be
-\frac{ip_-}{96} \left(\psi\gamma^{jk}\psi^\dagger\right)\left(\psi\gamma^{jk}\Pi\dot\psi^\dagger
	- \psi^\dagger\gamma^{jk}\Pi\dot\psi\right) 
& = & 	\frac{ip_-}{48}
	\left(\psi\gamma^{jk}\psi^\dagger\right)
		\left(\psi^\dagger\gamma^{jk}\Pi\dot\psi\right)~.
\ee
Terms of the form
\be
\label{fierz}
\frac{1}{4}\left(\dot x^- \right)
		\left(\psi\dot \psi^\dagger + \psi^\dagger\dot\psi\right)~,
\ee
however,
cannot be treated in the same manner.  
The presence of (\ref{fierz}) ultimately 
imposes a set of second-class constraints on the theory,
and we will eventually be lead to treat $\psi^\dagger$ 
as a constrained, dynamical degree of freedom in the Lagrangian.
The fermionic momenta therefore take the form
\be
\rho_\alpha & = & ip_-\psi^\dagger_\alpha + \frac{1}{R^2}\biggl\{
	\frac{i}{4}\left(\dot x^- + \frac{p_-}{2}(z^2 - y^2)\right)\psi^\dagger_\alpha
	- \frac{ip_-}{2}\tilde h^{00} \psi^\dagger_\alpha 
	- \frac{i{x'}^-}{4}\psi_\alpha 
\nn\\
& & 	- \frac{ip_-}{48}\left[ \left(\psi\gamma^{jk}\psi^\dagger\right)
		\left(\psi^\dagger\gamma^{jk}\Pi \right)_\alpha 
	- (j,k \rightleftharpoons j',k')\right]
	\biggr\}+{\cal O}(R^{-4}) 
\\
\rho^\dagger_\alpha & = & 
	\frac{1}{R^2}\biggl\{
	\frac{i}{4}\left(\dot x^- + \frac{p_-}{2}(z^2 - y^2)\right)\psi_\alpha
	- \frac{ip_-}{2}\tilde h^{00} \psi_\alpha 
 	- \frac{i{x'}^-}{4}\psi^\dagger_\alpha 
	\biggr\}+{\cal O}(R^{-4})~.
\ee
Using (\ref{a0}) and (\ref{a1}) to replace $\dot x^-$ and ${x'}^-$ at leading order
(in 16-component spinor notation), and using (\ref{h00}) to implement the appropriate
curvature corrections to the $h^{00}$ component of the worldsheet metric,
we find
\be
\label{rhoeqn}
\rho	& = & 	ip_- \psi^\dagger + \frac{1}{R^2}\biggl\{
	\frac{1}{4}y^2\rho + \frac{1}{8p_-^2}\left[ (p_A^2) + ({x'}^A)^2\right] \rho
	+ \frac{i}{4p_-}(p_A {x'}^A )\psi
	+ \frac{i}{4p_-}\left( \rho\Pi\psi \right) \rho 
\nn\\
& & 	- \frac{i}{8p_-}\left( \psi\rho' + \rho\psi' \right)\psi
	+ \frac{i}{8p_-}\left(\psi\psi' - \frac{1}{p_-^2}\rho\rho'\right)\rho
\nn\\
& &  	+ \frac{i}{48p_-}\left[ \left(\psi\gamma^{jk}\rho\right)\left(\rho\gamma^{jk}\Pi\right)
	- (j,k, \rightleftharpoons j',k') \right] 
	\biggr\}+{\cal O}(R^{-4})\ 
\\
\rho^\dagger & = & \frac{1}{R^2}\biggl\{
	\frac{i}{4}p_- y^2\psi + \frac{i}{8p_-}\left[ (p_A^2) + ({x'}^A)^2 \right]\psi
	+ \frac{1}{4p_-^2}\left( p_A {x'}^A \right)\rho
	-\frac{1}{4}\left( \rho\Pi\psi \right)\psi
\nn\\
& & 	-\frac{1}{8p_-^2}\left(\psi\rho' + \rho\psi' \right)\rho
	- \frac{1}{8}\left(\psi\psi' - \frac{1}{p_-^2}\rho\rho'\right)\psi
	\biggr\}+{\cal O}(R^{-4})~.
\ee
Denoting the ${\cal O}(1/R^2)$ corrections to $\rho$ in (\ref{rhoeqn}) by $\Phi$, 
the pp-wave Hamiltonian can be expressed in terms of canonical variables as
\be
{\cal H}_{\rm pp} & =  & - p_- \psi^\dagger \Pi \psi + \frac{i}{2}\psi\psi' 
	+ \frac{i}{2}\psi^\dagger {\psi'}^\dagger
\nn\\
& = & i\rho\Pi\psi + \frac{i}{2}\psi\psi' - \frac{i}{2p_-^2}\rho\rho'
	+ \frac{1}{R^2}\left\{ 
	\frac{i}{2p_-^2}\rho \Phi' + \frac{i}{2p_-^2}\Phi \rho' 
	- i \Phi \Pi \psi  \right\}~.
\ee
The ${\cal O}(1/R^2)$ correction to the Hamiltonian can also be expressed in terms
of canonical variables. The overall canonical Hamiltonian can conveniently be broken
into its BMN limit $({\cal H}_{\rm pp})$, pure bosonic $({\cal H}_{\rm BB})$,
pure fermionic $({\cal H}_{\rm FF})$ and boson-fermion $({\cal H}_{\rm BF})$
interacting subsectors:
\be
\label{Hppwave}
{\cal H}_{\rm pp} & = & 
	\frac{p_-}{2}(x^A)^2 + \frac{1}{2p_-}\left[(p_A)^2 + ({x'}^A)^2\right]
	+ {i}\rho\Pi\psi + \frac{i}{2}\psi\psi' - \frac{i}{2p_-^2} \rho \rho'
\ee
\be
\label{Hpurbos}
{\cal H}_{\rm BB} & = & \frac{1}{R^2}\biggl\{
	\frac{1}{4p_-}\left[ -y^2\left( p_z^2 + {z'}^2 + 2{y'}^2\right)
	+ z^2\left( p_{y}^2 + {y'}^2 + 2{z'}^2 \right)\right]
	+ \frac{p_-}{8}\left[ (x^A)^2 \right]^2
\nn\\
& & 	- \frac{1}{8p_-^3}\left\{  \left[ (p_A)^2\right]^2 + 2(p_A)^2({x'}^A)^2 
	+ \left[ ({x'}^A)^2\right]^2 \right\}
	 + \frac{1}{2p_-^3}\left({x'}^A p_A\right)^2
	\biggr\}
\ee
\be
\label{hbfeqn}
{\cal H}_{\rm FF} & = & -\frac{1}{4R^2}\biggl\{
	\frac{1}{p_-}\left(\rho\Pi\psi\right)^2
	+ \frac{1}{p_-^3}\left(\rho\Pi\psi\right)\rho\rho'
 	+ \frac{1}{2p_-^3}\left(\psi\psi' - \frac{1}{p_-^2}\rho\rho'\right)\rho\rho'
\nn\\
& & 	+\frac{1}{2p_-}\left(\psi\psi' - \frac{1}{p_-^2}\rho\rho'\right)(\rho\Pi\psi)
	+ \frac{1}{2p_-^3}\left(\psi\rho' + \rho\psi'\right)\rho'\psi
\nn\\
& & 	+ \frac{1}{12p_-^3}\left(\psi\gamma^{jk}\rho\right)
		\left(\rho\gamma^{jk}\Pi\rho'\right)
\nn\\
& & 	- \frac{1}{48p_-}\left(\psi\gamma^{jk}\psi - \frac{1}{p_-^2}\rho\gamma^{jk}\rho\right)
		\left(\rho'\gamma^{jk}\Pi\psi - \rho\gamma^{jk}\Pi\psi'\right)
	- (j,k \rightleftharpoons j',k')
	\biggr\}	
\ee
\be
\label{hffeqn}
{\cal H}_{\rm BF} & = & 
	\frac{1}{R^2}\biggl\{
	\frac{i}{4}z^2 \psi\psi' -\frac{i}{8p_-^2}\left[(p_A)^2 + ({x'}^A)^2\right]\psi\psi'
	+\frac{i}{4p_-^4}\left[(p_A)^2 + ({x'}^A)^2 + p_-^2(y^2 - z^2)\right]\rho\rho'
\nn\\
& & 	-\frac{i}{2p_-^2}\left( p_k^2 + {y'}^2 - p_-^2 z^2
		-\frac{1}{4}(p_A)^2 - \frac{1}{4}({x'}^A)^2
		-\frac{p_-^2}{2}y^2 \right)\rho\Pi\psi
\nn\\
& & 	+\frac{i}{4}(z'_j z_k)\left(\psi\gamma^{jk}\psi - \frac{1}{p_-^2}\rho\gamma^{jk}\rho\right)
	-\frac{i}{4}(y'_{j'} y_{k'})\left(\psi\gamma^{j'k'}\psi - \frac{1}{p_-^2}\rho\gamma^{j'k'}\rho\right)
\nn\\
& & 	-\frac{i}{8}(z'_k y_{k'} + z_k y'_{k'})
		\left(\psi\gamma^{kk'}\psi - \frac{1}{p_-^2}\rho\gamma^{kk'}\rho\right)
	+\frac{1}{4p_-}(p_k y_{k'} + z_k p_{k'} )\psi\gamma^{kk'}\rho
\nn\\
& & 	+\frac{1}{4p_-}(p_j z'_k)\left(\psi\gamma^{jk}\Pi\psi 
		+ \frac{1}{p_-^2}\rho\gamma^{jk}\Pi\rho\right)
	-\frac{1}{4p_-}(p_{j'} y'_{k'})\left(\psi\gamma^{j'k'}\Pi\psi 
		+ \frac{1}{p_-^2}\rho\gamma^{j'k'}\Pi\rho\right)
\nn\\
& & 	-\frac{1}{4p_-}(p_k y'_{k'} + z'_k p_{k'})
		\left(\psi\gamma^{kk'}\Pi\psi + \frac{1}{p_-^2}\rho\gamma^{kk'}\Pi\rho\right)
	-\frac{i}{2p_-^2}(p_kp_{k'} - z'_k y'_{k'})\psi\gamma^{kk'}\Pi\rho
\nn\\
& & 	-\frac{1}{4p_-^3}(p_A{x'}^A)(\rho\psi' + 2 \psi\rho' )
	\biggr\}~.
\ee

This Hamiltonian has one problem which we must resolve before attempting 
to extract its detailed consequences. At the end of section 2, we argued that
when the theory is restricted to the subspace of string zero-modes (i.e.~excitations 
of the string that are independent of the worldsheet coordinate 
$\sigma$), curvature corrections to the leading pp-wave Hamiltonian should 
vanish. The only terms in the Hamiltonian that survive in this limit are
those with no worldsheet spatial derivatives. Although ${\cal H}_{\rm BB}$ has 
no such terms, the fermionic pieces of the Hamiltonian do. For example,
${\cal H}_{\rm FF}$ contains a term $R^{-2}(\rho\Pi\psi)^2$ which would appear
to modify the zero-mode spectrum at ${\cal O}(1/R^2)$, contrary to 
expectation. In the end, we found that this problem can be traced to the 
presence of second-class constraints involving $\dot\psi^\dag$. As it turns
out, the constrained quantization procedure needed to handle second-class
constraints has the effect, among many others, of resolving the zero-mode
paradox just outlined. To see this, we must work out the appropriate 
constrained quantization procedure. 

The set of constraints that define canonical momenta are
known as primary constraints, and take the generic form $\chi = 0$. 
Primary constraints can be categorized as either first or
second class.  Second-class constraints arise when canonical momenta
do not have vanishing Poisson brackets with the primary
constraints themselves: $\left\{ \rho_\psi,\chi_\psi \right\} \neq
0$, $\left\{ \rho_{\psi^\dagger},\chi_{\psi^\dagger} \right\} \neq
0$. (First-class constraints are characterized by the more typical
condition $\left\{ \rho_{\psi^\dagger},\chi_{\psi^\dagger}\right\}
= \left\{ \rho_\psi,\chi_\psi \right\} = 0$.)
To the order of interest, the primary constraint equations are
\be
\label{con1}
\chi_\alpha^1 & = 0 = & \rho_\alpha - ip_-\psi^\dagger_\alpha
\nn\\
& & 	- \frac{ip_-}{8R^2}\biggl[
		2y^2 + \frac{1}{p_-^2}\left[ (p_A)^2 + ({x'}^A)^2\right]
		- 2(\psi^\dagger\Pi\psi) 
		+ \frac{i}{p_-}(\psi\psi' + \psi^\dagger{\psi'}^\dagger)
	\biggr] \psi_\alpha^\dagger
\nn\\
& & 	-\frac{i}{4p_-R^2}\biggl[
		(p_A {x'}^A) - \frac{ip_-}{2}(\psi{\psi'}^\dagger + \psi^\dagger\psi')
	\biggr]\psi_\alpha
	+ \frac{ip_-}{48R^2}(\psi\gamma^{jk}\psi^\dagger)(\psi^\dagger\gamma^{jk}\Pi)_\alpha
\\
\label{con2}
\chi_\alpha^2 & = 0 = & \rho^\dagger_\alpha
	- \frac{ip_-}{4R^2}\biggl[
		y^2 + \frac{1}{2p_-^2}\left[(p_A)^2 + ({x'}^A)^2 \right] - (\psi^\dagger\Pi\psi)
		+ \frac{i}{2p_-}(\psi\psi' + \psi^\dagger{\psi'}^\dagger)
	\biggr]\psi_\alpha
\nn\\
& & 	-\frac{i}{4p_-R^2}\biggl[
		(p_A{x'}^A)-\frac{ip_-}{2}(\psi{\psi'}^\dagger + \psi^\dagger\psi')
	\biggr]\psi^\dagger_\alpha\ .
\ee
It is clear that these constraints are second-class. In the presence of
second-class constraints, consistent quantization requires that
the quantum anticommutator of two fermionic fields be identified with their
Dirac bracket (which depends on the Poisson bracket algebra of the
constraints) rather than with their classical Poisson bracket.
The Dirac bracket is given in terms of Poisson brackets by 
(see, for example, \cite{Weinberg:mt}) 
\be
\{A,B\}_{\rm D} = \{A,B\}_{\rm P} - \{A,\chi_N\}_{\rm P} \left( C^{-1} \right)^{NM}
	\{\chi_M, B\}_{\rm P}~,
\ee
where
\be
C_{NM} \equiv \{\chi_N,\chi_M\}_{\rm P}\ .
\ee
The indices $N$ and $M$ denote both the spinor index $\alpha$ 
and the constraint label $a = 1,2$.
For Grassmanian fields $A$ and $B$, the Poisson bracket is defined by
\be
\{A,B\}_{\rm P} = -\left( \frac{\partial A}{\partial\psi^\alpha}\frac{\partial B}{\partial \rho_\alpha}
	+ \frac{\partial B}{\partial\psi^\alpha}\frac{\partial A}{\partial \rho_\alpha}\right) 
	-\left( \frac{\partial A}{\partial\psi^{\dagger\alpha} }\frac{\partial B}{\partial \rho^\dagger_\alpha}
	+ \frac{\partial B}{\partial\psi^{\dagger\alpha}}\frac{\partial A}{\partial \rho^\dagger_\alpha}\right)~.
\ee

As an example, the Dirac bracket $\{\rho_\alpha,\rho_\beta\}_{\rm D}$ 
is readily computed (to the order of interest) by noting
that the partial integration in (\ref{IBP}) 
introduces an asymmetry between $\psi$ and $\psi^\dag$
into the system.  Since  $\{\rho_\alpha,\rho_\beta\}_{\rm D}$ contains
\be
\{\rho_\alpha, \chi_{a\gamma} \} = {\cal O}(R^{-2}) \qquad 
	\{\chi_{b\eta},\rho_\beta \} = {\cal O}(R^{-2})~,
\ee
an immediate consequence of this asymmetry is that
$\{\rho_\alpha,\rho_\beta\}_{\rm D}$ vanishes to ${\cal O}(1/R^4)$.   
To compute $\{ \rho_\alpha, \psi_\beta \}_{\rm D}$, we note that 
\be
\{ \rho_\alpha, \chi_{(2\gamma)} \}_{\rm P} & = & -\delta_{\alpha\rho}
			\frac{\partial\chi_{(2\gamma)}}{\partial\psi_\rho} \nn\\
	& = & {\cal O}(R^{-2})~,
\ee
and, to leading order,
\be
(C^{-1})^{(2\gamma)(1\eta)} = -\frac{i}{p_-}\delta_{\gamma\eta} + {\cal O}(R^{-2})~,
\ee
such that
\be
\{ \rho_\alpha, \psi_\beta \}_{\rm D} & = & 
	-\delta_{\alpha\beta} -\frac{i}{p_-}\{ \rho_\alpha, \chi_{(2\beta)}\}_{\rm P}~.
\ee
Similar manipulations are required for $\{ \psi_\alpha, \psi_\beta \}_{\rm D}$,
which does exhibit ${\cal O}(1/R^2)$ corrections.
The second-class constraints on the fermionic sector of the system are removed
by enforcing
\be
\label{Dirac1}
\{ \rho_\alpha(\sigma),\psi_\beta(\sigma') \}_{\rm D} & = & 
	-\delta_{\alpha\beta}\delta(\sigma - \sigma')
	+ \frac{1}{4R^2}\delta(\sigma - \sigma')\biggl\{
	\frac{-i}{p_-}(\rho \Pi)_\alpha\psi_\beta 
	+ \frac{i}{p_-}(\rho\Pi\psi)\delta_{\alpha\beta}
\nn\\
& & 	+ \frac{i}{2p_-}\left[
		\left(\psi\psi'\delta_{\alpha\beta} 
		- \frac{1}{p_-}^2\rho\rho'\delta_{\alpha\beta}\right)
		+ \psi'_\alpha\psi_\beta
		+ \frac{1}{p_-^2}\rho'_\alpha\rho_\beta \right]
\nn\\
& & 	+ \frac{1}{2p_-^2}\left[ (p_A)^2 + ({x'}^A)^2\right]\delta_{\alpha\beta}
	+ y^2\delta_{\alpha\beta} \biggr\}
\nn\\
& & 	-\frac{i}{8p_-R^2}\left( \psi_\alpha\psi_\beta + \frac{1}{p_-^2}\rho_\alpha\rho_\beta\right)
	\frac{\partial}{\partial\sigma'}\delta(\sigma-\sigma')+{\cal O}(R^{-4})
\ee
\be
\label{Dirac2}
\{ \psi_\alpha(\sigma),\psi_\beta(\sigma') \}_{\rm D} & = & 
	\frac{i}{4p_-R^2}\delta(\sigma-\sigma')
	\biggl\{
	(\psi\Pi)_{(\alpha}\psi_{\beta)} 
	-\frac{1}{p_-^2}(p_A{x'}^A)\delta_{(\alpha\beta)}
\nn\\
& & 	+ \frac{1}{2p_-^2}\left[
		\psi'_{(\alpha}\rho_{\beta)} - \rho'_{(\alpha}\psi_{\beta)}
		+ (\psi\rho' + \rho\psi')\delta_{(\alpha\beta)}\right]
	\biggr\}
\nn\\
& & 	+ \frac{i}{8p_-^3R^2}\left( \rho_{(\alpha}\psi_{\beta)} 
	- \psi_{(\alpha}\rho_{\beta)}\right)
	\frac{\partial}{\partial\sigma'}\delta(\sigma-\sigma')+{\cal O}(R^{-4})
\ee
\be
\label{Dirac3}
\{\rho_\alpha(\sigma),\rho_\beta(\sigma')\}_{\rm D} & = & {\cal O}(R^{-4})~.
\ee
Identifying these Dirac brackets with the quantum anticommutators of the
fermionic fields in the theory naturally leads to additional 
${\cal O}(1/R^2)$ corrections to the energy spectrum.  One way to implement
these corrections is to retain the Fourier expansion of $\psi$ and $\psi^\dag$ 
given in (\ref{bdef1},\ref{bdef2}) while transforming the fermionic 
creation and annihilation operators
\be
b_n^\alpha \to c_n^\alpha \qquad\qquad
b_n^{\dag\alpha} \to c_n^{\dag\alpha}\ ,
\ee
such that 
$\{ \rho(c, c^\dag), \psi(c, c^\dag) \}_{\rm P}$,
for example, satisfies (\ref{Dirac1}).  This approach amounts to 
finding ${\cal O}(1/R^2)$ corrections to
$\{ c_n^\alpha , c_m^{\dag\beta} \}$ that allow the
usual anticommutators to be identified with the 
above Dirac brackets (\ref{Dirac1}-\ref{Dirac3}).  
In practice, extracting these solutions from
(\ref{Dirac1}-\ref{Dirac3}) can be circumvented by 
invoking a non-linear field redefinition 
$\psi \rightarrow \tilde\psi,\ \rho \rightarrow \tilde\rho$,
such that 
\be
\{ \rho(c,c^\dag),\psi(c,c^\dag) \}_{\rm P}
= \{ \tilde\rho(b,b^\dag),\tilde\psi(b,b^\dag) \}_{\rm P}~.
\ee
Both representations satisfy (\ref{Dirac1}), and the operators
$b_n^\alpha,b_m^{\dag\beta}$ are understood to obey the usual relations:
\be
\{ b_n^\alpha, b_m^{\dag\beta} \} = \delta^{\alpha\beta}\delta_{nm}~.
\ee
In general, the non-linear field redefinition 
$\tilde \psi(b,b^\dag) = \psi(b,b^\dag) + \dots $ 
contains corrections that are cubic in the 
fields $\rho(b,b^\dag)$, $\psi(b,b^\dag)$, $x^A(a,a^\dag)$ and $p_A(a,a^\dag)$.
Such correction terms can be written down by inspection, with matrix-valued 
coefficients to be solved for by comparing 
$\{ \tilde\rho(b,b^\dag),\tilde\psi(b,b^\dag) \}_{\rm P}$
and $\{ \tilde\psi(b,b^\dag),\tilde\psi(b,b^\dag) \}_{\rm P}$
with (\ref{Dirac1},\ref{Dirac2}).  A straightforward computation yields
\be
\label{redef1}
\rho_\alpha \rightarrow \tilde \rho_\alpha & = & \rho_\alpha \\
\label{redef2}
\psi_\beta \rightarrow \tilde \psi_\beta & = & \psi_\beta 
	+\frac{i}{8p_-R^2}\biggl\{
	(\psi'\psi)\psi_\beta 
	- 2(\rho\Pi\psi)\psi_\beta
	-\frac{1}{p_-^2}(\rho'\rho)\psi_\beta
	+ \frac{2}{p_-^2}(p_A {x'}^A)\rho_\beta
\nn\\
& & 	+\frac{1}{p_-^2}\left[ (\rho'\psi)\rho_\beta - (\rho\psi')\rho_\beta\right]
	+2ip_-\left[ y^2\psi_\beta 
	+ \frac{1}{2p_-^2}\left( (p_A)^2 + ({x'}^A)^2 \right)\psi_\beta\right]
	\biggr\}~.
\nn\\
& & 
\ee
This approach to enforcing the modified Dirac bracket structure amounts 
to adding ${\cal O}(1/R^2)$ correction terms to the Hamiltonian while keeping 
the standard commutation relations. It is much more convenient for calculating
matrix elements than the alternative approach of adding ${\cal O}(1/R^2)$ 
operator corrections to the fermi field anticommutators $\{ b, b^\dag \}$.

By invoking the redefinitions in (\ref{redef1},\ref{redef2}), the pieces
of the interaction Hamiltonian that involve fermions take the final forms
\be
\label{Hpurferm}
{\cal H}_{\rm FF} & = & 
	-\frac{1}{4p_-^3 R^2}\biggl\{
	p_-^2\left[ (\psi'\psi) + \frac{1}{p_-^2}(\rho\rho')\right](\rho\Pi\psi)
	-\frac{p_-^2}{2}(\psi'\psi)^2 - \frac{1}{2p_-^2}(\rho'\rho)^2 
	+ (\psi'\psi)(\rho'\rho)
\nn\\
& & 	+ (\rho\psi')(\rho'\psi)
	-\frac{1}{2}\left[ (\psi\rho')(\psi\rho') + (\psi'\rho)^2\right]
	+ \frac{1}{12}(\psi\gamma^{jk}\rho)(\rho\gamma^{jk}\Pi\rho')
\nn\\
& & 	-\frac{p_-^2}{48}
		\left(\psi\gamma^{jk}\psi - \frac{1}{p_-^2}\rho\gamma^{jk}\rho\right)
	\left(\rho'\gamma^{jk}\Pi\psi - \rho\gamma^{jk}\Pi\psi'\right)
	- (j,k \rightleftharpoons j',k')
	\biggr\}~,
\ee
\be
\label{Hmix}
{\cal H}_{\rm BF} & = & 
	\frac{1}{R^2}\biggl\{
	-\frac{i}{4p_-^2}\left[
	(p_A)^2+({x'}^A)^2 + p_-^2(y^2 - z^2)\right]\left(\psi\psi'-\frac{1}{p_-^2}\rho\rho'\right)
\nn\\
& & 	-\frac{1}{2p_-^3}(p_A{x'}^A)(\rho\psi' + \psi\rho' )
	-\frac{i}{2p_-^2}\left( p_k^2 + {y'}^2 - p_-^2 z^2 \right)\rho\Pi\psi
\nn\\
& & 	+\frac{i}{4}(z'_j z_k)\left(\psi\gamma^{jk}\psi - \frac{1}{p_-^2}\rho\gamma^{jk}\rho\right)
	-\frac{i}{4}(y'_{j'} y_{k'})\left(\psi\gamma^{j'k'}\psi - \frac{1}{p_-^2}\rho\gamma^{j'k'}\rho\right)
\nn\\
& & 	-\frac{i}{8}(z'_k y_{k'} + z_k y'_{k'})
		\left(\psi\gamma^{kk'}\psi - \frac{1}{p_-^2}\rho\gamma^{kk'}\rho\right)
	+\frac{1}{4p_-}(p_k y_{k'} +  z_k p_{k'} )\psi\gamma^{kk'}\rho
\nn\\
& & 	+\frac{1}{4p_-}(p_j z'_k)\left(\psi\gamma^{jk}\Pi\psi 
		+ \frac{1}{p_-^2}\rho\gamma^{jk}\Pi\rho\right)
	-\frac{1}{4p_-}(p_{j'} y'_{k'})\left(\psi\gamma^{j'k'}\Pi\psi 
		+ \frac{1}{p_-^2}\rho\gamma^{j'k'}\Pi\rho\right)
\nn\\
& & 	-\frac{1}{4p_-}(p_k y'_{k'} + z'_k p_{k'})
		\left(\psi\gamma^{kk'}\Pi\psi + \frac{1}{p_-^2}\rho\gamma^{kk'}\Pi\rho\right)
	-\frac{i}{2p_-^2}(p_kp_{k'} - z'_k y'_{k'})\psi\gamma^{kk'}\Pi\rho
	\biggr\}~.
\nn\\
& & 
\ee
The full Hamiltonian is the sum of these two terms plus the bosonic interaction term 
${\cal H}_{\rm BB}$ (\ref{Hpurbos}) and the free Hamiltonian ${\cal H}_{\rm pp}$ (\ref{Hppwave}). 
This system is quantized by imposing the standard (anti)commutator algebra for $x^A,\psi$ 
and their conjugate variables $p^A,\rho$. This will be done by expanding the field variables 
in creation and annihilation operators in a standard way.

Returning to the phenomenon that led us to explore second-class constraints in the first 
place, note that (\ref{Hpurferm}) manifestly vanishes on the subspace of string
zero-modes because all terms have at least one worldsheet spatial derivative.  
The bose-fermi mixing Hamiltonian (\ref{Hmix}) still has terms which can lead to 
curvature corrections to the string zero-mode energies, but their net effect 
vanishes by virtue of non-trivial cancellations between terms that split $SO(4)\times SO(4)$ 
indices and terms that span the entire $SO(8)$. How this comes about will be seen when we 
actually compute matrix elements of this Hamiltonian.

\section{Perturbative analysis of the string energy spectrum}

To compute the energy spectrum correct to first order in ${\cal O}(R^{-1})$, we
will do degenerate first-order perturbation theory on the Fock space of 
eigenstates of the free Hamiltonian ${\cal H}_{\rm pp}$. The
degenerate subspaces of the BMN theory are spanned by fixed numbers of
creation operators with specified mode indices (subject to the 
constraint that the mode indices sum up to zero) acting on the ground 
state $\ket{J}$, where $J=p_-R^2$ is the angular momentum (assumed large)
of the string center of mass in its motion around the 
equator of the $S^5$. In this paper we
restrict attention to ``two-impurity states'' generated by pairs of
creation operators of equal and opposite mode number. For each positive
mode number $n$, the 16 bosonic and fermionic creation operators can be combined 
in pairs to form the following 256 degenerate ``two-impurity'' states:
\be
\label{2impbasis}
a_n^{A\dagger} a_{-n}^{B\dagger} \ket{J} \qquad b_{n}^{\alpha\dagger}
	b_{-n}^{\beta\dagger}\ket{J}
\qquad a_n^{A\dagger} b_{-n}^{\alpha\dagger}\ket{J} \qquad
	a_{-n}^{A\dagger} b_{n}^{\alpha\dagger}\ket{J}~.
\ee
The creation operators are classified under the residual $SO(4)\times SO(4)$ 
symmetry to which the isometry group of the $AdS_5\times S^5$ target
space is broken by the lightcone gauge quantization procedure. 
The bosonic creation operators $a_n^{A\dag}$ decompose as 
$({\mathbf 4},{\mathbf 1})+ ({\mathbf 1},{\mathbf 4})$,
or, in the $SU(2)^2\times SU(2)^2$ notation introduced in
\cite{callan}, as $({\mathbf 2},{\mathbf 2};{\mathbf 1},{\mathbf 1}) + ({\mathbf
1},{\mathbf 1};{\mathbf 2},{\mathbf 2})$. Analogously, the fermionic 
operators $b_n^{\alpha\dag}$ decompose as $({\mathbf 2},{\mathbf
1};{\mathbf 2},{\mathbf 1}) + ({\mathbf 1},{\mathbf 2};{\mathbf1},{\mathbf 2})$ 
under the covering group. It is useful to note that the two fermion irreps are
eigenvectors, with opposite eigenvalue, of the $\Pi$ operator introduced in
(\ref{Pidef}). To find the perturbed energy spectrum, we must compute explicit 
matrix elements of ${\cal H}_{\rm int}$ in this basis and then diagonalize 
the resulting \mbox{$256\times 256$} matrix. We will compare the perturbed 
energy eigenvalues with general expectations from $PSU(2,2|4)$ as well as with
the large ${\cal R}$-charge limit of the anomalous dimensions of gauge theory 
operators with two ${\cal R}$-charge defects.
Higher-impurity string states can be treated in the same way, but we
defer such questions to a separate paper \cite{higher-imp}. 
Our purpose here is primarily to check that our methods (choice of action, 
light-cone gauge reduction, quantization rules, etc.)~are consistent and 
correct. Due to the algebraic complexity met with at each step, this 
check is far from trivial. Once reassured on these fundamental points, 
we can go on to examine a wider range of physically interesting issues.

The first step in carrying out this program is to expand ${\cal H}_{\rm int}$ 
in creation and annihilation operators using (\ref{adef},\ref{bdef1}) for 
$x^A,\psi$ and the related expansions for $p^A,\rho$. As an example, we quote 
the result for $H_{\rm BB}$ (keeping only terms with two creation and two 
annihilation operators):
\begin{eqnarray}
\label{Hcorrected}
{\cal H}_{\rm BB} & = &
    -\frac{1}{32 p_- R^2}\sum \frac{\delta(n+m+l+p)}{\xi}
    \times 
\nn\\
& & \biggl\{
    2 \biggl[ \xi^2 
	- (1 - k_l k_p k_n k_m )
     +  \omega_n \omega_m k_l k_p
      +  \omega_l \omega_p k_n k_m
    + 2 \omega_n \omega_l k_m k_p
\nn\\
& &     + 2 \omega_m \omega_p k_n k_l
    \biggr]
    a_{-n}^{\dagger A}a_{-m}^{\dagger A}a_l^B a_p^B
   +4 \biggl[ \xi^2 
	- (1 - k_l k_p k_n k_m )
     - 2 \omega_n \omega_m k_l k_p
     +  \omega_l \omega_m k_n k_p
\nn\\
& &   -  \omega_n \omega_l k_m k_p
    -  \omega_m \omega_p k_n k_l
    + \omega_n \omega_p k_m k_l \biggr]
    a_{-n}^{\dagger A}a_{-l}^{\dagger B}a_m^A a_p^B
     + 2  \biggl[8 k_l k_p
    a_{-n}^{\dagger i}a_{-l}^{\dagger j}a_m^i a_p^j
\nn\\
& &     + 2 (k_l k_p +k_n k_m)  
	a_{-n}^{\dagger i}a_{-m}^{\dagger i}a_l^j a_p^j
    +(\omega_l \omega_p+ k_l k_p -\omega_n 
	\omega_m- k_n k_m)a_{-n}^{\dagger i}a_{-m}^{\dagger i}a_l^{j'} a_p^{j'}
\nn\\
& &     -4 ( \omega_l \omega_p- k_l k_p)
	a_{-n}^{\dagger i}a_{-l}^{\dagger j'}a_m^i a_p^{j'} 
	-(i,j \rightleftharpoons i',j')
    \biggr]\biggr\}~,
\end{eqnarray}
with $\xi \equiv \sqrt{\omega_n \omega_m \omega_l \omega_p}$.
The expansion of the interaction terms involving fermi fields are too 
complicated to be worth writing down explicitly at this stage. 
Schematically, we organize the two-impurity matrix elements of the 
perturbing Hamiltonian as shown in Table \ref{blockform}.
\begin{table}[ht!]
\begin{eqnarray}
\begin{array}{|c|cccc|}
\hline
 ({\cal H})_{int} & a^{A\dagger}_n a^{B\dagger}_{-n} \ket{J} &
        b^{\alpha\dagger}_n b^{\beta\dagger}_{-n}\ket{J} &
        a^{A\dagger}_n b^{\alpha\dagger}_{-n} \ket{J} &
        a^{A\dagger}_{-n} b^{\alpha\dagger}_{n} \ket{J} \\
        \hline
\bra{J} a^{A}_n a^{B}_{-n} & {\cal H}_{\rm BB} & {\cal H}_{\rm BF} &0&0 \\
\bra{J} b^{\alpha}_n b^{\beta}_{-n} & {\cal H}_{\rm BF} & {\cal H}_{\rm
FF}&0&0\\ \bra{J} a^{A}_n b^{\alpha}_{-n} &0&0& {\cal H}_{\rm BF} & {\cal
H}_{\rm BF} \\ \bra{J} a^{A}_{-n} b^{\alpha}_n & 0 & 0 & {\cal H}_{\rm
BF} & {\cal H}_{\rm BF}\\
\hline
\end{array} \nonumber
\end{eqnarray}
\caption{Structure of the matrix of first-order energy
perturbations in the space of two-impurity string states}
\label{blockform}
\end{table}

To organize the perturbation theory, it is helpful to express everything in
terms of two parameters: $J$ and $\lambda'$. 
In the duality between 
Type IIB superstring theory on $AdS_5\times S^5$ and ${\cal N}=4$
$SU(N_c)$ super-Yang-Mills theory in four dimensions, we identify
\be
{\cal N}=4\ {\rm SYM} & \qquad & AdS_5\times S^5 \nn\\
SU(N_c) & \rightleftharpoons & \int_{S^5}F_5 = N_c \nn\\
g_{\rm YM}^2 N_c  & \rightleftharpoons &  {R^4} \nn\\
g_{\rm YM}^2 & \rightleftharpoons & g_{s} .
\ee
In the pp-wave limit, however, the AdS/CFT dictionary reads
\be
{\cal R} & \rightleftharpoons &  p_- R^2 = J \nn\\
 \frac{{\cal R}^2}{N_c} & \rightleftharpoons & g_{s} p_-^2 = g_2 \nn\\
{\cal R} \rightarrow \infty & \rightleftharpoons & p_- R^2, N_c \rightarrow \infty~.
\ee
The modified 't Hooft coupling
\be
\lambda' = \frac{g_{\rm YM}^2 N_c}{{\cal R}^2} \rightleftharpoons 
	\frac{1}{p_-^2}
\ee
is kept fixed in the ${\cal R},N_c \rightarrow \infty$ limit.  
(We have kept $\alpha' = \mu = 1$.)
Since the 
gauge theory is perturbative in $\lambda = g_{YM}^2 N_c$, 
and $p_-^2$ on the string side is mapped to ${\cal R}^2/(g_{YM}^2 N_c)$, 
we will expand string energies $\omega_q$ 
in powers of $1/p_-$, keeping terms up to some low order to correspond with
the loop expansion in the gauge theory.
This type of dictionary would be incorrect in the original coordinate system
characterized by the light-cone coordinates $t = x^+ - (x^-/2R^2)$ and 
$\phi = x^+ + (x^-/2R^2)$ given in (\ref{rescalePre}).  
In this case, one would calculate corrections
to ${\cal R} \rightleftharpoons p_- R^2$ appearing in the perturbing Hamiltonian 
(which amount to operator-valued corrections to $p_-$).
 
\subsection{Evaluating Fock space matrix elements of ${\cal H}_{\rm BB}$}
We now proceed to the construction of the perturbing Hamiltonian matrix
on the space of degenerate two-impurity states. To convey a sense of
what is involved, we display the matrix elements of ${\cal H}_{\rm BB}$ 
(\ref{Hpurbos}) between the bosonic two-impurity Fock space states:
\begin{eqnarray}
\label{bosonmatrix}
 \Braket{ J | a_n^A a_{-n}^B \left( {\cal H}_{\rm BB} \right)
        a_{-n}^{C \dagger} a_n^{D \dagger} | J } &  = &
        \left( N_{\rm BB}(n^2\lambda') - 2 n^2\lambda'\right)
	\frac{\delta^{ AD}\delta^{ BC}}{J} 
\qquad\qquad\qquad\nn\\
 +  &&  \frac{n^2\lambda'}{J(1+n^2\lambda')}      
	\left[ \delta^{ab}\delta^{cd}
        + \delta^{ad}\delta^{bc} - \delta^{ac}\delta^{bd} \right]
\nn\\
 - && \frac{n^2\lambda'}{J(1+n^2\lambda')}  	 
	\left[ \delta^{a'b'}\delta^{c'd'}
        + \delta^{a'd'}\delta^{b'c'} - \delta^{a'c'}\delta^{b'd'} \right]
\nn\\
 \approx  \left(n_{\rm BB}-2 \right)&&\kern-25pt   
	\frac{n^2\lambda'}{J} \delta^{ AD}\delta^{ BC}
	    + \frac{n^2\lambda'}{J}\left[ \delta^{ab}\delta^{cd}
        + \delta^{ad}\delta^{bc} - \delta^{ac}\delta^{bd} \right]
\nn\\
     - \frac{n^2\lambda'}{J}&&\left[ \delta^{a'b'}\delta^{c'd'}
        + \delta^{a'd'}\delta^{b'c'} - \delta^{a'c'}\delta^{b'd'} \right]
	+ {\cal O}({\lambda'}^{2})\ ,
\end{eqnarray}
where lower-case $SO(4)$ indices $a,b,c,d\in 1,\dots ,4$
indicate that $A,B,C,D$ are chosen from the first $SO(4)$, 
and $a',b',c',d'\in 5,\dots ,8$ indicate the second $SO(4)$ 
$(A,B,C,D \in 5,\dots ,8)$. We have also displayed the further expansion 
of these ${\cal O}(1/J)$ matrix elements in powers of $\lambda'$ (using
the basic BMN-limit energy eigenvalue condition
$\omega_n/p_- = \sqrt{1+\lambda' n^2}$).
This is to facilitate eventual contact with perturbative gauge theory 
via AdS/CFT duality. Note that ${\cal H}_{\rm BB}$ does not mix 
states built out of oscillators from different $SO(4)$ subgroups. 
There is a parallel no-mixing phenomenon in the gauge theory: two-impurity 
bosonic operators carrying spacetime vector indices do not mix with 
spacetime scalar bosonic operators carrying ${\cal R}$-charge vector indices. 

Due to operator ordering ambiguities, two-impurity matrix 
elements of ${\cal H}_{\rm BB}$ can differ by contributions
proportional to $\delta^{AD}\delta^{BC}$, depending on the particular
prescription chosen \cite{callan}. $N_{\rm BB}(n^2\lambda')$ is an 
arbitrary function of $n^2\lambda'$ which is included to account 
for such ambiguities (we will shortly succeed in fixing it).  To
match the dual gauge theory physics, it is best to expand $N_{\rm BB}$ as a 
power series in $\lambda'$. The zeroth-order term must vanish if 
the energy correction is to be perturbative in the gauge coupling. The 
next term in the expansion contributes one arbitrary constant 
(the $n_{\rm BB}$ term) and each higher term in the $\lambda'$ expansion
in principle contributes one additional arbitrary constant to this 
sector of the Hamiltonian. Simple general considerations will
fix them all.

\subsection{Evaluating Fock space matrix elements of ${\cal H}_{\rm FF}$}
The calculation of the two-impurity matrix elements of the parts of
${\cal H}_{\rm int}$ that involve fermionic fields is rather involved and 
we found it necessary to employ symbolic manipulation programs to 
keep track of the many different terms. The end results are fairly 
concise, however. For ${\cal H}_{FF}$ we find
\begin{eqnarray}
\label{fermimatrix}
\Braket{J| b_n^\alpha b_{-n}^\beta\left({\cal H}_{\rm FF}\right)
          b_{-n}^{\gamma\dagger} b_n^{\delta\dagger}|J} & = &
    \left(N_{\rm FF}(n^2\lambda')-2 {n^2\lambda'}\right) 
	\frac{\delta^{\alpha\delta}\delta^{\beta\gamma}}{J} \qquad\qquad\qquad\nn\\
     + \frac{n^2\lambda'}{24 J(1+n^2\lambda')} &&\kern-25pt
\left[ (\gamma^{ij})^{\alpha\delta}(\gamma^{ij})^{\beta\gamma}
        + (\gamma^{ij})^{\alpha\beta}(\gamma^{ij})^{\gamma\delta}
        - (\gamma^{ij})^{\alpha\gamma}(\gamma^{ij})^{\beta\delta} \right]\nn\\
    - \frac{n^2\lambda'}{24 J(1+n^2\lambda')} &&\kern-25pt
    \left[(\gamma^{i'j'})^{\alpha\delta}(\gamma^{i'j'})^{\beta\gamma}
        + (\gamma^{i'j'})^{\alpha\beta}(\gamma^{i'j'})^{\gamma\delta}
-
(\gamma^{i'j'})^{\alpha\gamma}(\gamma^{i'j'})^{\beta\delta}\right]
\nn\\
\approx	\left( n_{\rm FF}-2 \right) 
	&&\kern-25pt\frac{n^2\lambda'}{J}\delta^{\alpha\delta}\delta^{\beta\gamma} 
     	+ \frac{n^2\lambda'}{24 J} 
\left[ (\gamma^{ij})^{\alpha\delta}(\gamma^{ij})^{\beta\gamma}
        + (\gamma^{ij})^{\alpha\beta}(\gamma^{ij})^{\gamma\delta}
        - (\gamma^{ij})^{\alpha\gamma}(\gamma^{ij})^{\beta\delta} \right]
\nn\\
    - \frac{n^2\lambda'}{24 J} &&\kern-25pt
    \left[(\gamma^{i'j'})^{\alpha\delta}(\gamma^{i'j'})^{\beta\gamma}
        + (\gamma^{i'j'})^{\alpha\beta}(\gamma^{i'j'})^{\gamma\delta}
-
(\gamma^{i'j'})^{\alpha\gamma}(\gamma^{i'j'})^{\beta\delta}\right]
	+ {\cal O}({\lambda'}^2)~.
\nn\\ & & 
\end{eqnarray}
This sector has its own normal-ordering function $N_{\rm FF}$, with properties
similar those of $N_{\rm BB}$ described above. The index structure of the 
fermionic matrix elements is similar to that of its bosonic counterpart
(\ref{bosonmatrix}). 

We will now introduce some useful projection operators that
will help us understand the selection rules implicit in the index structure
of (\ref{fermimatrix}).
The original 16-component spinors $\psi$ were reduced to 8 components 
by the Weyl condition $\bar\gamma^9 \psi = \psi$. 
The remaining 8 components are further divided into spinors 
$\tilde\psi$ and $\hat\psi$ which are even or odd under the 
action of $\Pi$:
\be
\Pi \tilde\psi = -\tilde \psi & \qquad & 
	\Pi \tilde b^{\dag \alpha} = -\tilde b^{\dag\alpha} \nn\\
\Pi \hat\psi = \hat\psi & \qquad & \Pi\hat b^{\dag\alpha} = \hat b^{\dag\alpha}~.
\ee
The spinors $\hat \psi$ transform in the $({\bf 1,2; 1,2})$ of
$SO(4)\times SO(4)$, while $\tilde \psi$ transform in the $({\bf 2,1;2,1})$.
This correlation between $\Pi$-parity and $SO(4)\times SO(4)$ representation
will be very helpful for analyzing complicated fermionic matrix elements. 

We denote the $SU(2)$ generators of the active factors of the $({\bf 2,1;2,1})$
irrep as $\Sigma^+$ and $\Omega^+$, where the $\Sigma $ act on the SO(4) 
descended from the $AdS_5$, and the $\Omega$ act on the $SO(4)$ coming from 
the $S^5$. The $({\bf 1,2; 1,2})$ generators are similarly labeled by 
$\Sigma^-$ and $\Omega^-$.  Each set of spinors is annihilated by its 
counterpart set of $SU(2)$  generators:
\be
\label{su2def}
\Sigma^+ \hat b^{\dag\alpha} = \Omega^+ \hat b^{\dag\alpha} & = & 0 \nn\\
\Sigma^- \tilde b^{\dag\alpha} = \Omega^- \tilde b^{\dag\alpha} & = & 0~.
\ee
In terms of the projection operators 
\be
\Pi_+ = \frac{1}{2}(1+\Pi) \qquad \Pi_- = \frac{1}{2}(1-\Pi)\ ,
\ee
which select the disjoint $({\bf 1,2; 1,2}) $ and
$({\bf 2,1; 2,1}) $ irreps, respectively, we have
\be
\Pi_+ \psi = \hat \psi & \qquad & \Pi_+ \hat b^\alpha = \hat b^\alpha \nn\\
\label{PP1}
\Pi_- \psi = \tilde \psi & \qquad & \Pi_- \tilde b^\alpha = \tilde b^\alpha~.
\ee
The $\Pi_\pm$ projections commute with the $SO(4)$ generator matrices 
$\gamma^{ij}, \gamma^{i' j'}$, a fact which implies certain useful 
selection rules for the one-loop limit of (\ref{fermimatrix}). The rules are 
most succinctly stated using an obvious $\pm$ shorthand to indicate the 
representation content of states created by multiple fermionic creation 
operators. In brief, one finds that $++$ states connect only with $++$
and $--$ states connect only with $--$. The only subtle point
is the statement that all $++\to --$ matrix elements of (\ref{fermimatrix})
must vanish: this is the consequence of a simple cancellation between
two terms. This observation will simplify the matrix diagonalization 
we will eventually carry out.

\subsection{Evaluating Fock space matrix elements of ${\cal H}_{\rm BF}$}
The ${\cal H}_{\rm BF}$ sector in the Hamiltonian mediates mixing between
spacetime bosons of the two types (pure boson and bi-fermion) as well as between spacetime 
fermions (which of course contain both bosonic and fermionic oscillator excitations).  
The 64-dimensional boson mixing matrix
\be
\Braket{ J| b_{n}^\alpha b_{-n}^\beta \left( {\cal H}_{\rm BF} \right)
	a_{-n}^{A\dagger} a_{n}^{B\dagger} | J }\ ,
\nn
\ee
is an off-diagonal block in the bosonic sector of the perturbation matrix
in Table \ref{blockform}. The same methods used earlier in this section to
reduce Fock space matrix elements involving fermi fields can be used here
to obtain the simple explicit result (we omit the details) 
\begin{eqnarray}
\label{31}
\Braket{J| b_{n}^\alpha b_{-n}^\beta \left( {\cal H}_{\rm BF} \right)
        a_{-n}^{A\dagger} a_{n}^{B\dagger} |J} & = &
	\frac{n^2 {\lambda'}}{2J(1+n^2\lambda')}
	\biggl\{
	\sqrt{1+n^2\lambda'}\Bigl[
                \left( \gamma^{ab'} \right)^{\alpha\beta}
                - \left( \gamma^{a'b} \right)^{\alpha\beta} \Bigr]
\nn\\
& & 	+~ n\sqrt{ \lambda' }\left[
	\left( \gamma^{a'b'} \right)^{\alpha\beta}
	- \left( \gamma^{ab} \right)^{\alpha\beta}
	+ \left(\delta^{ab} - \delta^{a'b'}\right)
	\delta^{\alpha\beta} \right]
	\biggr\}
\nn\\
& \approx &
	\frac{ n^2 \lambda'}{2 J}\left[
                \left( \gamma^{ab'} \right)^{\alpha\beta}
                - \left( \gamma^{a'b} \right)^{\alpha\beta} \right]
	+{\cal O}({\lambda'}^{3/2})~. 
\end{eqnarray}
The complex conjugate of this matrix element gives the additional off-diagonal 
component of the upper $128\times 128$ block of spacetime bosons.  
We note that terms in the ${\cal H}_{\rm BF}$ sector split the $SO(8)$ group
(manifest in the pp-wave limit) into its
$SO(4)$ constituents such that states of the form 
$a_{-n}^{a'\dagger} a_{n}^{b'\dagger} \ket{J}$, for example, 
which descend strictly from the $S^5$ subspace, vanish in this subsector. 
This behavior is reproduced in the gauge theory, wherein two-boson states that are
either spacetime scalars or scalars of the ${\cal R}$-charge group do not mix with 
bi-fermionic scalars in either irrep.  

The 128-dimensional subsector of spacetime fermions is mixed by 
matrix elements of the same Hamiltonian taken between 
fermionic string states of the general form 
$b_n^{\alpha\dagger} a_{-n}^{A\dagger} \ket{J}$. Our standard methods 
yield the following simple results for the two independent types of 
spacetime fermion mixing matrix elements:
\begin{eqnarray}
\label{22}
\Braket{ J | b_{n}^\alpha a_{-n}^A \left( {\cal H}_{\rm BF} \right)
    b_{n}^{\beta\dagger} a_{-n}^{B\dagger}|J } & = &
	N_{\rm BF}(n^2\lambda')\frac{\delta^{AB}\delta^{\alpha\beta}}{J}
\nn\\
 	+ \frac{n^2\lambda'}{2J(1+n^2\lambda')}\biggl\{&&\kern-30pt
	\left( \gamma^{ab}\right)^{\alpha\beta}
        - \left( \gamma^{a'b'}\right)^{\alpha\beta} 
	- (3+4n^2\lambda') \delta^{ab} \delta^{\alpha\beta}
	- (5+4n^2\lambda')\delta^{a'b'}\delta^{\alpha\beta}
	\biggr\}
\nn\\
   \approx 
	 \frac{n^2 \lambda'}{2J}
	\biggl\{\left( \gamma^{ab}\right)^{\alpha\beta} 
        - \Bigl( &&\kern-30pt  \gamma^{a'b'}  \Bigr)^{\alpha\beta}
    + \left[ (2 n_{\rm BF}-3) \delta^{ab}+ (2 n_{\rm BF}-5) \delta^{a'b'}
	 \right]\delta^{\alpha\beta} \biggr\} 
	+ {\cal O}({\lambda'}^{2})~,
\nn\\
& & 
\end{eqnarray}
\begin{eqnarray}
\label{22a}
\Braket{J | b_{n}^\alpha a_{-n}^A \left({\cal H}_{\rm BF}\right)
        b_{-n}^{\beta\dagger} a_{n}^{B\dagger}|J } & = &
\frac{n^2\lambda'}{2J\sqrt{1+n^2\lambda'}}\biggl\{
	\left( \gamma^{ab}\right)^{\alpha\beta}
        - \left( \gamma^{a'b'}\right)^{\alpha\beta} 
\nn\\
 	-\frac{n {\lambda'}^{1/2}}{\sqrt{1+n^2\lambda'}}
	&&\kern-25pt\Bigl[
	\left( \gamma^{ab'}\right)^{\alpha\beta}
        - \left( \gamma^{a'b}\right)^{\alpha\beta} \Bigr]
	-\delta^{\alpha\beta}
	\left( \delta^{ab} - \delta^{a'b'} \right)
	\biggr\}
\nn\\
  \approx  \frac{n^2 \lambda'}{2 J}
	\biggl\{
         \left( \gamma^{ab}\right)^{\alpha\beta}&&\kern-25pt
        -  \left( \gamma^{a'b'}\right)^{\alpha\beta}
    - \left( \delta^{ab}- \delta^{a'b'} \right)\delta^{\alpha\beta} \biggr\}
	+ {\cal O}({\lambda^\prime}^{3/2})\ .
\end{eqnarray}

Equation (\ref{22}) involves yet another normal-ordering function. Since these
functions have a non-trivial effect on the spectrum, we must give them specific 
values before we can calculate actual numerical eigenvalues. 
The key point is that the structure of the perturbing Hamiltonian implies 
certain relations between all the normal-ordering functions.  
Because the interaction Hamiltonian
is quartic in oscillators, normal-ordering ambiguities give rise to terms
quadratic in oscillators, appearing as constant contributions to the
diagonal matrix elements.  There are
normal-ordering contributions from each sector of the theory: 
${\cal H}_{\rm BB}$ contributes a single term quadratic in
bosonic oscillators; ${\cal H}_{\rm FF}$ yields a term 
quadratic in fermionic oscillators;  ${\cal H}_{\rm BF}$
contributes one term quadratic in bosons and one quadratic in fermions.
The bosonic contributions multiply terms of the form $a^\dag a$,
which are collected into the function $N_{\rm BB}(n^2\lambda')$
with one contribution from ${\cal H}_{\rm BB}$ and one
contribution from ${\cal H}_{\rm BF}$.  Similarly, $N_{\rm FF}(n^2\lambda')$
collects terms multiplying $b^\dag b$, receiving one contribution 
from ${\cal H}_{\rm FF}$ and one contribution from ${\cal H}_{\rm BF}$.
Normal-ordering contributions from both $a^\dag a$ and $b^\dag b$ terms
are non-vanishing in the spacetime fermion subsector; all possible
normal-ordering ambiguities appear in this subspace.  The normal-ordering
function $N_{\rm BF}(n^2\lambda')$ therefore must satisfy
\be
N_{\rm BF}(n^2\lambda') = N_{\rm BB}(n^2\lambda') + N_{\rm FF}(n^2\lambda')\ .
\ee
The normal ordering functions are basically finite renormalizations
which must be adjusted so that the spectrum reflects the $PSU(2,2|4)$ 
global supersymmetry of the classical worldsheet action (a symmetry 
we want to preserve at the quantum level). 

As has been explained 
elsewhere \cite{Beisert:2002tn,callan} (and as we shall shortly
review), energy levels should be organized into multiplets obtained 
by acting on a `highest-weight' level with all possible combinations 
of the eight ${\cal R}$-charge raising supercharges. All the states 
obtained by acting with a total of $L$ supercharges have the same 
energy and we will refer to them as states at level $L$ in the 
supermultiplet. The levels of a multiplet run from $L=0$ to $L=8$.
A careful inspection of the way the normal ordering functions contribute
to the energies of states in the two-impurity sector shows that states
at levels $L=0,8$ are shifted by $N_{\rm BB}$ only. Similarly, levels $L=2,4,6$
are shifted by $N_{\rm FF}$ or $N_{\rm BB}$ and one must have 
$N_{\rm BB} = N_{\rm FF}$ if those levels are to remain internally 
degenerate. Finally, levels $L=1,3,5,7$ are shifted by $N_{\rm BF}$ only. 
By supersymmetry, the level 
spacing must be uniform throughout the supermultiplet and this is only
possible if we also set $N_{\rm BB} = N_{\rm BF}$. But then the constraint 
$N_{\rm BF} = N_{\rm BB} + N_{\rm FF}$ can only be met by setting
$N_{\rm BB} = N_{\rm FF} = N_{\rm BF} = 0$, which then eliminates
any normal-ordering ambiguity from the string theory. This is basically
an exercise in using global symmetry conditions to fix otherwise
undetermined finite renormalizations.

\subsection{Diagonalizing the one-loop perturbation matrix}
We are now ready to diagonalize the perturbing Hamiltonian and
examine whether the resulting energy shifts have the right multiplet structure
and whether the actual eigenvalues match gauge theory expectations.
To simplify the problem, we will begin by diagonalizing the perturbation 
matrix expanded to first nontrivial order 
in both $1/J$ and $\lambda'$. Our results should, by duality, match one-loop
gauge theory calculations and we will eventually return to the problem of
finding the string spectrum correct to higher orders in $\lambda'$.
From the structure of the results just obtained for the perturbation
matrices, we can see that the general structure of the
energy eigenvalues of two-impurity states must be
\be
\label{eigenformula}
E_{\rm int}(n) =  2 + n^2 \lambda'
	\left( 1 + \frac{\Lambda}{J}+ {\cal O}(J^{-2})\right)  + {\cal O}(\lambda'^{2})~,
\ee
where $\Lambda$ is dimensionless and the dependence on $1/J$, $\lambda'$ 
and mode number $n$ is given by (\ref{bosonmatrix},\ref{fermimatrix}).
The eigenvalues $\Lambda$ must meet certain conditions if the requirements of 
$PSU(2,2|4)$ symmetry are to be met, and we will state those conditions before 
solving the eigenvalue problem. 

Note that the eigenvalues in question are lightcone
energies and thus dual to the gauge theory quantity $\Delta=D-J$, the difference
between scaling dimension and ${\cal R}$-charge. Since conformal invariance is part
of the full symmetry group, states are organized into conformal multiplets built 
on conformal primaries. A supermultiplet will contain several conformal primaries 
having the same value of $\Delta$ and transforming into each other under the supercharges. 
All 16 supercharges increment the dimension of an operator by $1/2$, but only 8 
of them (call them $Q_\alpha$) also increment the ${\cal R}$-charge by $1/2$, so as to
leave $\Delta$ unchanged. These 8 supercharges act as `raising operators' on the 
conformal primaries of a supermultiplet: starting from a super-primary of lowest
${\cal R}$-charge, the other conformal primaries are created by acting on it in all 
possible ways with the eight $Q_\alpha$. Primaries obtained by acting with $L$ factors 
of $Q_\alpha$ on the super-primary are said to be at level $L$ in the supermultiplet (since 
the  $Q_\alpha$ anticommute, the range is $L=0$ to $L=8$). The multiplicities of 
states at the various levels are then determined: for every $L=0$ primary, 
there will in general be $C^8_L$ primaries at level $L$ (where $C^n_m$ is the 
binomial coefficient) and a total of $2^8=256$ conformal primaries summed over all $L$. 
If the $L=0$ conformal primary is not a singlet, the total number of conformal primary
states will be a multiple of $256$. Since the number of two-impurity string states
is exactly $256$, we expect the super-primary level to be a singlet (in both spacetime
and the residual $SO(4)$ ${\cal R}$-symmetry) and therefore necessarily a spacetime boson. 
This is the translation into string theory language of Beisert's careful analysis of 
supermultiplets of two-impurity BMN operators in ${\cal N}=4$ super Yang Mills theory 
\cite{Beisert:2002tn}.

These facts severely restrict the quantity $\Lambda$ in the general expression 
(\ref{eigenformula}) above. Although the two-impurity states in question
have the same $J$, they in fact belong to different levels $L$ in 
different supermultiplets. A state of given $L$ is a member of a supermultiplet 
built on a `highest-weight' or super-primary state with ${\cal R}=J-L/2$. 
Since all the primaries in a 
supermultiplet have the same $\Delta$, the joint dependence of eigenvalues on 
$\lambda,J,L$ must be of the form $\Delta(\lambda,J-L/2)$. The only way the 
expansion of (\ref{eigenformula}) can be consistent with this is if $\Lambda=L+c$,
where $c$ is a pure numerical constant (recall that $\lambda'=\lambda/J^2$). 
Successive spacetime boson (or successive spacetime fermion) members of
a supermultiplet must therefore have eigenvalues separated by exactly $2$.
We furthermore know that the multiplicity of the level-$L$ eigenvalue must be
$C^8_L=1,8,28,...,1$ for $L=0,1,2,...,8$. The representation content 
of the different levels under the $SO(3,1)$ spacetime and residual $SO(4)$ ${\cal R}$ 
symmetries can of course also be specified, if desired. Our program, then, is the
following: we will first verify that the quantization procedure preserves the 
$PSU(2,2|4)$ supersymmetry by showing that the eigenvalues $\Lambda$ satisfy 
the integer spacing and multiplicity rules just enumerated; in the process 
we will obtain specific values for $\Lambda$ which we will then compare
with what is known about one-loop gauge theory operator dimensions in order to 
check the gauge theory duality conjecture. The two issues are logically 
disconnected: the quantized string theory should make sense on its own
terms, whether or not it satisfies the more stringent requirements of
duality with four-dimensional gauge theory.

\subsection{Details of the one-loop diagonalization procedure.}
We now confront the problem of explicitly diagonalizing the 
first-order perturbation matrix $\Lambda$ (obtained by expanding the
relevant matrix elements to first order in $\lambda'$). 
The matrix block diagonalizes
on the spacetime boson and spacetime fermion subspaces, as indicated in 
Table~\ref{blockform}. Within these sub-blocks, there are further
block diagonalizations arising from special properties of the one-loop
form of the matrix elements of the perturbing Hamiltonian. For example, 
Fock space states built out of two bosonic creation operators that 
transform only under the internal $SO(4)$ mix only with themselves,
thus providing a $16\times 16$ dimensional diagonal sub-block. Within
such sub-blocks, symmetry considerations are often sufficient to
completely diagonalize the matrix or at least to reduce it to a
low-dimensional diagonalization problem. In short, the problem
reduces almost entirely to that of projecting the matrix elements 
of ${\cal H}_{\rm int}$ on subspaces of the two-impurity Fock space 
defined by various symmetry properties. Determining the $SO(4)\times SO(4)$
symmetry labels of each eigenstate in the diagonalization will furthermore
enable us to precisely match string states with gauge theory 
operators. In this subsection, we record for future reference the 
detailed arguments for the various special
cases that must be dealt with in order to fully diagonalize the
one-loop perturbation and characterize the irrep decomposition. 
Although the projections onto the various
invariant subspaces are matters of simple algebra, that algebra
is too complicated to be done by hand and we have resorted to
symbolic manipulation programs. The end result of the diagonalization is
quite simple and the reader willing to accept our results on faith
can skip ahead to the end of this subsection.
 
We begin with a discussion of the action of the purely bosonic
perturbation ${\cal H}_{\rm BB}$ on the $64$-dimensional Fock space 
created by pairs of bosonic creation operators. Part of this subspace 
connects via ${\cal H}_{\rm BF}$ to the Fock space of spacetime bosons
created by pairs of fermionic creation operators, and we will
deal with it later. There is, however, a subspace that only connects
to itself, through the purely bosonic perturbation ${\cal H}_{\rm BB}$. 
We will first deal with this purely bosonic block diagonalization,
leading to eigenvalues we will denote by $\Lambda_{\rm BB}$.
The 8 bosonic modes lie in the $SO(4) \times SO(4)$ representations 
$({\bf 2,2;1, 1})$ and $({\bf 1,1;2,2})$ (i.e.~they are vectors in
the $SO(4)$ subgroups descended from $AdS_5$ and $S^5$, respectively). The key fact 
about ${\cal H}_{\rm BB}$ is that the 16-dimensional spaces spanned
by two $({\bf 2,2;1, 1})$ oscillators or by two $({\bf 1,1;2,2})$
oscillators are closed under its action (it is also true that
${\cal H}_{\rm BF}$ annihilates both of these subspaces).
The $SO(4)$ representation content of the states created by such 
oscillator pairs is given by the formula $({\bf 2,2})\times ({\bf 2,2}) = 
({\bf 3,3})+({\bf 3,1})+({\bf 1, 3})+({\bf 1,1})$ (we use $SU(2)\times SU(2)$
notation, rather than $SO(4)$, since it is unavoidable when we discuss fermions). 
By projecting the ${\cal O}(\lambda')$ part of (\ref{bosonmatrix}) 
onto these subspaces, one can directly 
read off the eigenvalues $\Lambda_{\rm BB}$, with the 
results shown in Table~\ref{bosonspectrum}. The identification
of the representations associated with particular eigenvalues is
easy to do on the basis of multiplicity. In any event, projection onto
invariant subspaces is a simple matter of symmetrization or
antisymmetrization of oscillator indices and can be done directly.
The most important point to note is that the eigenvalues are
successive even integers, a simple result and one which is consistent
with our expectations from extended supersymmetry. It will
be straightforward to match these states to gauge theory operators
and compare eigenvalues with anomalous dimensions.
\begin{table}[ht!]
\begin{equation}
\begin{array}{|c|c|}\hline
 SO(4)_{AdS}\times  SO(4)_{S^5} & \Lambda_{\rm BB} \\
\hline 
({\bf 1,1;1, 1}) &  {-6  } \\ ({\bf 1,1; 3,3})
&{-2 } \\ ({\bf 1,1;3,1}) +({\bf 1,1;1,3})& {-4  } \\ \hline
\end{array}\qquad
\begin{array}{|c|c|}\hline
 SO(4)_{AdS}\times  SO(4)_{S^5} & \Lambda_{\rm BB} \\
\hline 
  ({\bf 1,1;1, 1}) & {\phantom +}2  \\
 ({\bf 3,3;1, 1}) & {-2 } \\
  ({\bf 3,1;1, 1}) + ({\bf 1,3 ;1, 1})& {\phantom +} 0  \\
\hline
\end{array} \nonumber
\end{equation}
\caption{Energy shifts at ${\cal O}(1/J)$ for unmixed bosonic modes}
\label{bosonspectrum}
\end{table}

The Fock space of spacetime bosons created by pairs of fermionic creation
operators contains a similar pair of $16\times 16$ diagonal sub-blocks.
The construction and application of the relevant projection operators
and the subsequent match-up with gauge theory operators is more complicated 
than on the bosonic side and we must develop some technical tools before 
we can obtain concrete results. 

Just as ${\cal H}_{\rm BB}$ is closed in the two 16-dimensional spaces of bosonic 
$({\bf 1,1;2,2})$ or $({\bf 2,2;1,1})$ states, ${\cal H}_{\rm FF}$ is 
closed on subspaces of bi-fermions spanned by a pair
of $({\bf 1,2;1,2})$ or a pair of $({\bf 2,1;2,1})$ fermionic oscillators
(i.e. $--$ or $++$ states, to use an obvious shorthand).
The complete spectrum of eigenvalues from these subsectors of the Hamiltonian 
can be computed by projecting out the $({\bf 2,1;2,1})$ and $({\bf 1,2;1,2})$
spinors in ${\cal H}_{\rm FF}$ (\ref{fermimatrix}). To do this, it will
be helpful to express the 8-component spinors of the string theory in a 
basis which allows us to define fermionic oscillators labeled by their 
$({\bf 2,1;2,1})$ and $({\bf 1,2;1,2})$ representation content. 

The original 32-component Majorana-Weyl spinors $\theta^I$ were reduced
by the Weyl projection and a light-cone gauge condition to an 8-component 
spinor $\psi^\alpha$ (transforming in the $8_s$ of $SO(8)$). The generators 
of the four $SU(2)$ factors (\ref{su2def}) of the manifest $SO(4)\times SO(4)$ 
symmetry can be expressed as $8\times 8$ $SO(8)$ matrices as follows:
\be
\label{xplctgen}
\Sigma_1^\pm = -\frac{1}{4i}(\gamma^2\gamma^3\pm\gamma^1\gamma^4) & \qquad & 
	\Omega_1^\pm = \frac{1}{4i}(-\gamma^6\gamma^7\pm\gamma^5) \nn\\
\Sigma_2^\pm = -\frac{1}{4i}(\gamma^3\gamma^1\pm\gamma^2\gamma^4) & \qquad & 
	\Omega_2^\pm = \frac{1}{4i}(-\gamma^7\gamma^5\pm\gamma^6) \nn\\
\Sigma_3^\pm = -\frac{1}{4i}(\gamma^1\gamma^2\pm\gamma^3\gamma^4) & \qquad & 
	\Omega_3^\pm = \frac{1}{4i}(-\gamma^5\gamma^6\pm\gamma^7)~.
\ee
We will use the representation for the $\gamma^A$ given in the Appendix 
(\ref{cliffmat}) when we need to make these generators explicit. 
The $8_s$ spinor may be further divided 
into its $({\bf 1,2; 1,2})$ and $({\bf 2,1; 2,1})$ components $\hat \psi$ 
and $\tilde \psi$, respectively, and this suggests a useful basis change
for the string creation operators: for the $({\bf 1,2; 1,2})$ spinor, we 
define four new objects $w,x,y,z$ by 
\be
\hat b^{\dag} = w 
	\left( \begin{array}{c}
	1 \\ 0 \\ 0 \\ -1 \\ 0 \\0 \\0 \\0 
	\end{array} \right) 
+ x \left( \begin{array}{c}
	0 \\ 1\\ 1 \\ 0 \\ 0 \\0 \\0 \\0 
	\end{array} \right)
+ y \left( \begin{array}{c}
	0 \\ 0\\ 0 \\ 0 \\ 1 \\0 \\0 \\1 
	\end{array} \right)
+ z \left( \begin{array}{c}
	0 \\ 1\\1 \\ 0 \\ 0 \\1 \\-1 \\0 
	\end{array} \right)~,
\ee
which we then organize in two different ways into two-component complex spinors:
\be
\label{20spinor}
\zeta = \left( \begin{array}{c}
	w+iy \\ z+ix \end{array} \right) \qquad 
\varphi = \left( \begin{array}{c}
	-z+ix \\ w-iy \end{array} \right)~~
\Leftarrow ~~ \Sigma_i^-
\nn\\
\bar \zeta = \left( \begin{array}{c}
	w+iy \\ -z+ix \end{array} \right) \qquad 
\bar \varphi = \left( \begin{array}{c}
	z+ix \\ w-iy \end{array} \right)~~
\Leftarrow ~~ \Omega_i^-~.
\ee
This organization into 2-spinors is meant to show how components of 
$\hat\psi$ transform under the two $SU(2)$ factors which act non-trivially
on them. As may be verified from the explicit forms of the $SU(2)$ generators 
obtained by substituting (\ref{cliffmat}) into (\ref{xplctgen}), the two-component 
spinors $\zeta$ and $\varphi$ transform as $({\bf 1,2})$ under the first $SO(4)$ 
and the spinors $\bar\zeta$ and $\bar\varphi$ transform as $({\bf 1,2})$ under the
second $SO(4)$ of $SO(4)\times SO(4)$. The explicit realization of the two 
$SU(2)$ factors involved here is found in this way to be
\be
\Sigma_1^- = \left( \begin{array}{cc} 0 & 1/2 \\  1/2 & 0 \end{array} \right) & &  
\Omega_1^- = \left( \begin{array}{cc} 0 & 1/2 \\ 1/2 & 0 \end{array} \right) 
\nn\\ 
\Sigma_2^- = \left( \begin{array}{cc} 0 & i/2 \\ -i/2 & 0 \end{array} \right) & &
\Omega_2^- = \left( \begin{array}{cc} 0 & -i/2 \\ i/2 & 0 \end{array} \right) 
\nn\\
\Sigma_3^- = \left( \begin{array}{cc} 1/2 & 0 \\ 0 & -1/2 \end{array} \right) & &
\Omega_3^- = \left( \begin{array}{cc} 1/2 & 0 \\ 0 & -1/2 \end{array} \right)\ .
\ee

One may similarly decompose $({\bf 2,1; 2,1})$ spinors 
and express the corresponding generators $\Sigma^+$ and $\Omega^+$. We decompose
$\tilde\psi$ into components $\bar w,\bar x,\bar y,\bar z$ according to
\be
\tilde b^\dag = \bar w \left( \begin{array}{c}
	1 \\ 0 \\ 0 \\ 1 \\ 0 \\0 \\0 \\0 
	\end{array} \right)
+ \bar x \left( \begin{array}{c}
	0 \\ 1\\ -1 \\ 0 \\ 0 \\0 \\0 \\0 
	\end{array} \right)
+ \bar y \left( \begin{array}{c}
	0 \\ 0\\ 0 \\ 0 \\ 1 \\0 \\0 \\-1 
	\end{array} \right)
+ \bar z \left( \begin{array}{c}
	0 \\ 0\\0 \\ 0 \\ 0 \\1 \\ 1 \\0
	\end{array}\right)~,
\ee
and rearrange them into two-component complex spinors:
\be
\xi = \left( \begin{array}{c}
	\bar z + i\bar x\\\bar w + i\bar y \end{array} \right) \qquad 
\eta = \left( \begin{array}{c}
	\bar w - i\bar y \\ -\bar z + i\bar x \end{array} \right)~~
\Leftarrow ~~\Sigma_i^+
\nn\\
\bar\xi = \left( \begin{array}{c}
	- \bar z + i\bar x \\ \bar w + i\bar y \end{array} \right) \qquad 
\bar\eta = \left( \begin{array}{c}
	\bar w - i\bar y \\ \bar z + i\bar x \end{array} \right)~~
\Leftarrow~~ \Omega_i^+.
\ee
The corresponding explicit $({\bf 2,1;2,1})$ generators are
given by
\be
\Sigma_1^+ = \left( \begin{array}{cc} 0 & -1/2 \\ -1/2 & 0 \end{array} \right) && 
\Omega_1^+ = \left( \begin{array}{cc} 0 & 1/2 \\ 1/2 & 0 \end{array} \right) 
\nn\\
\Sigma_2^+ = \left( \begin{array}{cc} 0 & i/2 \\ -i/2 & 0 \end{array} \right) &&  
\Omega_2^+ = \left( \begin{array}{cc} 0 & -i/2 \\ i/2 & 0 \end{array} \right) 
\nn\\
\Sigma_3^+ = \left( \begin{array}{cc} 1/2 & 0 \\ 0 & -1/2 \end{array} \right) && 
\Omega_3^+ = \left( \begin{array}{cc} 1/2 & 0 \\ 0 & -1/2 \end{array} \right)~.
\ee
These observations will make it possible to construct linear combinations of
products of components of $\psi^\alpha$ transforming in chosen irreps
of $SO(4)\times SO(4)$.

Let us now use this machinery to analyze the perturbation matrix on
spacetime bosons created by two fermionic creation operators (bi-fermions).
As explained in the discussion of (\ref{fermimatrix}), ${\cal H}_{\rm FF}$  
is block-diagonal on the 16-dimensional $++$ or $--$ bi-fermionic subspaces. 
To project out the $({\bf 2,1;2,1})$ or $++$ block of ${\cal H}_{\rm FF}$, 
we simply act on all indices of (\ref{fermimatrix}) with the $\Pi_+$
projection operator:
\be
\label{HFF--me}
\Braket{ J | \tilde b_n^\alpha  \tilde b_{-n}^\beta \left( {\cal H}_{\rm FF} \right) 
	  \tilde b_{-n}^{\gamma\dagger}  \tilde b_n^{\delta\dagger}| J }   & = &   
	-2 \frac{n^2\lambda'}{J} \Pi_+^{\alpha\delta}\Pi_+^{\beta\gamma}
 	+ \frac{n^2\lambda'}{24 J}\biggl\{
	\biggl[
	(\Pi_+\gamma^{ij}\Pi_+)^{\alpha\delta}(\Pi_+\gamma^{ij}\Pi_+)^{\beta\gamma}
\nn\\
& & 	+ (\Pi_+\gamma^{ij}\Pi_+)^{\alpha\beta}(\Pi_+\gamma^{ij}\Pi_+)^{\gamma\delta}
 	- (\Pi_+\gamma^{ij}\Pi_+)^{\alpha\gamma}(\Pi_+\gamma^{ij}\Pi_+)^{\beta\delta}
	\biggr]
\nn\\
& & 	- \biggl[
	(\Pi_+\gamma^{i'j'}\Pi_+)^{\alpha\delta}(\Pi_+\gamma^{i'j'}\Pi_+)^{\beta\gamma}
	+ (\Pi_+\gamma^{i'j'}\Pi_+)^{\alpha\beta}(\Pi_+\gamma^{i'j'}\Pi_+)^{\gamma\delta}
\nn\\
& & 	- (\Pi_+\gamma^{i'j'}\Pi_+)^{\alpha\gamma}(\Pi_+\gamma^{i'j'}\Pi_+)^{\beta\delta}
	\biggr]\biggr\}~.
\ee
The $SO(4)\times SO(4)$ representation content of this subspace is specified by 
$({\bf 2,1;2,1})\times ({\bf 2,1;2,1}) = ({\bf 1,1;1,1})\oplus ({\bf 1,1;3,1})\oplus
({\bf 3,1;1,1})\oplus ({\bf 3,1;3,1})$ and we must further project onto 
individual irreducible representations in order to identify the eigenvalues.

With the tools we have built up in the last few paragraphs, we are in a position
to directly project out some of the desired irreducible representations. 
Bi-fermions of $++$ type transforming as scalars under the first $SO(4)$ (i.e.
under $\Sigma_i^+$) are constructed by making $SU(2)$ invariants out of the 
two-component spinors $\xi$ and $\eta$. There are four such objects:
\be
\label{su2scalar}
\xi_{-n} \tau_2 \xi_n &\qquad & \xi_{-n} \tau_2 \eta_n \nn\\
\eta_{-n} \tau_2 \xi_n &\qquad & \eta_{-n} \tau_2 \eta_n~,
\ee
where $\tau_2$ is the second Pauli matrix. At the same time, they must
also comprise a {\bf 3} and a {\bf 1} under the second $SO(4)$ (i.e. 
under $\Omega_i^+$). To identify the irreducible linear combinations, 
one has to re-express the objects in (\ref{su2scalar}) in terms of the spinors 
$\bar\xi$ and $\bar\eta$ that transform simply under $\Omega_i^+$. Representative 
results for properly normalized creation operators of $++$ bi-fermion states 
in particular $SO(4)\times SO(4)$ irreps are
\be
\label{adsscalarminus}
\begin{array}{cc}
-\frac{1}{2}\left( \xi_{-n} \tau_2 \eta_n - \eta_{-n} \tau_2 \xi_n \right)
	&  \phantom{\Biggr\}}\ ({\bf 1,1;1,1}) \quad \Lambda_{\rm FF} = -2
\end{array}
\nn\\
\begin{array}{cc}
	\frac{1}{2}\left( \xi_{-n} \tau_2 \eta_n + \eta_{-n} \tau_2 \xi_n \right)
	& \\
	\frac{i}{2}\left( \xi_{-n} \tau_2 \xi_n + \eta_{-n} \tau_2 \eta_n \right)
	& \\
	-\frac{1}{2}\left( \xi_{-n} \tau_2 \xi_n - \eta_{-n} \tau_2 \eta_n \right)
	&  
\end{array}
  \Biggr\}\ ({\bf 1,1;3,1})\quad  \Lambda_{\rm FF} = 0~.
\ee
We simply have to re-express the $\xi,\eta$ bilinears in terms of the original
spinor creation operators $\tilde b$ in order to obtain an explicit projection
of the matrix elements ({\ref{HFF--me}) onto irreducible subspaces and to obtain
the eigenvalues $\Lambda_{FF}$ associated with each irrep. A parallel
analysis of states constructed by forming normalized $SU(2)$ invariants from 
$\bar\xi$ and $\bar\eta$ gives another irrep and eigenvalue:
\be
\label{sscalarminus}
\begin{array}{cc}
	\frac{1}{2}\left( \bar\xi_{-n} \tau_2 \bar\eta_n + \bar\eta_{-n} \tau_2 \bar\xi_n\right)
	&   \\
	\frac{i}{2}\left( \bar\xi_{-n} \tau_2 \bar\xi_n + \bar\eta_{-n} \tau_2 \bar\eta_n \right)
	&   \\
	-\frac{1}{2}\left( \bar\xi_{-n} \tau_2 \bar\xi_n - \bar\eta_{-n} \tau_2 \bar\eta_n \right)
	&
\end{array}
  \Biggr\}\ ({\bf 3,1;1,1}) \qquad  \Lambda_{\rm FF} = -4  ~.
\ee
By similar arguments, whose details we will omit, one can construct the creation
operator for the normalized ${\bf (3,1;3,1)}$ or $++$ bi-fermion and find the
eigenvalue $\Lambda_{FF}= -2$.

An exactly parallel analysis of 
$\Braket{ J | \hat b \hat b ({\cal H}_{\rm FF}) \hat b^\dag \hat b^\dag | J }$
on the 16-dimensional subspace spanned by $({\bf 1,2;1,2})$ bi-fermions
yields the same eigenvalue spectrum.  
The creation operators of irreducible states (built this time out of
$\zeta$ and $\phi$) and their eigenvalues are
\be
\label{adsscalarplus}
\begin{array}{c}
 -\frac{1}{2}\left( \zeta_{-n} \tau_2 \varphi_n - \varphi_{-n} \tau_2 \zeta_n \right)
	
\end{array}
\phantom{ \Biggr\}}\ ({\bf 1,1;1,1})  &\qquad \Lambda_{\rm FF} = -2
\nn\\
\begin{array}{cc}
	\frac{1}{2}\left( \zeta_{-n} \tau_2 \varphi_n + \varphi_{-n} \tau_2 \zeta_n \right)
	&  \\
	\frac{i}{2}\left( \zeta_{-n} \tau_2 \zeta_n + \varphi_{-n} \tau_2 \varphi_n \right)
	& \\
	-\frac{1}{2}\left( \zeta_{-n} \tau_2 \zeta_n - \varphi_{-n} \tau_2 \varphi_n \right)
	&  
\end{array}
\Biggr\}\ ({\bf 1,1;1,3}) &\qquad  \Lambda_{\rm FF} = 0 \\
\label{sscalarplus}
\begin{array}{cc}
\frac{1}{2}\left( \bar\zeta_{-n} \tau_2 \bar\varphi_n + \bar\varphi_{-n} \tau_2 \bar\zeta_n \right)
	&  \\
\frac{i}{2}\left( \bar\zeta_{-n} \tau_2 \bar\zeta_n + \bar\varphi_{-n} \tau_2 \bar\varphi_n \right)
	& \\
-\frac{1}{2}\left( \bar\zeta_{-n} \tau_2 \bar\zeta_n - \bar\varphi_{-n} \tau_2 \bar\varphi_n \right)
	&   
\end{array}
\Biggr\}\ ({\bf 1,3;1,1}) &\qquad  \Lambda_{\rm \rm FF} = -4~.
\ee
The overall results for this sector are displayed in Table \ref{FFmult1}.
\begin{table}[ht!]
\begin{equation}
\begin{array}{|c|c|}\hline
 SO(4)_{AdS}\times SO(4)_{S^5} & \Lambda_{\rm FF} \\
\hline ({\bf 1},{\bf 1};{\bf 1},{\bf 1})& {-2 }
\\ ({\bf 1},{\bf 1};{\bf 3},{\bf 1})& {\phantom +}0 \\
({\bf 3},{\bf 1};{\bf 1},{\bf 1})& {-4 } \\ ({\bf
3},{\bf 1};{\bf 3},{\bf 1})& {-2 } \\ \hline
\end{array} \qquad
\begin{array}{|c|c|}\hline
 SO(4)_{AdS}\times SO(4)_{S^5} & \Lambda_{\rm FF} \\
\hline 
({\bf 1},{\bf 1};{\bf 1},{\bf 1}) & {-2 }
\\ 
({\bf 1},{\bf 1};{\bf 1},{\bf 3}) & {\phantom +}0 \\
({\bf 1},{\bf 3};{\bf 1},{\bf 1}) & {-4 } \\ 
({\bf 1},{\bf 3};{\bf 1},{\bf 3}) & {-2 } \\ \hline
\end{array} \nonumber
\end{equation}
\caption{Energy shifts of states created by two fermions in 
({\bf 2},{\bf 1};{\bf 2},{\bf 1}) or
({\bf 1},{\bf 2};{\bf 1},{\bf 2}) } 
\label{FFmult1}
\end{table}

To this point, we have been able to study specific projections of the ${\cal H}_{\rm BB}$
and ${\cal H}_{\rm FF}$ subsectors by choosing states that are not mixed by ${\cal H}_{\rm BF}$.
We now must deal with the subspace of spacetime boson two-impurity states
that is not annihilated by ${\cal H}_{\rm BF}$. This 64-dimensional
space is spanned by pairs of bosonic creation operators taken from 
different $SO(4)$ subgroups and pairs of fermionic creation operators of opposite
$\Pi$-parity. The representation content of these creation-operator pairs
is such that the states in this sector all belong to $({\bf 2,2;2,2})$ irreps.
This space is of course also acted on by ${\cal H}_{\rm BB}$
and ${\cal H}_{\rm FF}$, so we will need the matrix elements of all three 
pieces of the Hamiltonian as they act on this subspace. By applying
the appropriate projections to the general one-loop matrix elements, 
we obtain the expressions
\be
\label{bosonmatrixP}
\Braket{ J | a_n^A a_{-n}^B \left( {\cal H}_{\rm BB} \right) a_{-n}^{C \dagger} a_n^{D \dagger} | J }   \to 
	-2 \frac{n^2 \lambda'}{J}
	\left(
	\delta^{ a d' }\delta^{ b' c } 
	+\delta^{ a' d }\delta^{ b c' } 
	+\delta^{ a d }\delta^{ b' c' } 
	+\delta^{ a' d' }\delta^{ b c } 
	\right)
\ee
\be
\Braket{ J| b_{n}^\alpha b_{-n}^\beta \left( {\cal H}_{\rm BF} \right)
	a_{-n}^{A\dagger} a_{n}^{B\dagger} | J }  &\to& 
	\frac{n^2 \lambda'}{2 J}\Bigl[ 
		\left( \Pi_+\gamma^{ab'} \Pi_-\right)^{\alpha\beta}
		- \left( \Pi_+\gamma^{a'b} \Pi_- \right)^{\alpha\beta}
		+\left( \Pi_-\gamma^{ab'} \Pi_+\right)^{\alpha\beta}
\nn\\
&&\kern+140pt	- \left( \Pi_-\gamma^{a'b} \Pi_+ \right)^{\alpha\beta}
	\Bigr]~
\ee
\be
\label{fermimatrixP}
&&\kern-25pt\Braket{ J |  b_n^\alpha  b_{-n}^\beta \left( {\cal H}_{\rm FF} \right) 
	  b_{-n}^{\gamma\dagger}  b_n^{\delta\dagger}| J }  \to 
	-2  \frac{n^2\lambda'}{J} \left(
	\Pi_+^{\alpha\delta}\Pi_-^{\beta\gamma}
	+ \Pi_-^{\alpha\delta}\Pi_+^{\beta\gamma}
	\right)	
\nn\\
&&\kern+45pt	+ \frac{n^2\lambda'}{24 J}\biggl\{
	\Bigl[
	(\Pi_+\gamma^{ij}\Pi_+)^{\alpha\delta}(\Pi_-\gamma^{ij}\Pi_-)^{\beta\gamma}
 	+ (\Pi_+\gamma^{ij}\Pi_-)^{\alpha\beta}(\Pi_-\gamma^{ij}\Pi_+)^{\gamma\delta}
\nn\\
&&\kern+45pt	- (\Pi_+\gamma^{ij}\Pi_-)^{\alpha\gamma}(\Pi_-\gamma^{ij}\Pi_+)^{\beta\delta}
	\Bigr]
	- \Bigl[
	(\Pi_+\gamma^{i'j'}\Pi_+)^{\alpha\delta}(\Pi_-\gamma^{i'j'}\Pi_-)^{\beta\gamma}
\nn\\
&&\kern+45pt	+ (\Pi_+\gamma^{i'j'}\Pi_-)^{\alpha\beta}(\Pi_-\gamma^{i'j'}\Pi_+)^{\gamma\delta}
	- (\Pi_+\gamma^{i'j'}\Pi_-)^{\alpha\gamma}(\Pi_-\gamma^{i'j'}\Pi_+)^{\beta\delta}
	\Bigr]
\nn\\
&&\kern+45pt	+ \Bigl[
	(\Pi_-\gamma^{ij}\Pi_-)^{\alpha\delta}(\Pi_+\gamma^{ij}\Pi_+)^{\beta\gamma}
	+ (\Pi_-\gamma^{ij}\Pi_+)^{\alpha\beta}(\Pi_+\gamma^{ij}\Pi_-)^{\gamma\delta}
\nn\\
&&\kern+45pt	- (\Pi_-\gamma^{ij}\Pi_+)^{\alpha\gamma}(\Pi_+\gamma^{ij}\Pi_-)^{\beta\delta}
	\Bigr]
	- \Bigl[
	(\Pi_-\gamma^{i'j'}\Pi_-)^{\alpha\delta}(\Pi_+\gamma^{i'j'}\Pi_+)^{\beta\gamma}
\nn\\
&&\kern+45pt	+ (\Pi_-\gamma^{i'j'}\Pi_+)^{\alpha\beta}(\Pi_+\gamma^{i'j'}\Pi_-)^{\gamma\delta}
	- (\Pi_-\gamma^{i'j'}\Pi_+)^{\alpha\gamma}(\Pi_+\gamma^{i'j'}\Pi_-)^{\beta\delta}
	\Bigr] \biggr\}~.
\ee
Since the 64-dimensional space must contain four copies of the $({\bf 2,2;2,2})$ irrep,
the diagonalization problem is really only $4\times 4$ and quite easy to solve.
The results for the eigenvalues appear in Table \ref{mixspectrum}. 
\begin{table}[ht!]
\begin{equation}
\begin{array}{|c|c|}\hline
 SO(4)_{AdS}\times SO(4)_{S^5} & \Lambda_{\rm BF} \\
\hline ({\bf 2},{\bf 2};{\bf 2},{\bf 2})& -4 \\ ({\bf 2},{\bf
2};{\bf 2},{\bf 2})\times 2 & -2 \\ ({\bf 2},{\bf 2};{\bf
2},{\bf 2})& 0  \\ \hline
\end{array} \nonumber
\end{equation}
\caption{String eigenstates in the subspace for which ${\cal H}_{\rm BF}$ has
non-zero matrix elements} \label{mixspectrum}
\end{table}
Collecting the above results, we present the
complete $SO(4)_{AdS}\times SO(4)_{S^5}$ decomposition of
spacetime boson two-impurity states in Table \ref{specfinal}. 
\begin{table}[ht!]
\begin{equation}
\begin{array}{|c|c|c|}\hline
	& SO(4)_{AdS}\times SO(4)_{S^5}   & 	 \Lambda   \\
\hline
{\cal H}_{\rm BB}  & ({\bf 1},{\bf 1};{\bf 1},{\bf 1})  	 & -6 \\
		&  ({\bf 1},{\bf 1};{\bf 1},{\bf 1})  		& 2 \\
		& ({\bf 1},{\bf 1};{\bf 3},{\bf 1}) +({\bf 1},{\bf 1};{\bf 1},{\bf 3}) 	& -4 \\
		& ({\bf 3},{\bf 1};{\bf 1},{\bf 1}) +({\bf 1},{\bf 3};{\bf 1},{\bf 1})  &  0 \\
		& ({\bf 1},{\bf 1};{\bf 3},{\bf 3}) 		& -2 \\
		& ({\bf 3},{\bf 3};{\bf 1},{\bf 1}) 		& -2 \\
\hline
\end{array} \qquad
\begin{array}{|c|c|c|}\hline
	& SO(4)_{AdS}\times SO(4)_{S^5}   & 	 \Lambda   \\
\hline
{\cal H}_{\rm FF} 	& ({\bf 1},{\bf 1};{\bf 1},{\bf 1}) 	& -2 \\
			& ({\bf 1},{\bf 1};{\bf 1},{\bf 1}) 	& -2 \\
			&  ({\bf 1},{\bf 1};{\bf 3},{\bf 1}) +({\bf 1},{\bf 1};{\bf 1},{\bf 3}) & 0 \\
			&  ({\bf 3},{\bf 1};{\bf 1},{\bf 1}) +({\bf 1},{\bf 3};{\bf 1},{\bf 1}) & -4 \\
			&  ({\bf 3},{\bf 1};{\bf 3},{\bf 1}) +({\bf 1},{\bf 3};{\bf 1},{\bf 3})	& -2 \\
\hline
 {\cal H}_{\rm BF} 	&	({\bf 2},{\bf 2};{\bf 2},{\bf 2}) 	& 0 \\
		&  	({\bf 2},{\bf 2};{\bf 2},{\bf 2}) \times 2  		& -2 \\
			&	({\bf 2},{\bf 2};{\bf 2},{\bf 2})  		& -4 \\ \hline
\end{array} \nn
\end{equation}
\caption{Group decomposition of the 128 two-impurity spacetime bosons}
\label{specfinal}
\end{table}

By projecting out closed subspaces of the one-loop Hamiltonian
we have successfully classified each of the energy levels in the bosonic
Fock space with an $SO(4)\times SO(4)$ symmetry label.  Similar arguments
can be applied to the fermionic Fock space, where two-impurity string states
mix individual bosonic and fermionic oscillators (we omit the details).
A summary of these results for all states, including spacetime fermions, is given in 
Table~\ref{allshifts}. The important fact to note is that the $\Lambda$ eigenvalues
and their multiplicities are exactly as required for consistency with the
full $PSU(2,2|4)$ symmetry of the theory. This is a non-trivial result since the
quantization procedure does not make the full symmetry manifest. It is also a very
satisfying check of the overall correctness of the extremely complicated set
of procedures we were forced to use. We can now proceed to a comparison with
gauge theory anomalous dimensions.

\begin{table}[ht!]
\begin{equation}
\begin{array}{|c|ccccc|}\hline
{\rm Level} & 0 & 2 & 4 & 6 & 8 \\
\hline
{\rm Mult.} & 1 & 28 & 70 & 28 & 1 \\
\hline
\Lambda_{\rm Bose}   & -6 & -4 & -2 & 0 & 2 \\ \hline
\end{array}
\qquad\qquad
\begin{array}{|c|cccc|} \hline
{\rm Level} & 1 & 3 & 5 & 7 \\
\hline
{\rm Mult.} & 8 & 56 & 56 & 8  \\
\hline
\Lambda_{\rm Fermi}   & -5 & -3 & -1 & 1 \\ \hline
\end{array} \nn
\end{equation}
\caption{First-order energy shift summary: complete two-impurity string multiplet} \label{allshifts}
\end{table}

\subsection{Gauge theory comparisons}

The most comprehensive analysis of one-loop anomalous dimensions of
BMN operators and their organization into supersymmetry multiplets
was given in \cite{Beisert:2002tn}. As stated in our previous summary 
publication \cite{callan}, the above string theory calculations are in 
perfect agreement with the one-loop gauge theory predictions. For
completeness, we present a summary of the spectrum of dimensions 
of gauge theory operators along with a sampling of information about
their group transformation properties.

The one-loop formula for operator dimensions takes the generic form
\be
\Delta_n^{\cal R} = 2 + \frac{g_{YM}^2 N_c}{{\cal R}^2} n^2
	\left( 1 + \frac{\bar\Lambda}{{\cal R}} + {\cal O}({\cal R}^{-2}) \right)\ .
\ee
The ${\cal O}({\cal R}^{-1})$ correction $\bar\Lambda$ 
for the set of two-impurity operators is predicted to match 
the corresponding ${\cal O}(J^{-1})$ energy correction to 
two-impurity string states, labeled above by $\Lambda$.
Part of the motivation for performing the special projections on two-impurity
string states detailed above was to emerge with specific symmetry labels for
each of the string eigenstates.  String states of a certain representation content
of the residual $SO(4)\times SO(4)$ symmetry of $AdS_5\times S^5$ are
expected, by duality, to map to gauge theory operators with the same
representation labels in the $SL(2,{\bf C})$ Lorentz and $SU(4)$ 
${\cal R}$-charge sectors of the gauge theory.  Knowing the symmetry content
of the string eigenstates therefore allows us to test this mapping
in detail.

The bosonic sector of the gauge theory, characterized by single-trace operators
with two bosonic insertions in the trace, appears in Table \ref{BBmatch}.
\begin{table}[ht!]
\begin{eqnarray}
\begin{array}{|c|c|c|}\hline
{\rm Operator} &  SO(4)_{AdS}\times SO(4)_{S^5} &\bar\Lambda\\
\hline
\Sigma_A\tr\left(\phi^AZ^p\phi^AZ^{R-p}\right)
 &({\bf 1,1;1, 1}) & -6 \\
\tr\left(\phi^{(i}Z^p\phi^{j)}Z^{R-p}\right)
 &({\bf 1,1; 3,3}) & -2\\
\tr\left(\phi^{[i}Z^p\phi^{j]}Z^{R-p}\right)
 &({\bf 1,1;3,1}) +({\bf 1,1;1,3})&-4\\
\hline
\tr\left(\nabla_\mu Z Z^p\nabla^\mu Z Z^{R-2-p}\right)
 &({\bf 1,1;1, 1})& 2\\
\tr\left(\nabla_{(\mu}ZZ^p\nabla_{\nu)}Z Z^{R-2-p}\right) &
 ({\bf 3,3;1, 1}) & -2\\
\tr\left(\nabla_{[\mu}ZZ^p\nabla_{\nu]}Z Z^{R-2-p}\right) & ({\bf
3,1;1, 1}) + ({\bf 1,3 ;1, 1}) & 0 \\ \hline
\end{array} \nonumber
\end{eqnarray}
\caption{Bosonic gauge theory operators: either spacetime or ${\cal R}$-charge singlet.}\label{BBmatch}
\end{table}
The set of operators comprising Lorentz scalars clearly agree with the
corresponding pure-boson string states in Table~\ref{specfinal} 
which are scalars in $AdS_5$.  Operators containing pairs of spacetime derivatives
correspond to string theory states that are scalars of the $S^5$ subspace.
The bi-fermion sector of the string theory corresponds to the set of two-gluino operators
in the gauge theory.  A few of these operators are listed in Table~\ref{FFmatch}.
\begin{table}[ht!]
\begin{eqnarray}
\begin{array}{|c|c|c|}\hline
{\rm Operator} & SO(4)_{AdS}\times SO(4)_{S^5}&\bar\Lambda\\ \hline
\tr\left(\chi^{[\alpha}Z^p\chi^{\beta]}Z^{R-1-p}\right)
    &({\bf 1},{\bf 1};{\bf 1},{\bf 1})& -2 \\
\tr\left(\chi^{(\alpha}Z^p\chi^{\beta)}Z^{R-1-p}\right)
    &({\bf 1},{\bf 1};{\bf 3},{\bf 1})& 0 \\
\tr\left(\chi[\sigma_{\mu},\tilde\sigma_\nu] Z^p\chi Z^{R-1-p}\right)
    &({\bf 3},{\bf 1};{\bf 1},{\bf 1})& -4 \\
\hline
\end{array} \nonumber
\end{eqnarray}
\caption{Bosonic gauge theory operators with two gluino impurities.} \label{FFmatch}
\end{table}
These states, which form either spacetime or ${\cal R}$-charge scalars,
clearly agree with their string theory counterparts which were constructed
explicitly above. The string states appearing in the $({\bf 2,2;2,2})$ representation
(listed in Table~\ref{mixspectrum}) correspond to the operators listed in 
Table~\ref{BFmatch}.
\begin{table}[ht!]
\begin{eqnarray}
\begin{array}{|c|c|c|}\hline
{\rm Operator} & SO(4)_{AdS}\times SO(4)_{S^5}&\bar\Lambda\\ \hline
\tr\left(\phi^i Z^p \nabla_\mu Z Z^{R-1-p}\right) + \dots
        &({\bf 2},{\bf 2};{\bf 2},{\bf 2})&  -4 \\ \hline
\tr\left(\phi^i Z^p \nabla_\mu Z Z^{R-1-p}\right)
        &({\bf 2},{\bf 2};{\bf 2},{\bf 2})& -2 \\ \hline
\tr\left(\phi^i Z^p \nabla_\mu Z Z^{R-1-p}\right) +\dots
        &({\bf 2},{\bf 2};{\bf 2},{\bf 2})& 0 \\ \hline
\end{array} \nonumber
\end{eqnarray}
\caption{Bosonic gauge theory operators: spacetime and ${\cal R}$-charge non-singlets } \label{BFmatch}
\end{table}
Finally, the complete supermultiplet spectrum of two-impurity gauge theory operators
appears in Table~\ref{smultiplicity}.
\begin{table}[ht!]
\begin{eqnarray}
\begin{array}{|l|l|l|l|l|l|l|l|l|l|}\hline
{\rm Level} & 0& 1& 2& 3& 4& 5& 6& 7& 8 \\ \hline
{\rm Multiplicity} & 1& 8& 28& 56& 70& 56& 28& 8& 1 \\ \hline
\delta E\times ({\cal R}^2/g_{YM}^2N_c n^2) & -{6}/{\cal R} & -{5}/{\cal R} &
-{4}/{\cal R} & -{3}/{\cal R} & -{2}/{\cal R} & -{1}/{\cal R} &
0  & {1}/{\cal R} & {2}/{\cal R} \\ \hline
\end{array} \nonumber
\end{eqnarray}
\caption{Anomalous dimensions of two-impurity operators}
\label{smultiplicity}
\end{table}
The extended supermultiplet spectrum is in perfect agreement with the complete
one-loop string theory spectrum in Table~\ref{allshifts} above.

\section{Energy spectrum at all loops in $\lambda'$} 
To make comparisons with gauge theory dimensions at one loop
in $\lambda = g_{YM}^2 N_c$, we have expanded all string
energies in powers of the modified 't Hooft coupling 
$\lambda' = g_{YM}^2 N_c/{\cal R}^2$. The string theory analysis is exact to all
orders in $\lambda'$, however, and it is possible to extract
a formula for the ${\cal O}(1/J)$ string energy corrections which is 
exact in $\lambda'$ and suitable for comparison with 
higher-order corrections to operator dimensions in the gauge theory.  
In practice, it is slightly more difficult to diagonalize the string
Hamiltonian when the matrix elements are not expanded in 
small $\lambda'$.  This is mainly because, beyond leading order,
${\cal H}_{\rm BF}$ acquires additional terms that mix bosonic
indices in the same $SO(4)$ and also mix bi-fermionic indices in the 
same $({\bf 1},{\bf 2};{\bf 1},{\bf 2})$ or $({\bf 2},{\bf 1};{\bf 2},{\bf 1})$
representation.  Instead of a direct diagonalization of the entire 
128-dimensional subspace of spacetime bosons, for example, we find
it more convenient to exploit the `dimension reduction' that can
be achieved by projecting the full Hamiltonian onto individual
irreps. 

For example, the $({\bf 1,1;1,1})$ irrep appears four times in 
Table~\ref{specfinal} and is present at levels $L=0,4,8$ in the
supermultiplet. To get the exact eigenvalues for this irrep, we 
will have to diagonalize a $4\times 4$ matrix. The basis vectors
of this bosonic sector comprise singlets of the two $SO(4)$ subgroups 
($a^{\dag a} a^{\dag a}\ket{J}$ and $a^{\dag a'}a^{\dag a'}\ket{J}$)
plus two bi-fermion singlets constructed from the $({\bf 2,1;2,1})$ 
and $({\bf 1,2;1,2})$ creation operators 
($\hat b^{\dag\alpha} \hat b^{\dag\alpha}\ket{J}$ and 
$\tilde b^{\dag\alpha}\tilde b^{\dag\alpha}\ket{J}$).
The different Hamiltonian matrix elements that enter the $4\times 4$ 
matrix are symbolically indicated in Table~\ref{singlet}. It is a 
simple matter to project the general expressions for matrix elements 
of ${\cal H}_{\rm BB}$, etc., onto singlet states and so obtain the matrix
as an explicit function of $\lambda', n$. The matrix can be exactly
diagonalized and yields the following energies:
\be
E_0(n,J) & = & 2\sqrt{1 + \lambda'n^2} - \frac{n^2\lambda'}{J}\left[ 2 + \frac{4}{\sqrt{1+n^2\lambda'}}\right]
	+ {\cal O}(1/J^2) 
\nn\\
E_4(n,J) & = & 2\sqrt{1 + \lambda'n^2} - \frac{2 n^2\lambda'}{J} + {\cal O}(1/J^2)
\nn\\
E_8(n,J) & = & 2\sqrt{1 + \lambda'n^2} - \frac{n^2\lambda'}{J}\left[ 2 - \frac{4}{\sqrt{1+n^2\lambda'}}\right]
	+ {\cal O}(1/J^2)~.
\ee
The subscript $L=0,4,8$ indicates the supermultiplet level to which the eigenvalue
connects in the weak coupling limit. The middle eigenvalue (L=4) is doubly
degenerate, as it was in the one-loop limit.

\begin{table}[ht!]
\begin{eqnarray}
\begin{array}{|c|cccc|}\hline
	& a^{\dag a}a^{\dag a}\ket{J} & a^{\dag a'}a^{\dag a'}\ket{J} & 
	\hat b^{\dag\alpha} \hat b^{\dag\alpha}\ket{J} & \tilde b^{\dag\alpha}\tilde b^{\dag\alpha}\ket{J}
\\ \hline
\bra{J}a^{a}a^{a} & {\cal H}_{\rm BB} & {\cal H}_{\rm BB} & {\cal H}_{\rm BF}  & {\cal H}_{\rm BF}
\\ 
\bra{J}a^{a'}a^{a'} & {\cal H}_{\rm BB} & {\cal H}_{\rm BB} & {\cal H}_{\rm BF}  & {\cal H}_{\rm BF}
\\ 
\bra{J}\hat b^{\alpha} \hat b^{\alpha} & {\cal H}_{\rm BF}  & {\cal H}_{\rm BF} & 
		{\cal H}_{\rm FF}  & {\cal H}_{\rm FF}
\\ 
\bra{J}\tilde b^{\alpha} \tilde b^{\alpha} & {\cal H}_{\rm BF}  & {\cal H}_{\rm BF} & 
		{\cal H}_{\rm FF}  & {\cal H}_{\rm FF}
\\ \hline
\end{array} \nn
\end{eqnarray}
\caption{Singlet projection at finite $\lambda'$ }
\label{singlet}
\end{table}

There are two independent $2\times 2$ matrices that mix states at levels 
$L=2,6$. According to Table~\ref{specfinal}, one can project out the 
antisymmetric bosonic and antisymmetric bi-fermionic states in the irrep
$({\bf 1,1;3,1})+({\bf 1,1;1,3})$ or in the irrep 
$({\bf 3,1;1,1})+({\bf 1,3;1,1})$. The results of eqns.~(\ref{adsscalarminus},\ref{sscalarminus},
\ref{adsscalarplus},\ref{sscalarplus}) can be used to carry out the needed 
projections and obtain explicit forms for the matrix elements of the perturbing 
Hamiltonian. The actual $2\times 2$ diagonalization is trivial to do and both
problems give the same result. The final result for the energy levels (using the
same notation as before) is
\be
E_2(n,J) & = & 2\sqrt{1 + \lambda'n^2} - \frac{n^2\lambda'}{J}\left[ 2 + \frac{2}{\sqrt{1+n^2\lambda'}}\right]
	+ {\cal O}(1/J^2) 
\nn\\
E_6(n,J) & = & 2\sqrt{1 + \lambda'n^2} - \frac{n^2\lambda'}{J}\left[ 2 - \frac{2}{\sqrt{1+n^2\lambda'}}\right]
	+ {\cal O}(1/J^2)~.
\ee
We can carry out similar diagonalizations for the remaining irreps of Table~\ref{specfinal},
but no new eigenvalues are encountered: the energies already listed are the exact energies 
of the $L=0,2,4,6,8$ levels. It is also easy to see that the degeneracy structure 
of the exact levels is the same as the one-loop degeneracy.

The odd levels of the supermultiplet are populated by the 128-dimensional 
spacetime fermions, and this sector of the theory can be diagonalized directly. 
Proceeding in a similar fashion as in the bosonic sector, we find exact energy
eigenvalues for the $L=1,3,5,7$ levels (with unchanged multiplicities). We refrain
from stating the individual results because the entire supermultiplet spectrum,
bosonic and fermionic, can be written in terms of a single concise formula:
to leading order in $1/J$ and all orders in $\lambda'$, the energies of the 
two-impurity multiplet are given by
\begin{eqnarray}
\label{stringfinal}
E_L(n,J) & = & 2\sqrt{1+\lambda' n^2} 
	-\frac{n^2\lambda'}{J}\left[
	2+\frac{(4-L)}{\sqrt{1+n^2\lambda'}}\right]+O(1/J^2)~,
\end{eqnarray}
where $L=0,1,\ldots,8$ indicates the level within the supermultiplet.
The degeneracies and irrep content are identical to what we found at
one loop in $\lambda'$. This expression can be rewritten, correct
to order $J^{-2}$, as follows:
\begin{eqnarray}
\label{stringxpnd}
E_L(n,J)  \simeq 2\sqrt{1+\frac{\lambda n^2}{(J-{L}/{2})^2}}
	-\frac{n^2\lambda}{(J-L/2)^3}\left[
	2+\frac{4}{\sqrt{1+{\lambda n^2}/{(J-L/2)^2}}}\right] ~.
\end{eqnarray}
This shows that, within this expansion, the joint dependence on $J$ 
and $L$ is exactly what is required for extended supersymmetry
multiplets. This is a rather nontrivial functional requirement,
and a stringent check on the correctness of our quantization 
procedure (independent of any comparison with gauge theory).

In order to make contact with gauge theory we expand (\ref{stringfinal}) in 
$\lambda'$, obtaining
\begin{eqnarray}
\label{stringfinalexp}
E_L(n,J)  & \approx& \left[ 2 + \lambda' n^2 -
	 \frac{1}{4}(\lambda' n^2)^2 + \frac{1}{8}(\lambda' n^2)^3 
	+\dots \right]  \nonumber\\ 
	&~ & + \frac{1}{J}\left[ n^2\lambda'(L-6)+
	(n^2\lambda')^2\left(\frac{4-L}{2}\right)+
	(n^2\lambda')^3\left(\frac{3L-12}{8}\right) +\ldots\right]~.
\end{eqnarray}

We can now address the comparison with higher-loop results 
on gauge theory operator dimensions. Beisert, Kristjansen and 
Staudacher \cite{Beisert:2003tq} computed the two-loop correction to 
the anomalous dimensions of a convenient class of operators 
lying at level four in the supermultiplet. The operators in question
lie in a symmetric-traceless irrep of an $SO(4)$ subgroup of the
${\cal R}$-charge and are guaranteed by group theory not to mix
with any other fields \cite{Beisert:2003tq}. The following 
expression for the two-loop anomalous dimension was found: 
\begin{eqnarray}
\label{twoloopgauge}
\delta\Delta_n^{\cal R} & = & -\frac{g_{YM}^4 N_c^2}{\pi^4}
	\sin^4\frac{n\pi}{{\cal R}+1}\left(
	\frac{1}{4}+\frac{\cos^2\frac{n\pi}{{\cal R}+1}}{{\cal R}+1}\right)~.
\end{eqnarray} 
As explained above, ${\cal N}=4$ supersymmetry insures that the dimensions 
of operators at other levels of the supermultiplet will be obtained
by making the substitution ${\cal R} \to {\cal R}+2-{L}/{2}$ in the
expression for the dimension of the $L=4$ operator. Making that substitution
and taking the large-${\cal R}$ limit we obtain a general formula for the
two-loop, large-${\cal R}$ correction to the anomalous dimension of the general
two-impurity operator:
\begin{eqnarray}
\label{twoloopgaugeL}
\delta\Delta_n^{{\cal R},L} & = & -\frac{g_{YM}^4 N_c^2}{\pi^4}
	\sin^4\frac{n\pi}{{\cal R}+3-L/2}~\left(
	\frac{1}{4}+\frac{\cos^2\frac{n\pi}{{\cal R}+3-L/2}}
		{{\cal R}+3-L/2}\right)~\nonumber\\
	&\approx &  -\frac{1}{4}(\lambda' n^2)^2 + 
	\frac{1}{2}(\lambda' n^2)^2~ \frac{4-L}{{\cal R}} + O(1/{\cal R}^2) ~, 
\end{eqnarray}
Using the identification ${\cal R}\rightleftharpoons J$ specified by 
duality, we see that this expression matches the corresponding string result
in (\ref{stringfinalexp}) to ${\cal O}(1/J)$, confirming the AdS/CFT
correspondence to two loops in the gauge coupling.  

The three-loop correction to the dimension of this same class of $L=4$ gauge theory 
operators has recently been definitively determined \cite{Beisert:2003ys}. The
calculation involves a remarkable interplay between gauge theory and integrable 
spin chain models
\cite{Beisert:2003jb,Beisert:2003tq,Beisert:2003jj,Klose:2003qc}. The final
result is
\be
\delta\Delta_n^{\cal R} & = & 
	\left(\frac{\lambda}{\pi^2}\right)^3
	\sin^6 \frac{n\pi}{{\cal R}+1}
	\left[
	\frac{1}{8} + \frac{
			\cos^2 \frac{n\pi}{{\cal R}+1}}{4({\cal R}+1)^2}
	\left( 3{\cal R}+2({\cal R}+6)\cos^2 \frac{n\pi}{{\cal R}+1}
	\right)
	\right]\ .
\label{3LP}
\ee
If we apply to this expression the same logic applied to 
the two-loop gauge theory result (\ref{twoloopgauge}), we obtain the 
following three-loop correction to the anomalous dimension of the 
general level of the two-impurity operator supermultiplet:
\begin{eqnarray}
\label{threeloopgaugeL}
\delta\Delta_n^{{\cal R},L} \approx  
 \frac{1}{8}(\lambda' n^2)^3 - 
	\frac{1}{8}(\lambda' n^2)^2~ \frac{8-3L}{{\cal R}} + O(1/{\cal R}^2) ~.
\end{eqnarray}
We see that this expression differs from the third-order contribution 
to the string result (\ref{stringfinalexp}) for the corresponding quantity.
The difference is a constant shift and one might hope to absorb it in a 
normal-ordering constant. However, our discussion of the normal-ordering 
issue earlier in the paper seems to exclude any such freedom.

\section{Discussion and conclusions }

As a complement to the work presented in \cite{callan}, we have given
a detailed account of the quantization of the first curvature correction 
to type IIB superstring theory in the plane-wave limit of $AdS_5\times S^5$. 
We have presented the detailed diagonalization of the resulting perturbing 
Hamiltonian on the degenerate subspace of two-impurity states, obtaining 
string energy corrections that can be compared with higher-loop anomalous
dimensions of gauge theory operators. Beyond the Penrose limit, the 
holographic mapping between each side of the correspondence is 
intricate and nontrivial, and works perfectly to two loops in the gauge 
coupling.  The agreement, however, appears to break down at three loops.  
(A similar three-loop disagreement appeared more recently in a semiclassical 
string analysis presented in \cite{Serban:2004jf}.) This troubling issue was 
first observed in \cite{callan}, at which time the third-order gauge theory 
anomalous dimension was somewhat conjectural. In the intervening time, the 
third-order result (\ref{3LP}) has acquired a solid basis, thus confirming 
the mismatch. Several questions arise about this mismatch: is it due to
a failure of the AdS/CFT correspondence itself, does it signal the need
to modify the worldsheet string action, or is it simply that the
perturbative approach to the gauge theory anomalous dimensions is not
adequate in the relevant limits? Despite vigorous investigation from 
several directions, all these questions remain open. One very promising
development has been the recognition of integrable structures on both sides 
of the duality. They have been the focus of many recent studies, and one may
hope that integrability will be a guide to recovering the correspondence 
beyond two loops or at least achieving some positive understanding of
the mismatch. Certainly the final solution will augment 
our understanding of integrability and of the AdS/CFT mechanism in general.
  
\section*{Acknowledgments}
We would like to thank John Schwarz for helpful guidance and
discussions, and for reading the manuscript.  We also
thank Hok Kong Lee and Xinkai Wu for many useful comments. 
This work was supported in part by US Department of Energy 
grants DE-FG02-91ER40671 (Princeton) and DE-FG03-92-ER40701 (Caltech).

\renewcommand{\thefigure}{A-\arabic{equation}} 
\setcounter{equation}{0}
\appendix 
\section{Notation and conventions}  

The various indices are chosen to represent the following:
\begin{eqnarray}
\mu,\nu,\rho = 0,\dots,9  & \qquad &	SO(9,1)\ {\rm vectors} 
\nonumber \\
\alpha,\beta,\gamma,\delta = 1,\dots,16  & \qquad & SO(9,1)\ {\rm spinors} 
\nonumber \\
A,B = 1,\dots,8 & \qquad & 	SO(8)\ {\rm vectors}
\nonumber \\
i,j,k = 1,\dots,4 & \qquad & 	SO(4)\ {\rm vectors}
\nonumber \\
i',j',k' = 5,\dots,8 & \qquad & 	SO(4)'\ {\rm vectors}
\nonumber \\
a,b = 0,1 	& \qquad & 	{\rm worldsheet\ coordinates}\ (\tau,\sigma)
\nonumber \\
I,J,K,L = 1,2	& \qquad & 	{\rm label\ two\ MW\ spinors\ of\ equal\ chirality}.
\end{eqnarray}

The $32\times 32$ Dirac gamma matrices are decomposed into a $16\times 16$
representation according to
\begin{eqnarray}
(\Gamma^\mu)_{32\times 32} = \left( \begin{array}{cc}
    0   &   \gamma^\mu  \\
    \bar\gamma^\mu &    0
    \end{array}  \right)\
&  \qquad &
\gamma^\mu \bar\gamma^\nu
+ \gamma^\nu \bar\gamma^\mu = 2\eta^{\mu\nu}
\nonumber \\
	\gamma^\mu = (1,\gamma^A, \gamma^9)
&  \qquad  &
	\bar\gamma^\mu = (-1,\gamma^A, \gamma^9)
\nonumber \\
	\gamma^+ = 1+\gamma^9 
& \qquad & 
	\bar\gamma^+ = -1 + \gamma^9~.
\end{eqnarray}
In particular, the notation $\bar \gamma^\mu$ lowers the $SO(9,1)$ spinor indices
$\alpha,\beta$:
\be
\gamma^\mu = (\gamma^\mu)^{\alpha\beta} \qquad 
\bar \gamma^\mu = (\gamma^\mu)_{\alpha\beta}~.
\ee
These conventions are chosen to match those of Metsaev in \cite{Metsaev:2001bj}.
By invoking $\kappa$-symmetry,
\be
\bar\gamma^+\theta = 0 & \Longrightarrow & \bar\gamma^9\theta = \theta \\
\bar\gamma^- = 1+\bar\gamma^9 & \Longrightarrow & \bar\gamma^-\theta = 2\theta~.
\ee
The antisymmetric product $\gamma^{\mu\nu}$ is given by
\be
(\gamma^{\mu\nu})^\alpha_{\phantom{\alpha}\beta} & \equiv &
	\frac{1}{2}(\gamma^\mu \bar\gamma^\nu)^\alpha_{\phantom{\alpha}\beta}
	- (\mu \rightleftharpoons \nu) 
\nonumber \\
(\bar \gamma^{\mu\nu})^\alpha_{\phantom{\alpha}\beta} & \equiv &
	\frac{1}{2}(\bar \gamma^\mu \gamma^\nu)_\alpha^{\phantom{\alpha}\beta}
	- (\mu \rightleftharpoons \nu)~.
\ee
We form the matrices $\Pi$ and $\tilde\Pi$ according to:  
\be
\Pi & \equiv & \gamma^1 \bar\gamma^2 \gamma^3 \bar\gamma^4 
\nn \\
\tilde \Pi & \equiv & \gamma^5 \bar\gamma^6 \gamma^7 \bar\gamma^8\ .
\ee
These form the projection operators $(\Pi^2 = \tilde\Pi^2 = 1)$
\be
\Pi_+ \equiv \frac{1}{2}(1+\Pi) & \qquad & \Pi_- \equiv \frac{1}{2}(1-\Pi) \nn\\
\tilde\Pi_+ \equiv \frac{1}{2}(1+\tilde\Pi) & \qquad & \tilde\Pi_- \equiv \frac{1}{2}(1-\tilde\Pi)~.
\ee

The spinors $\theta^I$ represent two 32-component Majorana-Weyl spinors
of $SO(9,1)$ with equal chirality.  The 32-component Weyl condition is
$\Gamma_{11}\theta = \theta$, with 
\be
\Gamma_{11} = \Gamma^0\dots \Gamma^9 = \left(
\begin{array}{cc}
{ 1} & 0 \\
0 & -{ 1} 
\end{array}  \right)_{32\times 32}~.
\ee
The Weyl condition is used to select the top 16 components of $\theta$ to form 
the 16-component spinors
\begin{eqnarray}
\theta^I & = & \left( {\theta^\alpha \atop 0} \right)^I\ .
\end{eqnarray}
It is useful to form a single complex 16-component spinor $\psi$ from 
the real spinors $\theta^1$ and $\theta^2$:
\be
\psi & = & \sqrt{2} (\theta^1 + i \theta^2 )\ .
\ee
The 16-component Weyl condition $\gamma^9 \theta = \theta $ selects the upper
8 components of $\theta $, with
\be
\gamma^9 = \gamma^1\dots \gamma^8 = \left(
\begin{array}{cc}
{ 1} & 0 \\
0 & -{ 1} 
\end{array}  \right)_{16\times 16}~.
\ee
The 16-component Dirac matrices $\gamma^\mu$ can, in turn, be constructed
from the familiar Spin(8) Clifford algebra, wherein (in terms of $SO(8)$ vector indices)
\be
(\gamma^A)_{16\times 16} =  \left(
	\begin{array}{cc}
	0 & \gamma^A \\
	(\gamma^A)^T & 0 
	\end{array} \right)~,
\ee
and
\be
\left\{ \gamma^A,\gamma^B \right\}_{16\times 16} = 2\delta^{AB} &\qquad &
\left( \gamma^A (\gamma^B)^T + \gamma^B(\gamma^A)^T = 2\delta^{AB}\right)_{8\times 8}~.
\ee
The Spin(8) Clifford algebra may be constructed explicitly in terms of 8 real matrices
\be
\label{cliffmat}
\gamma^1 = \epsilon\times\epsilon\times\epsilon & \qquad & 
	\gamma^5 = \tau_3\times\epsilon\times 1 \nn\\
\gamma^2 = 1\times \tau_1\times\epsilon & \qquad & 
	\gamma^6 = \epsilon\times 1\times\tau_1 \nn\\
\gamma^3 = 1\times \tau_3\times\epsilon & \qquad & 
	\gamma^7 = \epsilon\times 1\times\tau_3 \nn\\
\gamma^4 = \tau_1\times\epsilon\times 1 & \qquad & 
	\gamma^8 = 1\times 1\times 1~,
\ee
with
\be
\epsilon = \left( \begin{array}{cc}
		0 & 1 \\
		-1 & 0 \end{array}\right) \qquad 
\tau_1 = \left( \begin{array}{cc}
		0 & 1 \\
		1 & 0 \end{array}\right) \qquad 
\tau_3 = \left( \begin{array}{cc}
		1 & 0 \\
		0 & -1 \end{array}\right)~.
\ee




\end{document}